%% file: root.tex
\documentclass[11pt,a4paper]{article}
\usepackage{authblk}
\usepackage{a4wide}
\usepackage{latexsym,amsmath}
\usepackage{isabelle,isabellesym}

\usepackage{pdfsetup}

\isabellestyle{it}

\begin{document}

\title{Axiomatizing Category Theory in Free Logic}
\author[1]{Christoph Benzm\"uller}
\author[2]{Dana S. Scott}
\affil[1]{University of Luxemburg, Luxemburg \& Freie Universit\"at Berlin, Germany}
\affil[2]{Visiting Scholar at University of Califormia, Berkeley,  USA}
\date{\today}                     %% if you don't need date to appear
\setcounter{Maxaffil}{0}
\renewcommand\Affilfont{\itshape\small}

% \institue{
% Freie Universit\"at Berlin, Germany\\
% \email{c.benzmueller@fu-berlin.de},\\ 
% \texttt{http://www.christoph-benzmueller.de}
% \and
% Visiting Scholar at University of Califormia, Berkeley,  USA\\
% \email{dana.scott@cs.cmu.edu },\\ 
% \texttt{http://www.cs.cmu.edu/~scott/}
% }
\maketitle

%\tableofcontents

% sane default for proof documents
\parindent 0pt\parskip 0.5ex

% generated text of all theories
\input{session}

% optional bibliography
\bibliographystyle{plain}
\bibliography{root}

\end{document}

%% file: session.tex
\input{AxiomaticCategoryTheory.tex}

%%% Local Variables:
%%% mode: latex
%%% TeX-master: "root"
%%% End:

%% file: AxiomaticCategoryTheory.tex
%
\begin{isabellebody}%
\setisabellecontext{AxiomaticCategoryTheory}%
\isadelimtheory
\endisadelimtheory
\isatagtheory
\endisatagtheory
{\isafoldtheory}%
\isadelimtheory
\endisadelimtheory
\begin{isamarkuptext}%
\begin{abstract}
Starting from a generalization of the standard axioms for a monoid we present a stepwise development 
of various, mutually equivalent foundational axiom systems for category theory. 
Our axiom sets have been formalized in the Isabelle/HOL interactive proof assistant, and this
formalization utilizes a semantically correct embedding of free logic 
in classical higher-order logic. The modeling and formal analysis of our axiom sets has been 
significantly supported  by series of experiments with automated reasoning tools integrated 
with Isabelle/HOL. We also address the relation of our axiom systems to alternative proposals 
from the literature, including an axiom set proposed by Freyd and Scedrov for which we reveal 
a technical issue (when encoded in free logic): either all operations, e.g. morphism composition, 
are total or their axiom system is inconsistent. The repair for this problem is quite 
straightforward, however. 
\end{abstract}%
\end{isamarkuptext}\isamarkuptrue%
\isamarkupsection{Introduction%
}
\isamarkuptrue%
\begin{isamarkuptext}%
We present a stepwise development of axiom systems for category theory by generalizing 
the standard axioms for a monoid to a partial composition operation. Our purpose is not to make or
claim any contribution to category theory but rather to show how formalizations involving the kind 
of logic required (free logic) can be validated within modern proof assistants. 

A total of eight different axiom systems is studied. The systems I-VI are shown to 
be equivalent. The axiom system VII slightly modifies axiom system VI to obtain (modulo 
notational transformation) the set of axioms as proposed by  Freyd and Scedrov in their textbook
 ``Categories, Allegories'' \cite{FreydScedrov90}, published in 1990; 
see also Subsection \ref{subsec:FreydNotation} where we present their original system.
While the axiom systems I-VI are shown to be  consistent, a constricted inconsistency result is 
obtained for system VII (when encoded in free logic where free variables range over all 
objects): We can prove \isa{{\isacharparenleft}{\isasymexists}x{\isachardot}\ \isactrlbold {\isasymnot}{\isacharparenleft}E\ x{\isacharparenright}{\isacharparenright}\ \isactrlbold {\isasymrightarrow}\ False}, where \isa{E} is the existence 
predicate. Read this as: If there are undefined objects, e.g. the value of an undefined composition 
\isa{x{\isasymcdot}y}, then we have falsity.
By contraposition, all objects (and thus all compositions) must exist. But when we assume the latter,
then the axiom system VII essentially reduces categories to monoids.
We note that axiom system V, which avoids this problem, corresponds to a set of axioms proposed 
by Scott \cite{Scott79} in the 1970s. The problem can also be avoided by restricting the variables 
in axiom system VII to range only over existing objects and by postulating strictness conditions. 
This gives us axiom system VIII.

Our exploration has been significantly supported by series of experiments in which automated 
reasoning tools  have been called from within the proof assistant Isabelle/HOL \cite{Isabelle} 
via the Sledgehammer tool \cite{Sledgehammer}. Moreover, we have obtained very useful feedback 
at various stages from the model finder Nitpick \cite{Nitpick} saving us from making several
 mistakes.

At the conceptual level this paper exemplifies a new style of explorative mathematics which rests 
on a significant amount of human-machine interaction with integrated interactive-auto\-ma\-ted 
theorem proving technology. The experiments we have conducted are such that the required 
reasoning is often too tedious and time-consuming for humans to be carried out repeatedly with 
highest level of precision. It is here where cycles of formalization and experimentation efforts in 
Isabelle/HOL provided  significant support. Moreover, the technical inconsistency issue for
axiom system VII was discovered by automated theorem provers, which further emphasises the added 
value of automated theorem proving in this area. 

To enable our experiments we have exploited an embedding of free logic \cite{Scott67} 
in classical higher-order logic, which we have recently presented in a related paper \cite{C57}.

We also want to emphasize that this paper has been written entirely within the Isabelle 
framework by utilizing the Isabelle ``build'' tool; cf. \cite{IsabelleManual2016}, Section~2. 
It is thus an example of a formally verified mathematical document, where the PDF document as 
presented here has been generated directly from the verified source files mentioned above.
We also note that once the proofs have been mechanically checked, they are generally easy 
to find by hand using paper and pencil.%
\end{isamarkuptext}\isamarkuptrue%
\isamarkupsection{Embedding of Free Logic in HOL%
}
\isamarkuptrue%
\begin{isamarkuptext}%
Free logic models partial functions as total functions over a ``raw domain'' \isa{D}. 
A subset \isa{E} of \isa{D} is used to characterize the subdomain of ``existing'' objects;
 cf. \cite{Scott67} for further details.

The experiments presented in the subsequent sections exploit our embedding of free logic in 
HOL \cite{C57}. This embedding is trivial for the standard Boolean connectives. The interesting
aspect is that free logic quantifiers are guarded in the embedding by an explicit existence 
predicate \isa{E} (associated with the subdomain \isa{E} of \isa{D}), so 
that quantified variables range only over existing objects, while free 
variables and arbitrary terms may also denote undefined/non-existing objects outside of \isa{E}. 
This way we obtain an 
elegant treatment of partiality resp. undefinednes as required in category theory. In our related 
paper \cite{C57} we also show how definite description can be appropriately modeled in this
approach. However, the definite description is not required for purposes of this paper, so we omit 
it. Note that the connectives and quantifiers of free logic are displayed below in bold-face fonts. 
Normal,  non-bold-face connectives and quantifiers in contrast belong to the meta-logic HOL. The 
prefix ``f'', e.g. in \isa{fNot}, stands for ``free''.%
\end{isamarkuptext}\isamarkuptrue%
\isacommand{typedecl}\isamarkupfalse%
\ i\ %
\isamarkupcmt{Type for individuals%
}
\isanewline
\isacommand{consts}\isamarkupfalse%
\ fExistence{\isacharcolon}{\isacharcolon}\ {\isachardoublequoteopen}i{\isasymRightarrow}bool{\isachardoublequoteclose}\ {\isacharparenleft}{\isachardoublequoteopen}E{\isachardoublequoteclose}{\isacharparenright}\ %
\isamarkupcmt{Existence/definedness predicate in free logic%
}
\isanewline
\isanewline
\isacommand{abbreviation}\isamarkupfalse%
\ fNot\ {\isacharparenleft}{\isachardoublequoteopen}\isactrlbold {\isasymnot}{\isachardoublequoteclose}{\isacharparenright}\ %
\isamarkupcmt{Free negation%
}
\ \ \ \ \ \ \ \ \ \ \ \ \ \ \ \ \ \ \ \ \ \ \ \ \ \ \isanewline
\ \isakeyword{where}\ {\isachardoublequoteopen}\isactrlbold {\isasymnot}{\isasymphi}\ {\isasymequiv}\ {\isasymnot}{\isasymphi}{\isachardoublequoteclose}\ \ \ \ \ \isanewline
\isacommand{abbreviation}\isamarkupfalse%
\ fImplies\ {\isacharparenleft}\isakeyword{infixr}\ {\isachardoublequoteopen}\isactrlbold {\isasymrightarrow}{\isachardoublequoteclose}\ {\isadigit{1}}{\isadigit{3}}{\isacharparenright}\ %
\isamarkupcmt{Free implication%
}
\ \ \ \ \ \ \ \ \isanewline
\ \isakeyword{where}\ {\isachardoublequoteopen}{\isasymphi}\ \isactrlbold {\isasymrightarrow}\ {\isasympsi}\ {\isasymequiv}\ {\isasymphi}\ {\isasymlongrightarrow}\ {\isasympsi}{\isachardoublequoteclose}\isanewline
\isacommand{abbreviation}\isamarkupfalse%
\ fIdentity\ {\isacharparenleft}\isakeyword{infixr}\ {\isachardoublequoteopen}\isactrlbold {\isacharequal}{\isachardoublequoteclose}\ {\isadigit{1}}{\isadigit{3}}{\isacharparenright}\ %
\isamarkupcmt{Free identity%
}
\ \ \ \ \ \ \ \ \isanewline
\ \isakeyword{where}\ {\isachardoublequoteopen}l\ \isactrlbold {\isacharequal}\ r\ {\isasymequiv}\ l\ {\isacharequal}\ r{\isachardoublequoteclose}\isanewline
\isacommand{abbreviation}\isamarkupfalse%
\ fForall\ {\isacharparenleft}{\isachardoublequoteopen}\isactrlbold {\isasymforall}{\isachardoublequoteclose}{\isacharparenright}\ %
\isamarkupcmt{Free universal quantification guarded by existence 
                                predicate \isa{E}%
}
\ \ \ \ \ \ \ \ \ \ \ \ \ \ \ \ \ \ \isanewline
\ \isakeyword{where}\ {\isachardoublequoteopen}\isactrlbold {\isasymforall}{\isasymPhi}\ {\isasymequiv}\ {\isasymforall}x{\isachardot}\ E\ x\ {\isasymlongrightarrow}\ {\isasymPhi}\ x{\isachardoublequoteclose}\ \ \ \isanewline
\isacommand{abbreviation}\isamarkupfalse%
\ fForallBinder\ {\isacharparenleft}\isakeyword{binder}\ {\isachardoublequoteopen}\isactrlbold {\isasymforall}{\isachardoublequoteclose}\ {\isacharbrackleft}{\isadigit{8}}{\isacharbrackright}\ {\isadigit{9}}{\isacharparenright}\ %
\isamarkupcmt{Binder notation%
}
\ \isanewline
\ \isakeyword{where}\ {\isachardoublequoteopen}\isactrlbold {\isasymforall}x{\isachardot}\ {\isasymphi}\ x\ {\isasymequiv}\ \isactrlbold {\isasymforall}{\isasymphi}{\isachardoublequoteclose}%
\begin{isamarkuptext}%
Further free logic connectives can now be defined as usual.%
\end{isamarkuptext}\isamarkuptrue%
\isacommand{abbreviation}\isamarkupfalse%
\ fOr\ {\isacharparenleft}\isakeyword{infixr}\ {\isachardoublequoteopen}\isactrlbold {\isasymor}{\isachardoublequoteclose}\ {\isadigit{1}}{\isadigit{1}}{\isacharparenright}\ \ \ \ \ \ \ \ \ \ \ \ \ \ \ \ \ \ \ \ \ \ \ \ \ \ \ \ \ \ \ \ \ \isanewline
\ \isakeyword{where}\ {\isachardoublequoteopen}{\isasymphi}\ \isactrlbold {\isasymor}\ {\isasympsi}\ {\isasymequiv}\ {\isacharparenleft}\isactrlbold {\isasymnot}{\isasymphi}{\isacharparenright}\ \isactrlbold {\isasymrightarrow}\ {\isasympsi}{\isachardoublequoteclose}\ \isanewline
\isacommand{abbreviation}\isamarkupfalse%
\ fAnd\ {\isacharparenleft}\isakeyword{infixr}\ {\isachardoublequoteopen}\isactrlbold {\isasymand}{\isachardoublequoteclose}\ {\isadigit{1}}{\isadigit{2}}{\isacharparenright}\ \ \ \ \ \ \ \ \ \ \ \ \ \ \ \ \ \ \ \ \ \ \ \ \ \ \ \ \ \ \ \ \isanewline
\ \isakeyword{where}\ {\isachardoublequoteopen}{\isasymphi}\ \isactrlbold {\isasymand}\ {\isasympsi}\ {\isasymequiv}\ \isactrlbold {\isasymnot}{\isacharparenleft}\isactrlbold {\isasymnot}{\isasymphi}\ \isactrlbold {\isasymor}\ \isactrlbold {\isasymnot}{\isasympsi}{\isacharparenright}{\isachardoublequoteclose}\ \ \ \ \isanewline
\isacommand{abbreviation}\isamarkupfalse%
\ fImplied\ {\isacharparenleft}\isakeyword{infixr}\ {\isachardoublequoteopen}\isactrlbold {\isasymleftarrow}{\isachardoublequoteclose}\ {\isadigit{1}}{\isadigit{3}}{\isacharparenright}\ \ \ \ \ \ \ \isanewline
\ \isakeyword{where}\ {\isachardoublequoteopen}{\isasymphi}\ \isactrlbold {\isasymleftarrow}\ {\isasympsi}\ {\isasymequiv}\ {\isasympsi}\ \isactrlbold {\isasymrightarrow}\ {\isasymphi}{\isachardoublequoteclose}\ \isanewline
\isacommand{abbreviation}\isamarkupfalse%
\ fEquiv\ {\isacharparenleft}\isakeyword{infixr}\ {\isachardoublequoteopen}\isactrlbold {\isasymleftrightarrow}{\isachardoublequoteclose}\ {\isadigit{1}}{\isadigit{5}}{\isacharparenright}\ \ \ \ \ \ \ \ \ \ \ \ \ \ \ \ \ \ \ \ \ \ \ \ \ \ \ \ \ \isanewline
\ \isakeyword{where}\ {\isachardoublequoteopen}{\isasymphi}\ \isactrlbold {\isasymleftrightarrow}\ {\isasympsi}\ {\isasymequiv}\ {\isacharparenleft}{\isasymphi}\ \isactrlbold {\isasymrightarrow}\ {\isasympsi}{\isacharparenright}\ \isactrlbold {\isasymand}\ {\isacharparenleft}{\isasympsi}\ \isactrlbold {\isasymrightarrow}\ {\isasymphi}{\isacharparenright}{\isachardoublequoteclose}\ \ \isanewline
\isacommand{abbreviation}\isamarkupfalse%
\ fExists\ {\isacharparenleft}{\isachardoublequoteopen}\isactrlbold {\isasymexists}{\isachardoublequoteclose}{\isacharparenright}\ \ \ \ \ \ \ \ \ \ \ \ \ \ \ \ \ \ \ \ \ \ \ \ \ \ \ \ \ \ \ \ \ \ \ \ \ \ \ \isanewline
\ \isakeyword{where}\ {\isachardoublequoteopen}\isactrlbold {\isasymexists}{\isasymPhi}\ {\isasymequiv}\ \isactrlbold {\isasymnot}{\isacharparenleft}\isactrlbold {\isasymforall}{\isacharparenleft}{\isasymlambda}y{\isachardot}\ \isactrlbold {\isasymnot}{\isacharparenleft}{\isasymPhi}\ y{\isacharparenright}{\isacharparenright}{\isacharparenright}{\isachardoublequoteclose}\isanewline
\isacommand{abbreviation}\isamarkupfalse%
\ fExistsBinder\ {\isacharparenleft}\isakeyword{binder}\ {\isachardoublequoteopen}\isactrlbold {\isasymexists}{\isachardoublequoteclose}\ {\isacharbrackleft}{\isadigit{8}}{\isacharbrackright}{\isadigit{9}}{\isacharparenright}\ \ \ \ \ \ \ \ \ \ \ \ \ \ \ \ \ \ \ \ \ \isanewline
\ \isakeyword{where}\ {\isachardoublequoteopen}\isactrlbold {\isasymexists}x{\isachardot}\ {\isasymphi}\ x\ {\isasymequiv}\ \isactrlbold {\isasymexists}{\isasymphi}{\isachardoublequoteclose}%
\begin{isamarkuptext}%
In this framework partial and total functions are modelled as follows: 
A function \isa{f} is total if and only if for all \isa{x} we have \isa{E\ x\ \isactrlbold {\isasymrightarrow}\ E{\isacharparenleft}f\ x{\isacharparenright}}. 
For partial functions \isa{f} we may have some \isa{x} such that \isa{E\ x} but not
 \isa{E{\isacharparenleft}f\ x{\isacharparenright}}. A function \isa{f} is strict  if  and only if for all \isa{x} 
we have \isa{E{\isacharparenleft}f\ x{\isacharparenright}\ \isactrlbold {\isasymrightarrow}\ E\ x}.%
\end{isamarkuptext}\isamarkuptrue%
\isamarkupsection{Preliminaries%
}
\isamarkuptrue%
\begin{isamarkuptext}%
Morphisms in the category are objects of type \isa{i}. We introduce three partial 
functions, \isa{dom} (domain), \isa{cod} (codomain), and \isa{{\isasymcdot}} (morphism composition). 
Partiality of composition is handled exactly as expected: we generally may have 
non-existing compositions \isa{x{\isasymcdot}y} (i.e.~\isa{\isactrlbold {\isasymnot}{\isacharparenleft}E{\isacharparenleft}x{\isasymcdot}y{\isacharparenright}{\isacharparenright}}) for some existing  
morphisms \isa{x} and \isa{y} (i.e.~\isa{E\ x} and \isa{E\ y}).%
\end{isamarkuptext}\isamarkuptrue%
\isacommand{consts}\isamarkupfalse%
\ \isanewline
\ domain{\isacharcolon}{\isacharcolon}\ {\isachardoublequoteopen}i{\isasymRightarrow}i{\isachardoublequoteclose}\ {\isacharparenleft}{\isachardoublequoteopen}dom\ {\isacharunderscore}{\isachardoublequoteclose}\ {\isacharbrackleft}{\isadigit{1}}{\isadigit{0}}{\isadigit{8}}{\isacharbrackright}\ {\isadigit{1}}{\isadigit{0}}{\isadigit{9}}{\isacharparenright}\ \isanewline
\ codomain{\isacharcolon}{\isacharcolon}\ {\isachardoublequoteopen}i{\isasymRightarrow}i{\isachardoublequoteclose}\ {\isacharparenleft}{\isachardoublequoteopen}cod\ {\isacharunderscore}{\isachardoublequoteclose}\ {\isacharbrackleft}{\isadigit{1}}{\isadigit{1}}{\isadigit{0}}{\isacharbrackright}\ {\isadigit{1}}{\isadigit{1}}{\isadigit{1}}{\isacharparenright}\ \isanewline
\ composition{\isacharcolon}{\isacharcolon}\ {\isachardoublequoteopen}i{\isasymRightarrow}i{\isasymRightarrow}i{\isachardoublequoteclose}\ {\isacharparenleft}\isakeyword{infix}\ {\isachardoublequoteopen}{\isasymcdot}{\isachardoublequoteclose}\ {\isadigit{1}}{\isadigit{1}}{\isadigit{0}}{\isacharparenright}%
\begin{isamarkuptext}%
For composition \isa{{\isasymcdot}} we assume set-theoretical composition here (i.e., functional 
composition from right to left). This means that
\[\isa{{\isacharparenleft}cod\ x{\isacharparenright}{\isasymcdot}{\isacharparenleft}x{\isasymcdot}{\isacharparenleft}dom\ x{\isacharparenright}{\isacharparenright}\ {\isasymcong}\ x}\] and that \[\isa{{\isacharparenleft}x{\isasymcdot}y{\isacharparenright}a\ {\isasymcong}\ x{\isacharparenleft}y\ a{\isacharparenright}}\quad \text{when}\quad
\isa{dom\ x\ {\isasymsimeq}\ cod\ y}\] 
The equality symbol \isa{{\isasymcong}} denotes Kleene equality and it
is defined as follows (where \isa{{\isacharequal}} is identity on all objects, existing or non-existing, 
of type \isa{i}):%
\end{isamarkuptext}\isamarkuptrue%
\isacommand{abbreviation}\isamarkupfalse%
\ KlEq\ {\isacharparenleft}\isakeyword{infixr}\ {\isachardoublequoteopen}{\isasymcong}{\isachardoublequoteclose}\ {\isadigit{5}}{\isadigit{6}}{\isacharparenright}\ %
\isamarkupcmt{Kleene equality%
}
\isanewline
\ \isakeyword{where}\ {\isachardoublequoteopen}x\ {\isasymcong}\ y\ {\isasymequiv}\ {\isacharparenleft}E\ x\ \isactrlbold {\isasymor}\ E\ y{\isacharparenright}\ \isactrlbold {\isasymrightarrow}\ x\ \isactrlbold {\isacharequal}\ y{\isachardoublequoteclose}%
\begin{isamarkuptext}%
Reasoning tools in Isabelle quickly confirm that \isa{{\isasymcong}} is an equivalence relation. 
But existing identity \isa{{\isasymsimeq}}, in contrast, is only symmetric and transitive, and lacks 
reflexivity. It is defined as:%
\end{isamarkuptext}\isamarkuptrue%
\isacommand{abbreviation}\isamarkupfalse%
\ ExId\ {\isacharparenleft}\isakeyword{infixr}\ {\isachardoublequoteopen}{\isasymsimeq}{\isachardoublequoteclose}\ {\isadigit{5}}{\isadigit{6}}{\isacharparenright}\ %
\isamarkupcmt{Existing identity%
}
\ \ \isanewline
\ \isakeyword{where}\ {\isachardoublequoteopen}x\ {\isasymsimeq}\ y\ {\isasymequiv}\ E\ x\ \isactrlbold {\isasymand}\ E\ y\ \isactrlbold {\isasymand}\ x\ \isactrlbold {\isacharequal}\ y{\isachardoublequoteclose}%
\begin{isamarkuptext}%
We have:%
\end{isamarkuptext}\isamarkuptrue%
\isacommand{lemma}\isamarkupfalse%
\ {\isachardoublequoteopen}x\ {\isasymcong}\ x\ \isactrlbold {\isasymand}\ {\isacharparenleft}x\ {\isasymcong}\ y\ \isactrlbold {\isasymrightarrow}\ y\ {\isasymcong}\ x{\isacharparenright}\ \isactrlbold {\isasymand}\ {\isacharparenleft}{\isacharparenleft}x\ {\isasymcong}\ y\ \isactrlbold {\isasymand}\ y\ {\isasymcong}\ z{\isacharparenright}\ \isactrlbold {\isasymrightarrow}\ x\ {\isasymcong}\ z{\isacharparenright}{\isachardoublequoteclose}\ \isanewline
\isadelimproof
\ \ %
\endisadelimproof
\isatagproof
\isacommand{by}\isamarkupfalse%
\ blast%
\endisatagproof
{\isafoldproof}%
\isadelimproof
\isanewline
\endisadelimproof
\isacommand{lemma}\isamarkupfalse%
\ {\isachardoublequoteopen}x\ {\isasymsimeq}\ x{\isachardoublequoteclose}\ %
\isamarkupcmt{This does not hold; Nitpick finds a countermodel.\footnote{The keyword
  ``oops'' in Isabelle/HOL indicates a failed/incomplete proof attempt; the respective (invalid) 
  conjecture is then not made available for further use. The simplest countermodel for the 
  conjecture given here consists of single, non-existing element. }%
}
\isanewline
\ \ \isacommand{nitpick}\isamarkupfalse%
\ {\isacharbrackleft}user{\isacharunderscore}axioms{\isacharcomma}\ show{\isacharunderscore}all{\isacharcomma}\ format\ {\isacharequal}\ {\isadigit{2}}{\isacharcomma}\ expect\ {\isacharequal}\ genuine{\isacharbrackright}%
\isadelimproof
\ %
\endisadelimproof
\isatagproof
\isacommand{oops}\isamarkupfalse%
\endisatagproof
{\isafoldproof}%
\isadelimproof
\endisadelimproof
\ \ \isanewline
\isacommand{lemma}\isamarkupfalse%
\ {\isachardoublequoteopen}\ {\isacharparenleft}x\ {\isasymsimeq}\ y\ \isactrlbold {\isasymrightarrow}\ y\ {\isasymsimeq}\ x{\isacharparenright}\ \isactrlbold {\isasymand}\ {\isacharparenleft}{\isacharparenleft}x\ {\isasymsimeq}\ y\ \isactrlbold {\isasymand}\ y\ {\isasymsimeq}\ z{\isacharparenright}\ \isactrlbold {\isasymrightarrow}\ x\ {\isasymsimeq}\ z{\isacharparenright}{\isachardoublequoteclose}\ \isanewline
\isadelimproof
\ \ %
\endisadelimproof
\isatagproof
\isacommand{by}\isamarkupfalse%
\ blast%
\endisatagproof
{\isafoldproof}%
\isadelimproof
\isanewline
\endisadelimproof
\isacommand{lemma}\isamarkupfalse%
\ {\isachardoublequoteopen}x\ {\isasymsimeq}\ y\ \isactrlbold {\isasymrightarrow}\ x\ {\isasymcong}\ y{\isachardoublequoteclose}\ \isanewline
\isadelimproof
\ \ %
\endisadelimproof
\isatagproof
\isacommand{by}\isamarkupfalse%
\ simp%
\endisatagproof
{\isafoldproof}%
\isadelimproof
\isanewline
\endisadelimproof
\isacommand{lemma}\isamarkupfalse%
\ {\isachardoublequoteopen}x\ {\isasymsimeq}\ y\ \isactrlbold {\isasymleftarrow}\ x\ {\isasymcong}\ y{\isachardoublequoteclose}\ %
\isamarkupcmt{Nitpick finds a countermodel%
}
\isanewline
\ \ \isacommand{nitpick}\isamarkupfalse%
\ {\isacharbrackleft}user{\isacharunderscore}axioms{\isacharcomma}\ show{\isacharunderscore}all{\isacharcomma}\ format\ {\isacharequal}\ {\isadigit{2}}{\isacharcomma}\ expect\ {\isacharequal}\ genuine{\isacharbrackright}%
\isadelimproof
\ %
\endisadelimproof
\isatagproof
\isacommand{oops}\isamarkupfalse%
\endisatagproof
{\isafoldproof}%
\isadelimproof
\endisadelimproof
\begin{isamarkuptext}%
Next, we define the identity morphism predicate \isa{I} as follows:%
\end{isamarkuptext}\isamarkuptrue%
\isacommand{abbreviation}\isamarkupfalse%
\ I\ \isakeyword{where}\ {\isachardoublequoteopen}I\ i\ {\isasymequiv}\ {\isacharparenleft}\isactrlbold {\isasymforall}x{\isachardot}\ E{\isacharparenleft}i{\isasymcdot}x{\isacharparenright}\ \isactrlbold {\isasymrightarrow}\ i{\isasymcdot}x\ {\isasymcong}\ x{\isacharparenright}\ \isactrlbold {\isasymand}\ {\isacharparenleft}\isactrlbold {\isasymforall}x{\isachardot}\ E{\isacharparenleft}x{\isasymcdot}i{\isacharparenright}\ \isactrlbold {\isasymrightarrow}\ x{\isasymcdot}i\ {\isasymcong}\ x{\isacharparenright}{\isachardoublequoteclose}%
\begin{isamarkuptext}%
This definition was suggested by an exercise in \cite{FreydScedrov90} on p.~4.
In earlier experiments we used a longer definition which can be proved equivalent on the basis
of the other axioms. For monoids, where composition is total, \isa{I\ i} means \isa{i} is
a two-sided identity — and such are unique. For categories the property is much weaker.%
\end{isamarkuptext}\isamarkuptrue%
\isamarkupsection{Axiom Set I%
}
\isamarkuptrue%
\begin{isamarkuptext}%
Axiom Set I is our most basic axiom set for category theory generalizing the 
axioms for a monoid to a partial composition operation. Remember that a monoid is an 
algebraic structure $(S,\circ)$, where $\circ$ is a binary operator on set $S$, 
satisfying the following properties:

\begin{tabular}{ll}
Closure: & $ \forall a,b \in S.\ a \circ b \in S$ \\
Associativity: & $\forall a,b,c \in S.\ a \circ (b \circ c) = (a \circ b) \circ c$ \\
Identity: & $\exists id_S \in S. \forall a\in S.\ id_S\circ a = a = a \circ id_S$
\end{tabular}

That is, a monoid is a semigroup with a two-sided identity element.

Our first axiom set for category theory employs a partial, strict binary 
composition operation \isa{{\isasymcdot}}, and the existence of left and right identity elements is 
addressed in the last two axioms. The notions of \isa{dom} (Domain) and \isa{cod} (Codomain)
abstract from their common meaning in the context of sets. In category theory we
work with just a single type of objects (the type \isa{i} of morphisms) and therefore 
identity morphisms are employed to suitably characterize their meanings.%
\end{isamarkuptext}\isamarkuptrue%
\ S\isactrlsub i{\isacharcolon}\ %
\isamarkupcmt{\makebox[2cm][l]{Strictness:}%
}
\ {\isachardoublequoteopen}E{\isacharparenleft}x{\isasymcdot}y{\isacharparenright}\ \isactrlbold {\isasymrightarrow}\ {\isacharparenleft}E\ x\ \isactrlbold {\isasymand}\ E\ y{\isacharparenright}{\isachardoublequoteclose}\ \isakeyword{and}\isanewline
\ E\isactrlsub i{\isacharcolon}\ %
\isamarkupcmt{\makebox[2cm][l]{Existence:}%
}
\ {\isachardoublequoteopen}E{\isacharparenleft}x{\isasymcdot}y{\isacharparenright}\ \isactrlbold {\isasymleftarrow}\ {\isacharparenleft}E\ x\ \isactrlbold {\isasymand}\ E\ y\ \isactrlbold {\isasymand}\ {\isacharparenleft}\isactrlbold {\isasymexists}z{\isachardot}\ z{\isasymcdot}z\ {\isasymcong}\ z\ \isactrlbold {\isasymand}\ x{\isasymcdot}z\ {\isasymcong}\ x\ \isactrlbold {\isasymand}\ z{\isasymcdot}y\ {\isasymcong}\ y{\isacharparenright}{\isacharparenright}{\isachardoublequoteclose}\ \isakeyword{and}\isanewline
\ A\isactrlsub i{\isacharcolon}\ %
\isamarkupcmt{\makebox[2cm][l]{Associativity:}%
}
\ {\isachardoublequoteopen}x{\isasymcdot}{\isacharparenleft}y{\isasymcdot}z{\isacharparenright}\ {\isasymcong}\ {\isacharparenleft}x{\isasymcdot}y{\isacharparenright}{\isasymcdot}z{\isachardoublequoteclose}\ \isakeyword{and}\isanewline
\ C\isactrlsub i{\isacharcolon}\ %
\isamarkupcmt{\makebox[2cm][l]{Codomain:}%
}
\ {\isachardoublequoteopen}\isactrlbold {\isasymforall}y{\isachardot}\isactrlbold {\isasymexists}i{\isachardot}\ I\ i\ \isactrlbold {\isasymand}\ i{\isasymcdot}y\ {\isasymcong}\ y{\isachardoublequoteclose}\ \isakeyword{and}\isanewline
\ D\isactrlsub i{\isacharcolon}\ %
\isamarkupcmt{\makebox[2cm][l]{Domain:}%
}
\ {\isachardoublequoteopen}\isactrlbold {\isasymforall}x{\isachardot}\isactrlbold {\isasymexists}j{\isachardot}\ I\ j\ \isactrlbold {\isasymand}\ x{\isasymcdot}j\ {\isasymcong}\ x{\isachardoublequoteclose}%
\begin{isamarkuptext}%
Nitpick confirms that this axiom set is consistent.%
\end{isamarkuptext}\isamarkuptrue%
\ \ \isacommand{lemma}\isamarkupfalse%
\ True\ \ %
\isamarkupcmt{Nitpick finds a model%
}
\isanewline
\ \ \ \ \isacommand{nitpick}\isamarkupfalse%
\ {\isacharbrackleft}satisfy{\isacharcomma}\ user{\isacharunderscore}axioms{\isacharcomma}\ show{\isacharunderscore}all{\isacharcomma}\ format\ {\isacharequal}\ {\isadigit{2}}{\isacharcomma}\ expect\ {\isacharequal}\ genuine{\isacharbrackright}%
\isadelimproof
\ %
\endisadelimproof
\isatagproof
\isacommand{oops}\isamarkupfalse%
\endisatagproof
{\isafoldproof}%
\isadelimproof
\endisadelimproof
\begin{isamarkuptext}%
Even if we assume there are non-existing objects we get consistency (which is e.g. not the
case for Axiom Set VII below).%
\end{isamarkuptext}\isamarkuptrue%
\isacommand{lemma}\isamarkupfalse%
\ \isakeyword{assumes}\ {\isachardoublequoteopen}{\isasymexists}x{\isachardot}\ \isactrlbold {\isasymnot}{\isacharparenleft}E\ x{\isacharparenright}{\isachardoublequoteclose}\ \isakeyword{shows}\ True\ \ %
\isamarkupcmt{Nitpick finds a model\footnote{To display the models 
 or countermodels from Nitpick in the Isabelle/HOL system interface simply put the mouse on the 
 expression "nitpick".}%
}
\ \isanewline
\ \ \ \ \isacommand{nitpick}\isamarkupfalse%
\ {\isacharbrackleft}satisfy{\isacharcomma}\ user{\isacharunderscore}axioms{\isacharcomma}\ show{\isacharunderscore}all{\isacharcomma}\ format\ {\isacharequal}\ {\isadigit{2}}{\isacharcomma}\ expect\ {\isacharequal}\ genuine{\isacharbrackright}%
\isadelimproof
\ %
\endisadelimproof
\isatagproof
\isacommand{oops}\isamarkupfalse%
\endisatagproof
{\isafoldproof}%
\isadelimproof
\endisadelimproof
\begin{isamarkuptext}%
We may also assume an existing and a non-existing object and still get consistency.%
\end{isamarkuptext}\isamarkuptrue%
\ \ \isacommand{lemma}\isamarkupfalse%
\ \isakeyword{assumes}\ {\isachardoublequoteopen}{\isacharparenleft}{\isasymexists}x{\isachardot}\ \isactrlbold {\isasymnot}{\isacharparenleft}E\ x{\isacharparenright}{\isacharparenright}\ {\isasymand}\ {\isacharparenleft}{\isasymexists}x{\isachardot}\ {\isacharparenleft}E\ x{\isacharparenright}{\isacharparenright}{\isachardoublequoteclose}\ \isakeyword{shows}\ True\ \ %
\isamarkupcmt{Nitpick finds a model%
}
\ \isanewline
\ \ \ \ \isacommand{nitpick}\isamarkupfalse%
\ {\isacharbrackleft}satisfy{\isacharcomma}\ user{\isacharunderscore}axioms{\isacharcomma}\ show{\isacharunderscore}all{\isacharcomma}\ format\ {\isacharequal}\ {\isadigit{2}}{\isacharcomma}\ expect\ {\isacharequal}\ genuine{\isacharbrackright}%
\isadelimproof
\ %
\endisadelimproof
\isatagproof
\isacommand{oops}\isamarkupfalse%
\endisatagproof
{\isafoldproof}%
\isadelimproof
\endisadelimproof
\begin{isamarkuptext}%
The left-to-right direction of existence axiom \isa{E\isactrlsub i} is implied.%
\end{isamarkuptext}\isamarkuptrue%
\ \ \isacommand{lemma}\isamarkupfalse%
\ E\isactrlsub iImplied{\isacharcolon}\ {\isachardoublequoteopen}E{\isacharparenleft}x{\isasymcdot}y{\isacharparenright}\ \isactrlbold {\isasymrightarrow}\ {\isacharparenleft}E\ x\ \isactrlbold {\isasymand}\ E\ y\ \isactrlbold {\isasymand}\ {\isacharparenleft}\isactrlbold {\isasymexists}z{\isachardot}\ z{\isasymcdot}z\ {\isasymcong}\ z\ \isactrlbold {\isasymand}\ x{\isasymcdot}z\ {\isasymcong}\ x\ \isactrlbold {\isasymand}\ z{\isasymcdot}y\ {\isasymcong}\ y{\isacharparenright}{\isacharparenright}{\isachardoublequoteclose}\ \isanewline
\isadelimproof
\ \ \ \ %
\endisadelimproof
\isatagproof
\isacommand{by}\isamarkupfalse%
\ {\isacharparenleft}metis\ A\isactrlsub i\ C\isactrlsub i\ S\isactrlsub i{\isacharparenright}%
\endisatagproof
{\isafoldproof}%
\isadelimproof
\endisadelimproof
\begin{isamarkuptext}%
We can prove that the \isa{i} in axiom \isa{C\isactrlsub i} is unique. The proofs can be 
   found automatically by Sledgehammer.\footnote{In our initial experiments proof reconstruction 
   of the external ATP proofs failed in Isabelle/HOL. The SMT reasoner Z3 \cite{Z3}, which is 
   employed in the \isa{smt} tactic by default, was too weak. Therefore we first introduced 
   further lemmata, which helped. 
   However, an alternative way out, which we discovered later, has been to replace Z3 by 
   CVC4 \cite{CVC4} in Isabelle's \isa{smt} 
   tactic (this can be done by stating ``\isa{declare\ {\isacharbrackleft}{\isacharbrackleft}\ smt{\isacharunderscore}solver\ {\isacharequal}\ cvc{\isadigit{4}}{\isacharbrackright}{\isacharbrackright}}'' in the 
   source document).
   In the latest version of the proof document we now suitably switch between the two SMT solvers 
   to obtain best results.}%
\end{isamarkuptext}\isamarkuptrue%
\ \ \isacommand{lemma}\isamarkupfalse%
\ UC\isactrlsub i{\isacharcolon}\ {\isachardoublequoteopen}\isactrlbold {\isasymforall}y{\isachardot}\isactrlbold {\isasymexists}i{\isachardot}\ I\ i\ \isactrlbold {\isasymand}\ i{\isasymcdot}y\ {\isasymcong}\ y\ \isactrlbold {\isasymand}\ {\isacharparenleft}\isactrlbold {\isasymforall}j{\isachardot}{\isacharparenleft}I\ j\ \isactrlbold {\isasymand}\ j{\isasymcdot}y\ {\isasymcong}\ y{\isacharparenright}\ \isactrlbold {\isasymrightarrow}\ i\ {\isasymcong}\ j{\isacharparenright}{\isachardoublequoteclose}\ \isanewline
\isadelimproof
\ \ \ \ %
\endisadelimproof
\isatagproof
\isacommand{by}\isamarkupfalse%
\ {\isacharparenleft}smt\ A\isactrlsub i\ C\isactrlsub i\ S\isactrlsub i{\isacharparenright}%
\endisatagproof
{\isafoldproof}%
\isadelimproof
\endisadelimproof
\begin{isamarkuptext}%
Analogously, the provers quickly show that \isa{j} in axiom \isa{D} is unique.%
\end{isamarkuptext}\isamarkuptrue%
\ \ \isacommand{lemma}\isamarkupfalse%
\ UD\isactrlsub i{\isacharcolon}\ {\isachardoublequoteopen}\isactrlbold {\isasymforall}x{\isachardot}\isactrlbold {\isasymexists}j{\isachardot}\ I\ j\ \isactrlbold {\isasymand}\ x{\isasymcdot}j\ {\isasymcong}\ x\ \isactrlbold {\isasymand}\ {\isacharparenleft}\isactrlbold {\isasymforall}i{\isachardot}{\isacharparenleft}I\ i\ \isactrlbold {\isasymand}\ x{\isasymcdot}i\ {\isasymcong}\ x{\isacharparenright}\ \isactrlbold {\isasymrightarrow}\ j\ {\isasymcong}\ i{\isacharparenright}{\isachardoublequoteclose}\ \ \isanewline
\isadelimproof
\ \ \ \ %
\endisadelimproof
\isatagproof
\isacommand{by}\isamarkupfalse%
\ {\isacharparenleft}smt\ A\isactrlsub i\ D\isactrlsub i\ S\isactrlsub i{\isacharparenright}%
\endisatagproof
{\isafoldproof}%
\isadelimproof
\endisadelimproof
\begin{isamarkuptext}%
However, the \isa{i} and \isa{j} need not be equal. Using the Skolem 
   function symbols \isa{C} and \isa{D} this can be encoded in
   our formalization as follows:%
\end{isamarkuptext}\isamarkuptrue%
\ \isacommand{lemma}\isamarkupfalse%
\ {\isachardoublequoteopen}{\isacharparenleft}{\isasymexists}C\ D{\isachardot}\ {\isacharparenleft}\isactrlbold {\isasymforall}y{\isachardot}\ I\ {\isacharparenleft}C\ y{\isacharparenright}\ \isactrlbold {\isasymand}\ {\isacharparenleft}C\ y{\isacharparenright}{\isasymcdot}y\ {\isasymcong}\ y{\isacharparenright}\ \isactrlbold {\isasymand}\ {\isacharparenleft}\isactrlbold {\isasymforall}x{\isachardot}\ I\ {\isacharparenleft}D\ x{\isacharparenright}\ \isactrlbold {\isasymand}\ x{\isasymcdot}{\isacharparenleft}D\ x{\isacharparenright}\ {\isasymcong}\ x{\isacharparenright}\ \isactrlbold {\isasymand}\ \isactrlbold {\isasymnot}{\isacharparenleft}D\ \isactrlbold {\isacharequal}\ C{\isacharparenright}{\isacharparenright}{\isachardoublequoteclose}\isanewline
\ \ \ \isacommand{nitpick}\isamarkupfalse%
\ {\isacharbrackleft}satisfy{\isacharcomma}\ user{\isacharunderscore}axioms{\isacharcomma}\ show{\isacharunderscore}all{\isacharcomma}\ format\ {\isacharequal}\ {\isadigit{2}}{\isacharcomma}\ expect\ {\isacharequal}\ genuine{\isacharbrackright}%
\isadelimproof
\ %
\endisadelimproof
\isatagproof
\isacommand{oops}\isamarkupfalse%
\ %
\isamarkupcmt{Nitpick finds a model.%
}
\endisatagproof
{\isafoldproof}%
\isadelimproof
\endisadelimproof
\begin{isamarkuptext}%
Nitpick finds a model for cardinality \isa{i\ {\isacharequal}\ {\isadigit{2}}}. This model consists of two 
   non-existing objects \isa{i\isactrlsub {\isadigit{1}}} and \isa{i\isactrlsub {\isadigit{2}}}. \isa{C} maps both \isa{i\isactrlsub {\isadigit{1}}} and 
   \isa{i\isactrlsub {\isadigit{2}}} to \isa{i\isactrlsub {\isadigit{2}}}. \isa{D} maps \isa{i\isactrlsub {\isadigit{1}}} to \isa{i\isactrlsub {\isadigit{2}}}, and vice versa. The 
   composition \isa{i\isactrlsub {\isadigit{2}}{\isasymcdot}i\isactrlsub {\isadigit{2}}} is mapped to \isa{i\isactrlsub {\isadigit{2}}}. All other composition pairs are mapped 
   to \isa{i\isactrlsub {\isadigit{1}}}.%
\end{isamarkuptext}\isamarkuptrue%
\begin{isamarkuptext}%
Even if we require at least one existing object Nitpick still finds a model:%
\end{isamarkuptext}\isamarkuptrue%
\ \isacommand{lemma}\isamarkupfalse%
\ {\isachardoublequoteopen}{\isacharparenleft}{\isasymexists}x{\isachardot}\ E\ x{\isacharparenright}\ \isactrlbold {\isasymand}\ {\isacharparenleft}{\isasymexists}C\ D{\isachardot}\ {\isacharparenleft}\isactrlbold {\isasymforall}y{\isachardot}\ I\ {\isacharparenleft}C\ y{\isacharparenright}\ \isactrlbold {\isasymand}\ {\isacharparenleft}C\ y{\isacharparenright}{\isasymcdot}y\ {\isasymcong}\ y{\isacharparenright}\ \isactrlbold {\isasymand}\ {\isacharparenleft}\isactrlbold {\isasymforall}x{\isachardot}\ I\ {\isacharparenleft}D\ x{\isacharparenright}\ \isactrlbold {\isasymand}\ x{\isasymcdot}{\isacharparenleft}D\ x{\isacharparenright}\ {\isasymcong}\ x{\isacharparenright}\ \isactrlbold {\isasymand}\ \isactrlbold {\isasymnot}{\isacharparenleft}D\ \isactrlbold {\isacharequal}\ C{\isacharparenright}{\isacharparenright}{\isachardoublequoteclose}\isanewline
\ \ \ \isacommand{nitpick}\isamarkupfalse%
\ {\isacharbrackleft}satisfy{\isacharcomma}\ user{\isacharunderscore}axioms{\isacharcomma}\ show{\isacharunderscore}all{\isacharcomma}\ format\ {\isacharequal}\ {\isadigit{2}}{\isacharcomma}\ expect\ {\isacharequal}\ genuine{\isacharbrackright}%
\isadelimproof
\ %
\endisadelimproof
\isatagproof
\isacommand{oops}\isamarkupfalse%
\ \ %
\isamarkupcmt{Nitpick finds a model.%
}
\endisatagproof
{\isafoldproof}%
\isadelimproof
\endisadelimproof
\begin{isamarkuptext}%
Again the model is of cardinality \isa{i\ {\isacharequal}\ {\isadigit{2}}}, but now we have a non-existing \isa{i\isactrlsub {\isadigit{1}}} and 
  and an existing \isa{i\isactrlsub {\isadigit{2}}}. Composition \isa{{\isasymcdot}} and \isa{C} are as above, but 
  \isa{D} is now identity on all objects.%
\end{isamarkuptext}\isamarkuptrue%
\isamarkupsection{Axiom Set II%
}
\isamarkuptrue%
\begin{isamarkuptext}%
Axiom Set II is developed from Axiom Set I by Skolemization of \isa{i} and \isa{j} 
 in axioms \isa{C\isactrlsub i} and \isa{D\isactrlsub i}. We can argue semantically that every model of Axiom Set I has such 
 functions. Hence, we get a conservative extension of Axiom Set I. This could be done for any 
 theory with an ``\isa{{\isasymforall}x{\isachardot}{\isasymexists}i{\isachardot}}''-axiom. The strictness axiom \isa{S} is extended, 
 so that strictness is now also postulated for the new Skolem functions \isa{dom} 
 and \isa{cod}. Note: the values of Skolem functions outside E can just be given by 
the identity function.%
\end{isamarkuptext}\isamarkuptrue%
\ S\isactrlsub i\isactrlsub i{\isacharcolon}\ %
\isamarkupcmt{\makebox[2cm][l]{Strictness:}%
}
\ {\isachardoublequoteopen}{\isacharparenleft}E{\isacharparenleft}x{\isasymcdot}y{\isacharparenright}\ \isactrlbold {\isasymrightarrow}\ {\isacharparenleft}E\ x\ \isactrlbold {\isasymand}\ E\ y{\isacharparenright}{\isacharparenright}\ \isactrlbold {\isasymand}\ {\isacharparenleft}E{\isacharparenleft}dom\ x{\isacharparenright}\ \isactrlbold {\isasymrightarrow}\ E\ x{\isacharparenright}\ \isactrlbold {\isasymand}\ {\isacharparenleft}E{\isacharparenleft}cod\ y{\isacharparenright}\ \isactrlbold {\isasymrightarrow}\ E\ y{\isacharparenright}{\isachardoublequoteclose}\ \ \isakeyword{and}\isanewline
\ E\isactrlsub i\isactrlsub i{\isacharcolon}\ %
\isamarkupcmt{\makebox[2cm][l]{Existence:}%
}
\ {\isachardoublequoteopen}E{\isacharparenleft}x{\isasymcdot}y{\isacharparenright}\ \isactrlbold {\isasymleftarrow}\ {\isacharparenleft}E\ x\ \isactrlbold {\isasymand}\ E\ y\ \isactrlbold {\isasymand}\ {\isacharparenleft}\isactrlbold {\isasymexists}z{\isachardot}\ z{\isasymcdot}z\ {\isasymcong}\ z\ \isactrlbold {\isasymand}\ x{\isasymcdot}z\ {\isasymcong}\ x\ \isactrlbold {\isasymand}\ z{\isasymcdot}y\ {\isasymcong}\ y{\isacharparenright}{\isacharparenright}{\isachardoublequoteclose}\ \isakeyword{and}\isanewline
\ A\isactrlsub i\isactrlsub i{\isacharcolon}\ %
\isamarkupcmt{\makebox[2cm][l]{Associativity:}%
}
\ {\isachardoublequoteopen}x{\isasymcdot}{\isacharparenleft}y{\isasymcdot}z{\isacharparenright}\ {\isasymcong}\ {\isacharparenleft}x{\isasymcdot}y{\isacharparenright}{\isasymcdot}z{\isachardoublequoteclose}\ \isakeyword{and}\isanewline
\ C\isactrlsub i\isactrlsub i{\isacharcolon}\ %
\isamarkupcmt{\makebox[2cm][l]{Codomain:}%
}
\ {\isachardoublequoteopen}E\ y\ \isactrlbold {\isasymrightarrow}\ {\isacharparenleft}I{\isacharparenleft}cod\ y{\isacharparenright}\ \isactrlbold {\isasymand}\ {\isacharparenleft}cod\ y{\isacharparenright}{\isasymcdot}y\ {\isasymcong}\ y{\isacharparenright}{\isachardoublequoteclose}\ \isakeyword{and}\isanewline
\ D\isactrlsub i\isactrlsub i{\isacharcolon}\ %
\isamarkupcmt{\makebox[2cm][l]{Domain:}%
}
\ {\isachardoublequoteopen}E\ x\ \isactrlbold {\isasymrightarrow}\ {\isacharparenleft}I{\isacharparenleft}dom\ x{\isacharparenright}\ \isactrlbold {\isasymand}\ x{\isasymcdot}{\isacharparenleft}dom\ x{\isacharparenright}\ {\isasymcong}\ x{\isacharparenright}{\isachardoublequoteclose}%
\begin{isamarkuptext}%
As above, we first check for consistency.%
\end{isamarkuptext}\isamarkuptrue%
\ \ \isacommand{lemma}\isamarkupfalse%
\ True\ \ %
\isamarkupcmt{Nitpick finds a model%
}
\isanewline
\ \ \ \ \isacommand{nitpick}\isamarkupfalse%
\ {\isacharbrackleft}satisfy{\isacharcomma}\ user{\isacharunderscore}axioms{\isacharcomma}\ show{\isacharunderscore}all{\isacharcomma}\ format\ {\isacharequal}\ {\isadigit{2}}{\isacharcomma}\ expect\ {\isacharequal}\ genuine{\isacharbrackright}%
\isadelimproof
\ %
\endisadelimproof
\isatagproof
\isacommand{oops}\isamarkupfalse%
\endisatagproof
{\isafoldproof}%
\isadelimproof
\endisadelimproof
\isanewline
\ \ \isacommand{lemma}\isamarkupfalse%
\ \isakeyword{assumes}\ {\isachardoublequoteopen}{\isasymexists}x{\isachardot}\ \isactrlbold {\isasymnot}{\isacharparenleft}E\ x{\isacharparenright}{\isachardoublequoteclose}\ \isakeyword{shows}\ True\ \ %
\isamarkupcmt{Nitpick finds a model%
}
\ \ \isanewline
\ \ \ \ \isacommand{nitpick}\isamarkupfalse%
\ {\isacharbrackleft}satisfy{\isacharcomma}\ user{\isacharunderscore}axioms{\isacharcomma}\ show{\isacharunderscore}all{\isacharcomma}\ format\ {\isacharequal}\ {\isadigit{2}}{\isacharcomma}\ expect\ {\isacharequal}\ genuine{\isacharbrackright}%
\isadelimproof
\ %
\endisadelimproof
\isatagproof
\isacommand{oops}\isamarkupfalse%
\endisatagproof
{\isafoldproof}%
\isadelimproof
\endisadelimproof
\ \isanewline
\ \ \isacommand{lemma}\isamarkupfalse%
\ \isakeyword{assumes}\ {\isachardoublequoteopen}{\isacharparenleft}{\isasymexists}x{\isachardot}\ \isactrlbold {\isasymnot}{\isacharparenleft}E\ x{\isacharparenright}{\isacharparenright}\ {\isasymand}\ {\isacharparenleft}{\isasymexists}x{\isachardot}\ {\isacharparenleft}E\ x{\isacharparenright}{\isacharparenright}{\isachardoublequoteclose}\ \isakeyword{shows}\ True\ \ %
\isamarkupcmt{Nitpick finds a model%
}
\ \isanewline
\ \ \ \ \isacommand{nitpick}\isamarkupfalse%
\ {\isacharbrackleft}satisfy{\isacharcomma}\ user{\isacharunderscore}axioms{\isacharcomma}\ show{\isacharunderscore}all{\isacharcomma}\ format\ {\isacharequal}\ {\isadigit{2}}{\isacharcomma}\ expect\ {\isacharequal}\ genuine{\isacharbrackright}%
\isadelimproof
\ %
\endisadelimproof
\isatagproof
\isacommand{oops}\isamarkupfalse%
\endisatagproof
{\isafoldproof}%
\isadelimproof
\endisadelimproof
\begin{isamarkuptext}%
The left-to-right direction of existence axiom \isa{E\isactrlsub i\isactrlsub i} is implied.%
\end{isamarkuptext}\isamarkuptrue%
\ \ \isacommand{lemma}\isamarkupfalse%
\ E\isactrlsub i\isactrlsub iImplied{\isacharcolon}\ {\isachardoublequoteopen}E{\isacharparenleft}x{\isasymcdot}y{\isacharparenright}\ \isactrlbold {\isasymrightarrow}\ {\isacharparenleft}E\ x\ \isactrlbold {\isasymand}\ E\ y\ \isactrlbold {\isasymand}\ {\isacharparenleft}\isactrlbold {\isasymexists}z{\isachardot}\ z{\isasymcdot}z\ {\isasymcong}\ z\ \isactrlbold {\isasymand}\ x{\isasymcdot}z\ {\isasymcong}\ x\ \isactrlbold {\isasymand}\ z{\isasymcdot}y\ {\isasymcong}\ y{\isacharparenright}{\isacharparenright}{\isachardoublequoteclose}\ \isanewline
\isadelimproof
\ \ \ \ %
\endisadelimproof
\isatagproof
\isacommand{by}\isamarkupfalse%
\ {\isacharparenleft}metis\ A\isactrlsub i\isactrlsub i\ C\isactrlsub i\isactrlsub i\ S\isactrlsub i\isactrlsub i{\isacharparenright}%
\endisatagproof
{\isafoldproof}%
\isadelimproof
\endisadelimproof
\begin{isamarkuptext}%
Axioms \isa{C\isactrlsub i\isactrlsub i} and \isa{D\isactrlsub i\isactrlsub i}, together with \isa{S\isactrlsub i\isactrlsub i}, show that
\isa{dom} and \isa{cod} are total functions -- as intended.%
\end{isamarkuptext}\isamarkuptrue%
\isacommand{lemma}\isamarkupfalse%
\ domTotal{\isacharcolon}\ {\isachardoublequoteopen}E\ x\ \isactrlbold {\isasymrightarrow}\ E{\isacharparenleft}dom\ x{\isacharparenright}{\isachardoublequoteclose}\ \isanewline
\isadelimproof
\ \ %
\endisadelimproof
\isatagproof
\isacommand{by}\isamarkupfalse%
\ {\isacharparenleft}metis\ D\isactrlsub i\isactrlsub i\ S\isactrlsub i\isactrlsub i{\isacharparenright}%
\endisatagproof
{\isafoldproof}%
\isadelimproof
\ \isanewline
\endisadelimproof
\isacommand{lemma}\isamarkupfalse%
\ codTotal{\isacharcolon}\ {\isachardoublequoteopen}E\ x\ \isactrlbold {\isasymrightarrow}\ E{\isacharparenleft}cod\ x{\isacharparenright}{\isachardoublequoteclose}\ \isanewline
\isadelimproof
\ \ %
\endisadelimproof
\isatagproof
\isacommand{by}\isamarkupfalse%
\ {\isacharparenleft}metis\ C\isactrlsub i\isactrlsub i\ S\isactrlsub i\isactrlsub i{\isacharparenright}%
\endisatagproof
{\isafoldproof}%
\isadelimproof
\endisadelimproof
\begin{isamarkuptext}%
Axiom Set II implies Axiom Set I.\footnote{Axiom Set I also implies Axiom Set II. This can 
be shown by semantical means on the meta-level. We have also attempted to prove this equivalence 
within Isabelle/HOL, but so far without final success. However, we succeed to 
prove that the following holds: \isa{{\isasymexists}Cod\ Dom{\isachardot}\ {\isacharparenleft}{\isacharparenleft}{\isasymforall}x\ y{\isachardot}\ {\isacharparenleft}E{\isacharparenleft}x{\isasymcdot}y{\isacharparenright}\ \isactrlbold {\isasymrightarrow}\ {\isacharparenleft}E\ x\ \isactrlbold {\isasymand}\ E\ y{\isacharparenright}{\isacharparenright}{\isacharparenright}\ \isactrlbold {\isasymand}\ {\isacharparenleft}{\isasymforall}x\ y{\isachardot}\ E{\isacharparenleft}x{\isasymcdot}y{\isacharparenright}\ \isactrlbold {\isasymleftarrow}\ {\isacharparenleft}E\ x\ \isactrlbold {\isasymand}\ E\ y\ \isactrlbold {\isasymand}\ {\isacharparenleft}\isactrlbold {\isasymexists}z{\isachardot}\ z{\isasymcdot}z\ {\isasymcong}\ z\ \isactrlbold {\isasymand}\ x{\isasymcdot}z\ {\isasymcong}\ x\ \isactrlbold {\isasymand}\ z{\isasymcdot}y\ {\isasymcong}\ y{\isacharparenright}{\isacharparenright}{\isacharparenright}\ \isactrlbold {\isasymand}\ {\isacharparenleft}{\isasymforall}x\ y\ z{\isachardot}\ x{\isasymcdot}{\isacharparenleft}y{\isasymcdot}z{\isacharparenright}\ {\isasymcong}\ {\isacharparenleft}x{\isasymcdot}y{\isacharparenright}{\isasymcdot}z{\isacharparenright}\ \isactrlbold {\isasymand}\ {\isacharparenleft}\isactrlbold {\isasymforall}y{\isachardot}\ I\ {\isacharparenleft}Cod\ y{\isacharparenright}\ \isactrlbold {\isasymand}\ {\isacharparenleft}Cod\ y{\isacharparenright}{\isasymcdot}y\ {\isasymcong}\ y{\isacharparenright}\ \isactrlbold {\isasymand}\ {\isacharparenleft}\isactrlbold {\isasymforall}x{\isachardot}\ I\ {\isacharparenleft}Dom\ x{\isacharparenright}\ \isactrlbold {\isasymand}\ x{\isasymcdot}{\isacharparenleft}Dom\ x{\isacharparenright}\ {\isasymcong}\ x{\isacharparenright}\ {\isacharparenright}}. Note that the inclusion of strictness of \isa{Cod} and \isa{Dom} is still missing.}%
\end{isamarkuptext}\isamarkuptrue%
\ \ \isacommand{lemma}\isamarkupfalse%
\ S\isactrlsub iFromII{\isacharcolon}\ {\isachardoublequoteopen}E{\isacharparenleft}x{\isasymcdot}y{\isacharparenright}\ \isactrlbold {\isasymrightarrow}\ {\isacharparenleft}E\ x\ \isactrlbold {\isasymand}\ E\ y{\isacharparenright}{\isachardoublequoteclose}\ \ \isanewline
\isadelimproof
\ \ \ \ %
\endisadelimproof
\isatagproof
\isacommand{using}\isamarkupfalse%
\ S\isactrlsub i\isactrlsub i\ \isacommand{by}\isamarkupfalse%
\ blast%
\endisatagproof
{\isafoldproof}%
\isadelimproof
\isanewline
\endisadelimproof
\ \ \isacommand{lemma}\isamarkupfalse%
\ E\isactrlsub iFromII{\isacharcolon}\ {\isachardoublequoteopen}E{\isacharparenleft}x{\isasymcdot}y{\isacharparenright}\ \isactrlbold {\isasymleftarrow}\ {\isacharparenleft}E\ x\ \isactrlbold {\isasymand}\ E\ y\ \isactrlbold {\isasymand}\ {\isacharparenleft}\isactrlbold {\isasymexists}z{\isachardot}\ z{\isasymcdot}z\ {\isasymcong}\ z\ \isactrlbold {\isasymand}\ x{\isasymcdot}z\ {\isasymcong}\ x\ \isactrlbold {\isasymand}\ z{\isasymcdot}y\ {\isasymcong}\ y{\isacharparenright}{\isacharparenright}{\isachardoublequoteclose}\ \isanewline
\isadelimproof
\ \ \ \ %
\endisadelimproof
\isatagproof
\isacommand{using}\isamarkupfalse%
\ E\isactrlsub i\isactrlsub i\ \isacommand{by}\isamarkupfalse%
\ blast%
\endisatagproof
{\isafoldproof}%
\isadelimproof
\isanewline
\endisadelimproof
\ \ \isacommand{lemma}\isamarkupfalse%
\ A\isactrlsub iFromII{\isacharcolon}\ {\isachardoublequoteopen}x{\isasymcdot}{\isacharparenleft}y{\isasymcdot}z{\isacharparenright}\ {\isasymcong}\ {\isacharparenleft}x{\isasymcdot}y{\isacharparenright}{\isasymcdot}z{\isachardoublequoteclose}\ \isanewline
\isadelimproof
\ \ \ \ %
\endisadelimproof
\isatagproof
\isacommand{using}\isamarkupfalse%
\ A\isactrlsub i\isactrlsub i\ \isacommand{by}\isamarkupfalse%
\ blast%
\endisatagproof
{\isafoldproof}%
\isadelimproof
\isanewline
\endisadelimproof
\ \ \isacommand{lemma}\isamarkupfalse%
\ C\isactrlsub iFromII{\isacharcolon}\ {\isachardoublequoteopen}\isactrlbold {\isasymforall}y{\isachardot}\isactrlbold {\isasymexists}i{\isachardot}\ I\ i\ \isactrlbold {\isasymand}\ i{\isasymcdot}y\ {\isasymcong}\ y{\isachardoublequoteclose}\ \isanewline
\isadelimproof
\ \ \ \ %
\endisadelimproof
\isatagproof
\isacommand{by}\isamarkupfalse%
\ {\isacharparenleft}metis\ C\isactrlsub i\isactrlsub i\ S\isactrlsub i\isactrlsub i{\isacharparenright}%
\endisatagproof
{\isafoldproof}%
\isadelimproof
\isanewline
\endisadelimproof
\ \ \isacommand{lemma}\isamarkupfalse%
\ D\isactrlsub iFromII{\isacharcolon}\ {\isachardoublequoteopen}\isactrlbold {\isasymforall}x{\isachardot}\isactrlbold {\isasymexists}j{\isachardot}\ I\ j\ \isactrlbold {\isasymand}\ x{\isasymcdot}j\ {\isasymcong}\ x{\isachardoublequoteclose}\ \isanewline
\isadelimproof
\ \ \ \ %
\endisadelimproof
\isatagproof
\isacommand{by}\isamarkupfalse%
\ {\isacharparenleft}metis\ D\isactrlsub i\isactrlsub i\ S\isactrlsub i\isactrlsub i{\isacharparenright}%
\endisatagproof
{\isafoldproof}%
\isadelimproof
\endisadelimproof
\isamarkupsection{Axiom Set III%
}
\isamarkuptrue%
\begin{isamarkuptext}%
In Axiom Set III the existence  axiom  \isa{E} is simplified by taking advantage of 
  the two new Skolem functions \isa{dom} and \isa{cod}.%
\end{isamarkuptext}\isamarkuptrue%
\ S\isactrlsub i\isactrlsub i\isactrlsub i{\isacharcolon}\ %
\isamarkupcmt{\makebox[2cm][l]{Strictness:}%
}
\ {\isachardoublequoteopen}{\isacharparenleft}E{\isacharparenleft}x{\isasymcdot}y{\isacharparenright}\ \isactrlbold {\isasymrightarrow}\ {\isacharparenleft}E\ x\ \isactrlbold {\isasymand}\ E\ y{\isacharparenright}{\isacharparenright}\ \isactrlbold {\isasymand}\ {\isacharparenleft}E{\isacharparenleft}dom\ x\ {\isacharparenright}\ \isactrlbold {\isasymrightarrow}\ E\ x{\isacharparenright}\ \isactrlbold {\isasymand}\ {\isacharparenleft}E{\isacharparenleft}cod\ y{\isacharparenright}\ \isactrlbold {\isasymrightarrow}\ E\ y{\isacharparenright}{\isachardoublequoteclose}\ \ \isakeyword{and}\isanewline
\ E\isactrlsub i\isactrlsub i\isactrlsub i{\isacharcolon}\ %
\isamarkupcmt{\makebox[2cm][l]{Existence:}%
}
\ {\isachardoublequoteopen}E{\isacharparenleft}x{\isasymcdot}y{\isacharparenright}\ \isactrlbold {\isasymleftarrow}\ {\isacharparenleft}dom\ x\ {\isasymcong}\ cod\ y\ \isactrlbold {\isasymand}\ E{\isacharparenleft}cod\ y{\isacharparenright}{\isacharparenright}{\isachardoublequoteclose}\ \isakeyword{and}\isanewline
\ A\isactrlsub i\isactrlsub i\isactrlsub i{\isacharcolon}\ %
\isamarkupcmt{\makebox[2cm][l]{Associativity:}%
}
\ {\isachardoublequoteopen}x{\isasymcdot}{\isacharparenleft}y{\isasymcdot}z{\isacharparenright}\ {\isasymcong}\ {\isacharparenleft}x{\isasymcdot}y{\isacharparenright}{\isasymcdot}z{\isachardoublequoteclose}\ \isakeyword{and}\isanewline
\ C\isactrlsub i\isactrlsub i\isactrlsub i{\isacharcolon}\ %
\isamarkupcmt{\makebox[2cm][l]{Codomain:}%
}
\ {\isachardoublequoteopen}E\ y\ \isactrlbold {\isasymrightarrow}\ {\isacharparenleft}I{\isacharparenleft}cod\ y{\isacharparenright}\ \isactrlbold {\isasymand}\ {\isacharparenleft}cod\ y{\isacharparenright}{\isasymcdot}y\ {\isasymcong}\ y{\isacharparenright}{\isachardoublequoteclose}\ \isakeyword{and}\isanewline
\ D\isactrlsub i\isactrlsub i\isactrlsub i{\isacharcolon}\ %
\isamarkupcmt{\makebox[2cm][l]{Domain:}%
}
\ {\isachardoublequoteopen}E\ x\ \isactrlbold {\isasymrightarrow}\ {\isacharparenleft}I{\isacharparenleft}dom\ x{\isacharparenright}\ \isactrlbold {\isasymand}\ x{\isasymcdot}{\isacharparenleft}dom\ x{\isacharparenright}\ {\isasymcong}\ x{\isacharparenright}{\isachardoublequoteclose}%
\begin{isamarkuptext}%
The obligatory consistency check is positive.%
\end{isamarkuptext}\isamarkuptrue%
\ \ \isacommand{lemma}\isamarkupfalse%
\ True\ \ %
\isamarkupcmt{Nitpick finds a model%
}
\isanewline
\ \ \ \ \isacommand{nitpick}\isamarkupfalse%
\ {\isacharbrackleft}satisfy{\isacharcomma}\ user{\isacharunderscore}axioms{\isacharcomma}\ show{\isacharunderscore}all{\isacharcomma}\ format\ {\isacharequal}\ {\isadigit{2}}{\isacharcomma}\ expect\ {\isacharequal}\ genuine{\isacharbrackright}%
\isadelimproof
\ %
\endisadelimproof
\isatagproof
\isacommand{oops}\isamarkupfalse%
\endisatagproof
{\isafoldproof}%
\isadelimproof
\endisadelimproof
\isanewline
\ \ \isacommand{lemma}\isamarkupfalse%
\ \isakeyword{assumes}\ {\isachardoublequoteopen}{\isasymexists}x{\isachardot}\ \isactrlbold {\isasymnot}{\isacharparenleft}E\ x{\isacharparenright}{\isachardoublequoteclose}\ \isakeyword{shows}\ True\ \ %
\isamarkupcmt{Nitpick finds a model%
}
\ \isanewline
\ \ \ \ \isacommand{nitpick}\isamarkupfalse%
\ {\isacharbrackleft}satisfy{\isacharcomma}\ user{\isacharunderscore}axioms{\isacharcomma}\ show{\isacharunderscore}all{\isacharcomma}\ format\ {\isacharequal}\ {\isadigit{2}}{\isacharcomma}\ expect\ {\isacharequal}\ genuine{\isacharbrackright}%
\isadelimproof
\ %
\endisadelimproof
\isatagproof
\isacommand{oops}\isamarkupfalse%
\endisatagproof
{\isafoldproof}%
\isadelimproof
\endisadelimproof
\isanewline
\ \ \isacommand{lemma}\isamarkupfalse%
\ \isakeyword{assumes}\ {\isachardoublequoteopen}{\isacharparenleft}{\isasymexists}x{\isachardot}\ \isactrlbold {\isasymnot}{\isacharparenleft}E\ x{\isacharparenright}{\isacharparenright}\ {\isasymand}\ {\isacharparenleft}{\isasymexists}x{\isachardot}\ {\isacharparenleft}E\ x{\isacharparenright}{\isacharparenright}{\isachardoublequoteclose}\ \isakeyword{shows}\ True\ \ %
\isamarkupcmt{Nitpick finds a model%
}
\ \isanewline
\ \ \ \ \isacommand{nitpick}\isamarkupfalse%
\ {\isacharbrackleft}satisfy{\isacharcomma}\ user{\isacharunderscore}axioms{\isacharcomma}\ show{\isacharunderscore}all{\isacharcomma}\ format\ {\isacharequal}\ {\isadigit{2}}{\isacharcomma}\ expect\ {\isacharequal}\ genuine{\isacharbrackright}%
\isadelimproof
\ %
\endisadelimproof
\isatagproof
\isacommand{oops}\isamarkupfalse%
\endisatagproof
{\isafoldproof}%
\isadelimproof
\endisadelimproof
\begin{isamarkuptext}%
The left-to-right direction of existence axiom \isa{E\isactrlsub i\isactrlsub i\isactrlsub i} is implied.%
\end{isamarkuptext}\isamarkuptrue%
\ \ \isacommand{lemma}\isamarkupfalse%
\ E\isactrlsub i\isactrlsub i\isactrlsub iImplied{\isacharcolon}\ {\isachardoublequoteopen}E{\isacharparenleft}x{\isasymcdot}y{\isacharparenright}\ \isactrlbold {\isasymrightarrow}\ {\isacharparenleft}dom\ x\ {\isasymcong}\ cod\ y\ \isactrlbold {\isasymand}\ E{\isacharparenleft}cod\ y{\isacharparenright}{\isacharparenright}{\isachardoublequoteclose}\ \isanewline
\isadelimproof
\ \ \ \ %
\endisadelimproof
\isatagproof
\isacommand{by}\isamarkupfalse%
\ {\isacharparenleft}metis\ {\isacharparenleft}full{\isacharunderscore}types{\isacharparenright}\ A\isactrlsub i\isactrlsub i\isactrlsub i\ C\isactrlsub i\isactrlsub i\isactrlsub i\ D\isactrlsub i\isactrlsub i\isactrlsub i\ S\isactrlsub i\isactrlsub i\isactrlsub i{\isacharparenright}%
\endisatagproof
{\isafoldproof}%
\isadelimproof
\endisadelimproof
\begin{isamarkuptext}%
Moreover, Axiom Set II is implied.%
\end{isamarkuptext}\isamarkuptrue%
\ \ \isacommand{lemma}\isamarkupfalse%
\ S\isactrlsub i\isactrlsub iFromIII{\isacharcolon}\ {\isachardoublequoteopen}{\isacharparenleft}E{\isacharparenleft}x{\isasymcdot}y{\isacharparenright}\ \isactrlbold {\isasymrightarrow}\ {\isacharparenleft}E\ x\ \isactrlbold {\isasymand}\ E\ y{\isacharparenright}{\isacharparenright}\ \isactrlbold {\isasymand}\ {\isacharparenleft}E{\isacharparenleft}dom\ x\ {\isacharparenright}\ \isactrlbold {\isasymrightarrow}\ E\ x{\isacharparenright}\ \isactrlbold {\isasymand}\ {\isacharparenleft}E{\isacharparenleft}cod\ y{\isacharparenright}\ \isactrlbold {\isasymrightarrow}\ E\ y{\isacharparenright}{\isachardoublequoteclose}\ \ \isanewline
\isadelimproof
\ \ \ \ %
\endisadelimproof
\isatagproof
\isacommand{using}\isamarkupfalse%
\ S\isactrlsub i\isactrlsub i\isactrlsub i\ \isacommand{by}\isamarkupfalse%
\ blast%
\endisatagproof
{\isafoldproof}%
\isadelimproof
\isanewline
\endisadelimproof
\ \ \isacommand{lemma}\isamarkupfalse%
\ E\isactrlsub i\isactrlsub iFromIII{\isacharcolon}\ {\isachardoublequoteopen}E{\isacharparenleft}x{\isasymcdot}y{\isacharparenright}\ \isactrlbold {\isasymleftarrow}\ {\isacharparenleft}E\ x\ \isactrlbold {\isasymand}\ E\ y\ \isactrlbold {\isasymand}\ {\isacharparenleft}\isactrlbold {\isasymexists}z{\isachardot}\ z{\isasymcdot}z\ {\isasymcong}\ z\ \isactrlbold {\isasymand}\ x{\isasymcdot}z\ {\isasymcong}\ x\ \isactrlbold {\isasymand}\ z{\isasymcdot}y\ {\isasymcong}\ y{\isacharparenright}{\isacharparenright}{\isachardoublequoteclose}\ \isanewline
\isadelimproof
\ \ \ \ %
\endisadelimproof
\isatagproof
\isacommand{by}\isamarkupfalse%
\ {\isacharparenleft}metis\ A\isactrlsub i\isactrlsub i\isactrlsub i\ C\isactrlsub i\isactrlsub i\isactrlsub i\ D\isactrlsub i\isactrlsub i\isactrlsub i\ E\isactrlsub i\isactrlsub i\isactrlsub i\ S\isactrlsub i\isactrlsub i\isactrlsub i{\isacharparenright}%
\endisatagproof
{\isafoldproof}%
\isadelimproof
\isanewline
\endisadelimproof
\ \ \isacommand{lemma}\isamarkupfalse%
\ A\isactrlsub i\isactrlsub iFromIII{\isacharcolon}\ {\isachardoublequoteopen}x{\isasymcdot}{\isacharparenleft}y{\isasymcdot}z{\isacharparenright}\ {\isasymcong}\ {\isacharparenleft}x{\isasymcdot}y{\isacharparenright}{\isasymcdot}z{\isachardoublequoteclose}\ \isanewline
\isadelimproof
\ \ \ \ %
\endisadelimproof
\isatagproof
\isacommand{using}\isamarkupfalse%
\ A\isactrlsub i\isactrlsub i\isactrlsub i\ \isacommand{by}\isamarkupfalse%
\ blast%
\endisatagproof
{\isafoldproof}%
\isadelimproof
\isanewline
\endisadelimproof
\ \ \isacommand{lemma}\isamarkupfalse%
\ C\isactrlsub i\isactrlsub iFromIII{\isacharcolon}\ {\isachardoublequoteopen}E\ y\ \isactrlbold {\isasymrightarrow}\ {\isacharparenleft}I{\isacharparenleft}cod\ y{\isacharparenright}\ \isactrlbold {\isasymand}\ {\isacharparenleft}cod\ y{\isacharparenright}{\isasymcdot}y\ {\isasymcong}\ y{\isacharparenright}{\isachardoublequoteclose}\ \isanewline
\isadelimproof
\ \ \ \ %
\endisadelimproof
\isatagproof
\isacommand{using}\isamarkupfalse%
\ C\isactrlsub i\isactrlsub i\isactrlsub i\ \isacommand{by}\isamarkupfalse%
\ auto%
\endisatagproof
{\isafoldproof}%
\isadelimproof
\isanewline
\endisadelimproof
\ \ \isacommand{lemma}\isamarkupfalse%
\ D\isactrlsub i\isactrlsub iFromIII{\isacharcolon}\ {\isachardoublequoteopen}E\ x\ \isactrlbold {\isasymrightarrow}\ {\isacharparenleft}I{\isacharparenleft}dom\ x{\isacharparenright}\ \isactrlbold {\isasymand}\ x{\isasymcdot}{\isacharparenleft}dom\ x{\isacharparenright}\ {\isasymcong}\ x{\isacharparenright}{\isachardoublequoteclose}\ \isanewline
\isadelimproof
\ \ \ \ %
\endisadelimproof
\isatagproof
\isacommand{using}\isamarkupfalse%
\ D\isactrlsub i\isactrlsub i\isactrlsub i\ \isacommand{by}\isamarkupfalse%
\ auto%
\endisatagproof
{\isafoldproof}%
\isadelimproof
\endisadelimproof
\begin{isamarkuptext}%
A side remark on the experiments: All proofs above and all proofs in the rest of this paper 
 have been obtained fully automatically with the Sledgehammer tool in Isabelle/HOL. This
 tool interfaces to prominent first-order automated theorem provers such as CVC4 \cite{CVC4}, 
 Z3 \cite{Z3}, E \cite{E} and Spass \cite{Spass}. 
 Remotely, also provers such as Vampire \cite{Vampire}, or the higher-order provers 
 Satallax \cite{Satallax} and LEO-II \cite{LEO} 
 can be reached. For example, to prove lemma \isa{E\isactrlsub i\isactrlsub i\isactrlsub iFromII} we have called Sledgehammer on all 
 postulated axioms of the theory: \isa{sledgehammer\ {\isacharparenleft}S\isactrlsub i\isactrlsub i\ E\isactrlsub i\isactrlsub i\ A\isactrlsub i\isactrlsub i\ C\isactrlsub i\isactrlsub i\ D\isactrlsub i\isactrlsub i{\isacharparenright}}.  
 The provers then, via Sledgehammer, suggested to call trusted/verified tools in Isabelle/HOL
 with the exactly required dependencies they detected. In lemma \isa{E\isactrlsub i\isactrlsub i\isactrlsub iFromII}, for 
 example, all  axioms from Axiom Set II are required. With the provided dependency information 
 the trusted tools in Isabelle/HOL were then able to reconstruct the external proofs on their own.
 This way we obtain a verified Isabelle/HOL document in which all the proofs have nevertheless been contributed
 by automated theorem provers.%
\end{isamarkuptext}\isamarkuptrue%
\begin{isamarkuptext}%
Axiom Set II also implies Axiom Set III. Hence, both theories are
 equivalent. The only interesting case is lemma \isa{E\isactrlsub i\isactrlsub i\isactrlsub iFromII}, the other cases are 
 trivial.%
\end{isamarkuptext}\isamarkuptrue%
\isanewline
\ \ \isacommand{lemma}\isamarkupfalse%
\ S\isactrlsub i\isactrlsub i\isactrlsub iFromII{\isacharcolon}\ {\isachardoublequoteopen}{\isacharparenleft}E{\isacharparenleft}x{\isasymcdot}y{\isacharparenright}\ \isactrlbold {\isasymrightarrow}\ {\isacharparenleft}E\ x\ \isactrlbold {\isasymand}\ E\ y{\isacharparenright}{\isacharparenright}\ \isactrlbold {\isasymand}\ {\isacharparenleft}E{\isacharparenleft}dom\ x{\isacharparenright}\ \isactrlbold {\isasymrightarrow}\ E\ x{\isacharparenright}\ \isactrlbold {\isasymand}\ {\isacharparenleft}E{\isacharparenleft}cod\ y{\isacharparenright}\ \isactrlbold {\isasymrightarrow}\ E\ y{\isacharparenright}{\isachardoublequoteclose}\ \ \isanewline
\isadelimproof
\ \ \ \ %
\endisadelimproof
\isatagproof
\isacommand{using}\isamarkupfalse%
\ S\isactrlsub i\isactrlsub i\ \isacommand{by}\isamarkupfalse%
\ blast%
\endisatagproof
{\isafoldproof}%
\isadelimproof
\isanewline
\endisadelimproof
\ \ \isacommand{lemma}\isamarkupfalse%
\ E\isactrlsub i\isactrlsub i\isactrlsub iFromII{\isacharcolon}\ {\isachardoublequoteopen}E{\isacharparenleft}x{\isasymcdot}y{\isacharparenright}\ \isactrlbold {\isasymleftarrow}\ {\isacharparenleft}dom\ x\ {\isasymcong}\ cod\ y\ \isactrlbold {\isasymand}\ {\isacharparenleft}E{\isacharparenleft}cod\ y{\isacharparenright}{\isacharparenright}{\isacharparenright}{\isachardoublequoteclose}\ \isanewline
\isadelimproof
\ \ \ \ %
\endisadelimproof
\isatagproof
\isacommand{by}\isamarkupfalse%
\ {\isacharparenleft}metis\ C\isactrlsub i\isactrlsub i\ D\isactrlsub i\isactrlsub i\ E\isactrlsub i\isactrlsub i\ S\isactrlsub i\isactrlsub i{\isacharparenright}%
\endisatagproof
{\isafoldproof}%
\isadelimproof
\isanewline
\endisadelimproof
\ \ \isacommand{lemma}\isamarkupfalse%
\ A\isactrlsub i\isactrlsub i\isactrlsub iFromII{\isacharcolon}\ {\isachardoublequoteopen}x{\isasymcdot}{\isacharparenleft}y{\isasymcdot}z{\isacharparenright}\ {\isasymcong}\ {\isacharparenleft}x{\isasymcdot}y{\isacharparenright}{\isasymcdot}z{\isachardoublequoteclose}\ \isanewline
\isadelimproof
\ \ \ \ %
\endisadelimproof
\isatagproof
\isacommand{using}\isamarkupfalse%
\ A\isactrlsub i\isactrlsub i\ \isacommand{by}\isamarkupfalse%
\ blast%
\endisatagproof
{\isafoldproof}%
\isadelimproof
\isanewline
\endisadelimproof
\ \ \isacommand{lemma}\isamarkupfalse%
\ C\isactrlsub i\isactrlsub i\isactrlsub iFromII{\isacharcolon}\ {\isachardoublequoteopen}E\ y\ \isactrlbold {\isasymrightarrow}\ {\isacharparenleft}I{\isacharparenleft}cod\ y{\isacharparenright}\ \isactrlbold {\isasymand}\ {\isacharparenleft}cod\ y{\isacharparenright}{\isasymcdot}y\ {\isasymcong}\ y{\isacharparenright}{\isachardoublequoteclose}\ \isanewline
\isadelimproof
\ \ \ \ %
\endisadelimproof
\isatagproof
\isacommand{using}\isamarkupfalse%
\ C\isactrlsub i\isactrlsub i\ \isacommand{by}\isamarkupfalse%
\ auto%
\endisatagproof
{\isafoldproof}%
\isadelimproof
\isanewline
\endisadelimproof
\ \ \isacommand{lemma}\isamarkupfalse%
\ D\isactrlsub i\isactrlsub i\isactrlsub iFromII{\isacharcolon}\ {\isachardoublequoteopen}E\ x\ \isactrlbold {\isasymrightarrow}\ {\isacharparenleft}I{\isacharparenleft}dom\ x{\isacharparenright}\ \isactrlbold {\isasymand}\ x{\isasymcdot}{\isacharparenleft}dom\ x{\isacharparenright}\ {\isasymcong}\ x{\isacharparenright}{\isachardoublequoteclose}\ \isanewline
\isadelimproof
\ \ \ \ %
\endisadelimproof
\isatagproof
\isacommand{using}\isamarkupfalse%
\ D\isactrlsub i\isactrlsub i\ \isacommand{by}\isamarkupfalse%
\ auto%
\endisatagproof
{\isafoldproof}%
\isadelimproof
\endisadelimproof
\isamarkupsection{Axiom Set IV%
}
\isamarkuptrue%
\begin{isamarkuptext}%
Axiom Set IV simplifies the axioms \isa{C\isactrlsub i\isactrlsub i\isactrlsub i} and  \isa{D\isactrlsub i\isactrlsub i\isactrlsub i}. However, as it turned 
 out, these simplifications also require the existence axiom \isa{E\isactrlsub i\isactrlsub i\isactrlsub i} to be strengthened into
 an equivalence.%
\end{isamarkuptext}\isamarkuptrue%
\ S\isactrlsub i\isactrlsub v{\isacharcolon}\ %
\isamarkupcmt{\makebox[2cm][l]{Strictness:}%
}
\ {\isachardoublequoteopen}{\isacharparenleft}E{\isacharparenleft}x{\isasymcdot}y{\isacharparenright}\ \isactrlbold {\isasymrightarrow}\ {\isacharparenleft}E\ x\ \isactrlbold {\isasymand}\ E\ y{\isacharparenright}{\isacharparenright}\ \isactrlbold {\isasymand}\ {\isacharparenleft}E{\isacharparenleft}dom\ x{\isacharparenright}\ \isactrlbold {\isasymrightarrow}\ E\ x{\isacharparenright}\ \isactrlbold {\isasymand}\ {\isacharparenleft}E{\isacharparenleft}cod\ y{\isacharparenright}\ \isactrlbold {\isasymrightarrow}\ E\ y{\isacharparenright}{\isachardoublequoteclose}\ \ \isakeyword{and}\isanewline
\ E\isactrlsub i\isactrlsub v{\isacharcolon}\ %
\isamarkupcmt{\makebox[2cm][l]{Existence:}%
}
\ {\isachardoublequoteopen}E{\isacharparenleft}x{\isasymcdot}y{\isacharparenright}\ \isactrlbold {\isasymleftrightarrow}\ {\isacharparenleft}dom\ x\ {\isasymcong}\ cod\ y\ \isactrlbold {\isasymand}\ E{\isacharparenleft}cod\ y{\isacharparenright}{\isacharparenright}{\isachardoublequoteclose}\ \isakeyword{and}\isanewline
\ A\isactrlsub i\isactrlsub v{\isacharcolon}\ %
\isamarkupcmt{\makebox[2cm][l]{Associativity:}%
}
\ {\isachardoublequoteopen}x{\isasymcdot}{\isacharparenleft}y{\isasymcdot}z{\isacharparenright}\ {\isasymcong}\ {\isacharparenleft}x{\isasymcdot}y{\isacharparenright}{\isasymcdot}z{\isachardoublequoteclose}\ \isakeyword{and}\isanewline
\ C\isactrlsub i\isactrlsub v{\isacharcolon}\ %
\isamarkupcmt{\makebox[2cm][l]{Codomain:}%
}
\ {\isachardoublequoteopen}{\isacharparenleft}cod\ y{\isacharparenright}{\isasymcdot}y\ {\isasymcong}\ y{\isachardoublequoteclose}\ \isakeyword{and}\isanewline
\ D\isactrlsub i\isactrlsub v{\isacharcolon}\ %
\isamarkupcmt{\makebox[2cm][l]{Domain:}%
}
\ {\isachardoublequoteopen}x{\isasymcdot}{\isacharparenleft}dom\ x{\isacharparenright}\ {\isasymcong}\ x{\isachardoublequoteclose}%
\begin{isamarkuptext}%
The obligatory consistency check is again positive.%
\end{isamarkuptext}\isamarkuptrue%
\ \ \isacommand{lemma}\isamarkupfalse%
\ True\ \ %
\isamarkupcmt{Nitpick finds a model%
}
\isanewline
\ \ \ \ \isacommand{nitpick}\isamarkupfalse%
\ {\isacharbrackleft}satisfy{\isacharcomma}\ user{\isacharunderscore}axioms{\isacharcomma}\ show{\isacharunderscore}all{\isacharcomma}\ format\ {\isacharequal}\ {\isadigit{2}}{\isacharcomma}\ expect\ {\isacharequal}\ genuine{\isacharbrackright}%
\isadelimproof
\ %
\endisadelimproof
\isatagproof
\isacommand{oops}\isamarkupfalse%
\endisatagproof
{\isafoldproof}%
\isadelimproof
\endisadelimproof
\isanewline
\ \ \isacommand{lemma}\isamarkupfalse%
\ \isakeyword{assumes}\ {\isachardoublequoteopen}{\isasymexists}x{\isachardot}\ \isactrlbold {\isasymnot}{\isacharparenleft}E\ x{\isacharparenright}{\isachardoublequoteclose}\ \isakeyword{shows}\ True\ \ %
\isamarkupcmt{Nitpick finds a model%
}
\ \ \isanewline
\ \ \ \ \isacommand{nitpick}\isamarkupfalse%
\ {\isacharbrackleft}satisfy{\isacharcomma}\ user{\isacharunderscore}axioms{\isacharcomma}\ show{\isacharunderscore}all{\isacharcomma}\ format\ {\isacharequal}\ {\isadigit{2}}{\isacharcomma}\ expect\ {\isacharequal}\ genuine{\isacharbrackright}%
\isadelimproof
\ %
\endisadelimproof
\isatagproof
\isacommand{oops}\isamarkupfalse%
\endisatagproof
{\isafoldproof}%
\isadelimproof
\endisadelimproof
\ \isanewline
\ \ \isacommand{lemma}\isamarkupfalse%
\ \isakeyword{assumes}\ {\isachardoublequoteopen}{\isacharparenleft}{\isasymexists}x{\isachardot}\ \isactrlbold {\isasymnot}{\isacharparenleft}E\ x{\isacharparenright}{\isacharparenright}\ {\isasymand}\ {\isacharparenleft}{\isasymexists}x{\isachardot}\ {\isacharparenleft}E\ x{\isacharparenright}{\isacharparenright}{\isachardoublequoteclose}\ \isakeyword{shows}\ True\ \ %
\isamarkupcmt{Nitpick finds a model%
}
\ \isanewline
\ \ \ \ \isacommand{nitpick}\isamarkupfalse%
\ {\isacharbrackleft}satisfy{\isacharcomma}\ user{\isacharunderscore}axioms{\isacharcomma}\ show{\isacharunderscore}all{\isacharcomma}\ format\ {\isacharequal}\ {\isadigit{2}}{\isacharcomma}\ expect\ {\isacharequal}\ genuine{\isacharbrackright}%
\isadelimproof
\ %
\endisadelimproof
\isatagproof
\isacommand{oops}\isamarkupfalse%
\endisatagproof
{\isafoldproof}%
\isadelimproof
\endisadelimproof
\begin{isamarkuptext}%
The Axiom Set III is implied. The only interesting cases are 
 lemmata \isa{C\isactrlsub i\isactrlsub i\isactrlsub iFromIV} and \isa{D\isactrlsub i\isactrlsub i\isactrlsub iFromIV}. Note that the strengthened 
 axiom \isa{E\isactrlsub i\isactrlsub v} is used here.%
\end{isamarkuptext}\isamarkuptrue%
\ \ \isacommand{lemma}\isamarkupfalse%
\ S\isactrlsub i\isactrlsub i\isactrlsub iFromIV{\isacharcolon}\ {\isachardoublequoteopen}{\isacharparenleft}E{\isacharparenleft}x{\isasymcdot}y{\isacharparenright}\ \isactrlbold {\isasymrightarrow}\ {\isacharparenleft}E\ x\ \isactrlbold {\isasymand}\ E\ y{\isacharparenright}{\isacharparenright}\ \isactrlbold {\isasymand}\ {\isacharparenleft}E{\isacharparenleft}dom\ x{\isacharparenright}\ \isactrlbold {\isasymrightarrow}\ E\ x{\isacharparenright}\ \isactrlbold {\isasymand}\ {\isacharparenleft}E{\isacharparenleft}cod\ y{\isacharparenright}\ \isactrlbold {\isasymrightarrow}\ E\ y{\isacharparenright}{\isachardoublequoteclose}\ \ \isanewline
\isadelimproof
\ \ \ \ %
\endisadelimproof
\isatagproof
\isacommand{using}\isamarkupfalse%
\ S\isactrlsub i\isactrlsub v\ \isacommand{by}\isamarkupfalse%
\ blast%
\endisatagproof
{\isafoldproof}%
\isadelimproof
\isanewline
\endisadelimproof
\ \ \isacommand{lemma}\isamarkupfalse%
\ E\isactrlsub i\isactrlsub i\isactrlsub iFromIV{\isacharcolon}\ {\isachardoublequoteopen}E{\isacharparenleft}x{\isasymcdot}y{\isacharparenright}\ \isactrlbold {\isasymleftarrow}\ {\isacharparenleft}dom\ x\ {\isasymcong}\ cod\ y\ \isactrlbold {\isasymand}\ {\isacharparenleft}E{\isacharparenleft}cod\ y{\isacharparenright}{\isacharparenright}{\isacharparenright}{\isachardoublequoteclose}\ \isanewline
\isadelimproof
\ \ \ \ %
\endisadelimproof
\isatagproof
\isacommand{using}\isamarkupfalse%
\ E\isactrlsub i\isactrlsub v\ \isacommand{by}\isamarkupfalse%
\ blast%
\endisatagproof
{\isafoldproof}%
\isadelimproof
\isanewline
\endisadelimproof
\ \ \isacommand{lemma}\isamarkupfalse%
\ A\isactrlsub i\isactrlsub i\isactrlsub iFromIV{\isacharcolon}\ {\isachardoublequoteopen}x{\isasymcdot}{\isacharparenleft}y{\isasymcdot}z{\isacharparenright}\ {\isasymcong}\ {\isacharparenleft}x{\isasymcdot}y{\isacharparenright}{\isasymcdot}z{\isachardoublequoteclose}\ \isanewline
\isadelimproof
\ \ \ \ %
\endisadelimproof
\isatagproof
\isacommand{using}\isamarkupfalse%
\ A\isactrlsub i\isactrlsub v\ \isacommand{by}\isamarkupfalse%
\ blast%
\endisatagproof
{\isafoldproof}%
\isadelimproof
\isanewline
\endisadelimproof
\ \ \isacommand{lemma}\isamarkupfalse%
\ C\isactrlsub i\isactrlsub i\isactrlsub iFromIV{\isacharcolon}\ {\isachardoublequoteopen}E\ y\ \isactrlbold {\isasymrightarrow}\ {\isacharparenleft}I{\isacharparenleft}cod\ y{\isacharparenright}\ \isactrlbold {\isasymand}\ {\isacharparenleft}cod\ y{\isacharparenright}{\isasymcdot}y\ {\isasymcong}\ y{\isacharparenright}{\isachardoublequoteclose}\ \isanewline
\isadelimproof
\ \ \ \ %
\endisadelimproof
\isatagproof
\isacommand{by}\isamarkupfalse%
\ {\isacharparenleft}metis\ C\isactrlsub i\isactrlsub v\ D\isactrlsub i\isactrlsub v\ E\isactrlsub i\isactrlsub v{\isacharparenright}%
\endisatagproof
{\isafoldproof}%
\isadelimproof
\isanewline
\endisadelimproof
\ \ \isacommand{lemma}\isamarkupfalse%
\ D\isactrlsub i\isactrlsub i\isactrlsub iFromIV{\isacharcolon}\ {\isachardoublequoteopen}E\ x\ \isactrlbold {\isasymrightarrow}\ {\isacharparenleft}I{\isacharparenleft}dom\ x{\isacharparenright}\ \isactrlbold {\isasymand}\ x{\isasymcdot}{\isacharparenleft}dom\ x{\isacharparenright}\ {\isasymcong}\ x{\isacharparenright}{\isachardoublequoteclose}\isanewline
\isadelimproof
\ \ \ \ %
\endisadelimproof
\isatagproof
\isacommand{by}\isamarkupfalse%
\ {\isacharparenleft}metis\ C\isactrlsub i\isactrlsub v\ D\isactrlsub i\isactrlsub v\ E\isactrlsub i\isactrlsub v{\isacharparenright}%
\endisatagproof
{\isafoldproof}%
\isadelimproof
\endisadelimproof
\begin{isamarkuptext}%
Vice versa, Axiom Set III implies Axiom Set IV. Hence, both theories are
 equivalent. The interesting cases are lemmata \isa{E\isactrlsub i\isactrlsub vFromIII}, \isa{C\isactrlsub i\isactrlsub vFromIII}
 and \isa{D\isactrlsub i\isactrlsub vFromIII}.%
\end{isamarkuptext}\isamarkuptrue%
\isanewline
\ \ \isacommand{lemma}\isamarkupfalse%
\ S\isactrlsub i\isactrlsub vFromIII{\isacharcolon}\ {\isachardoublequoteopen}{\isacharparenleft}E{\isacharparenleft}x{\isasymcdot}y{\isacharparenright}\ \isactrlbold {\isasymrightarrow}\ {\isacharparenleft}E\ x\ \isactrlbold {\isasymand}\ E\ y{\isacharparenright}{\isacharparenright}\ \isactrlbold {\isasymand}\ {\isacharparenleft}E{\isacharparenleft}dom\ x\ {\isacharparenright}\ \isactrlbold {\isasymrightarrow}\ E\ x{\isacharparenright}\ \isactrlbold {\isasymand}\ {\isacharparenleft}E{\isacharparenleft}cod\ y{\isacharparenright}\ \isactrlbold {\isasymrightarrow}\ E\ y{\isacharparenright}{\isachardoublequoteclose}\ \ \isanewline
\isadelimproof
\ \ \ \ %
\endisadelimproof
\isatagproof
\isacommand{using}\isamarkupfalse%
\ S\isactrlsub i\isactrlsub i\isactrlsub i\ \isacommand{by}\isamarkupfalse%
\ blast%
\endisatagproof
{\isafoldproof}%
\isadelimproof
\isanewline
\endisadelimproof
\ \ \isacommand{lemma}\isamarkupfalse%
\ E\isactrlsub i\isactrlsub vFromIII{\isacharcolon}\ {\isachardoublequoteopen}E{\isacharparenleft}x{\isasymcdot}y{\isacharparenright}\ \isactrlbold {\isasymleftrightarrow}\ {\isacharparenleft}dom\ x\ {\isasymcong}\ cod\ y\ \isactrlbold {\isasymand}\ E{\isacharparenleft}cod\ y{\isacharparenright}{\isacharparenright}{\isachardoublequoteclose}\ \isanewline
\isadelimproof
\ \ \ \ %
\endisadelimproof
\isatagproof
\isacommand{by}\isamarkupfalse%
\ {\isacharparenleft}metis\ {\isacharparenleft}full{\isacharunderscore}types{\isacharparenright}\ A\isactrlsub i\isactrlsub i\isactrlsub i\ C\isactrlsub i\isactrlsub i\isactrlsub i\ D\isactrlsub i\isactrlsub i\isactrlsub i\ E\isactrlsub i\isactrlsub i\isactrlsub i\ S\isactrlsub i\isactrlsub i\isactrlsub i{\isacharparenright}%
\endisatagproof
{\isafoldproof}%
\isadelimproof
\isanewline
\endisadelimproof
\ \ \isacommand{lemma}\isamarkupfalse%
\ A\isactrlsub i\isactrlsub vFromIII{\isacharcolon}\ {\isachardoublequoteopen}x{\isasymcdot}{\isacharparenleft}y{\isasymcdot}z{\isacharparenright}\ {\isasymcong}\ {\isacharparenleft}x{\isasymcdot}y{\isacharparenright}{\isasymcdot}z{\isachardoublequoteclose}\ \isanewline
\isadelimproof
\ \ \ \ %
\endisadelimproof
\isatagproof
\isacommand{using}\isamarkupfalse%
\ A\isactrlsub i\isactrlsub i\isactrlsub i\ \isacommand{by}\isamarkupfalse%
\ blast%
\endisatagproof
{\isafoldproof}%
\isadelimproof
\isanewline
\endisadelimproof
\ \ \isacommand{lemma}\isamarkupfalse%
\ C\isactrlsub i\isactrlsub vFromIII{\isacharcolon}\ {\isachardoublequoteopen}{\isacharparenleft}cod\ y{\isacharparenright}{\isasymcdot}y\ {\isasymcong}\ y{\isachardoublequoteclose}\ \isanewline
\isadelimproof
\ \ \ \ %
\endisadelimproof
\isatagproof
\isacommand{using}\isamarkupfalse%
\ C\isactrlsub i\isactrlsub i\isactrlsub i\ S\isactrlsub i\isactrlsub i\isactrlsub i\ \isacommand{by}\isamarkupfalse%
\ blast%
\endisatagproof
{\isafoldproof}%
\isadelimproof
\isanewline
\endisadelimproof
\ \ \isacommand{lemma}\isamarkupfalse%
\ D\isactrlsub i\isactrlsub vFromIII{\isacharcolon}\ {\isachardoublequoteopen}x{\isasymcdot}{\isacharparenleft}dom\ x{\isacharparenright}\ {\isasymcong}\ x{\isachardoublequoteclose}\ \isanewline
\isadelimproof
\ \ \ \ %
\endisadelimproof
\isatagproof
\isacommand{using}\isamarkupfalse%
\ D\isactrlsub i\isactrlsub i\isactrlsub i\ S\isactrlsub i\isactrlsub i\isactrlsub i\ \isacommand{by}\isamarkupfalse%
\ blast%
\endisatagproof
{\isafoldproof}%
\isadelimproof
\endisadelimproof
\isamarkupsection{Axiom Set V%
}
\isamarkuptrue%
\begin{isamarkuptext}%
Axiom Set V has been proposed by Scott \cite{Scott79} in the 1970s. This set of
 axioms is equivalent to the axiom set presented by Freyd and Scedrov in their textbook
 ``Categories, Allegories'' \cite{FreydScedrov90} when encoded in free logic, corrected/adapted 
 and further simplified. Their axiom set is technically flawed when encoded in our given context. 
 This issue has been detected by automated theorem provers with the same technical infrastructure 
 as employed so far. See the subsequent section for more details. 
 We have modified the axioms of \cite{FreydScedrov90} by replacing the original Kleene 
 equality \isa{{\isasymcong}} in axiom S3 by the non-reflexive, existing identity \isa{{\isasymsimeq}}. Note that 
 the modified axiom \isa{S{\isadigit{3}}} is equivalent to \isa{E\isactrlsub i\isactrlsub v}; see the mutual proofs below.%
\end{isamarkuptext}\isamarkuptrue%
\ S{\isadigit{1}}{\isacharcolon}\ %
\isamarkupcmt{\makebox[2cm][l]{Strictness:}%
}
\ {\isachardoublequoteopen}E{\isacharparenleft}dom\ x{\isacharparenright}\ \isactrlbold {\isasymrightarrow}\ E\ x{\isachardoublequoteclose}\ \isakeyword{and}\isanewline
\ S{\isadigit{2}}{\isacharcolon}\ %
\isamarkupcmt{\makebox[2cm][l]{Strictness:}%
}
\ {\isachardoublequoteopen}E{\isacharparenleft}cod\ y{\isacharparenright}\ \isactrlbold {\isasymrightarrow}\ E\ y{\isachardoublequoteclose}\ \isakeyword{and}\isanewline
\ S{\isadigit{3}}{\isacharcolon}\ %
\isamarkupcmt{\makebox[2cm][l]{Existence:}%
}
\ {\isachardoublequoteopen}E{\isacharparenleft}x{\isasymcdot}y{\isacharparenright}\ \isactrlbold {\isasymleftrightarrow}\ dom\ x\ {\isasymsimeq}\ cod\ y{\isachardoublequoteclose}\ \isakeyword{and}\isanewline
\ S{\isadigit{4}}{\isacharcolon}\ %
\isamarkupcmt{\makebox[2cm][l]{Associativity:}%
}
\ {\isachardoublequoteopen}x{\isasymcdot}{\isacharparenleft}y{\isasymcdot}z{\isacharparenright}\ {\isasymcong}\ {\isacharparenleft}x{\isasymcdot}y{\isacharparenright}{\isasymcdot}z{\isachardoublequoteclose}\ \isakeyword{and}\isanewline
\ S{\isadigit{5}}{\isacharcolon}\ %
\isamarkupcmt{\makebox[2cm][l]{Domain:}%
}
\ {\isachardoublequoteopen}x{\isasymcdot}{\isacharparenleft}dom\ x{\isacharparenright}\ {\isasymcong}\ x{\isachardoublequoteclose}\ \isakeyword{and}\isanewline
\ S{\isadigit{6}}{\isacharcolon}\ %
\isamarkupcmt{\makebox[2cm][l]{Codomain:}%
}
\ {\isachardoublequoteopen}{\isacharparenleft}cod\ y{\isacharparenright}{\isasymcdot}y\ {\isasymcong}\ y{\isachardoublequoteclose}%
\begin{isamarkuptext}%
The obligatory consistency check is again positive.%
\end{isamarkuptext}\isamarkuptrue%
\ \ \isacommand{lemma}\isamarkupfalse%
\ True\ %
\isamarkupcmt{Nitpick finds a model%
}
\isanewline
\ \ \ \ \isacommand{nitpick}\isamarkupfalse%
\ {\isacharbrackleft}satisfy{\isacharcomma}\ user{\isacharunderscore}axioms{\isacharcomma}\ show{\isacharunderscore}all{\isacharcomma}\ format\ {\isacharequal}\ {\isadigit{2}}{\isacharcomma}\ expect\ {\isacharequal}\ genuine{\isacharbrackright}%
\isadelimproof
\ %
\endisadelimproof
\isatagproof
\isacommand{oops}\isamarkupfalse%
\endisatagproof
{\isafoldproof}%
\isadelimproof
\endisadelimproof
\ \isanewline
\ \ \isacommand{lemma}\isamarkupfalse%
\ \isakeyword{assumes}\ {\isachardoublequoteopen}{\isasymexists}x{\isachardot}\ \isactrlbold {\isasymnot}{\isacharparenleft}E\ x{\isacharparenright}{\isachardoublequoteclose}\ \isakeyword{shows}\ True\ \ %
\isamarkupcmt{Nitpick finds a model%
}
\ \ \isanewline
\ \ \ \ \isacommand{nitpick}\isamarkupfalse%
\ {\isacharbrackleft}satisfy{\isacharcomma}\ user{\isacharunderscore}axioms{\isacharcomma}\ show{\isacharunderscore}all{\isacharcomma}\ format\ {\isacharequal}\ {\isadigit{2}}{\isacharcomma}\ expect\ {\isacharequal}\ genuine{\isacharbrackright}%
\isadelimproof
\ %
\endisadelimproof
\isatagproof
\isacommand{oops}\isamarkupfalse%
\endisatagproof
{\isafoldproof}%
\isadelimproof
\endisadelimproof
\ \isanewline
\ \ \isacommand{lemma}\isamarkupfalse%
\ \isakeyword{assumes}\ {\isachardoublequoteopen}{\isacharparenleft}{\isasymexists}x{\isachardot}\ \isactrlbold {\isasymnot}{\isacharparenleft}E\ x{\isacharparenright}{\isacharparenright}\ {\isasymand}\ {\isacharparenleft}{\isasymexists}x{\isachardot}\ {\isacharparenleft}E\ x{\isacharparenright}{\isacharparenright}{\isachardoublequoteclose}\ \isakeyword{shows}\ True\ \ %
\isamarkupcmt{Nitpick finds a model%
}
\ \isanewline
\ \ \ \ \isacommand{nitpick}\isamarkupfalse%
\ {\isacharbrackleft}satisfy{\isacharcomma}\ user{\isacharunderscore}axioms{\isacharcomma}\ show{\isacharunderscore}all{\isacharcomma}\ format\ {\isacharequal}\ {\isadigit{2}}{\isacharcomma}\ expect\ {\isacharequal}\ genuine{\isacharbrackright}%
\isadelimproof
\ %
\endisadelimproof
\isatagproof
\isacommand{oops}\isamarkupfalse%
\endisatagproof
{\isafoldproof}%
\isadelimproof
\endisadelimproof
\begin{isamarkuptext}%
The Axiom Set IV is implied. The only interesting cases are 
 lemmata \isa{S\isactrlsub i\isactrlsub vFromV} and \isa{E\isactrlsub i\isactrlsub vFromV}.%
\end{isamarkuptext}\isamarkuptrue%
\ \ \isacommand{lemma}\isamarkupfalse%
\ S\isactrlsub i\isactrlsub vFromV{\isacharcolon}\ {\isachardoublequoteopen}{\isacharparenleft}E{\isacharparenleft}x{\isasymcdot}y{\isacharparenright}\ \isactrlbold {\isasymrightarrow}\ {\isacharparenleft}E\ x\ \isactrlbold {\isasymand}\ E\ y{\isacharparenright}{\isacharparenright}\ \isactrlbold {\isasymand}\ {\isacharparenleft}E{\isacharparenleft}dom\ x\ {\isacharparenright}\ \isactrlbold {\isasymrightarrow}\ E\ x{\isacharparenright}\ \isactrlbold {\isasymand}\ {\isacharparenleft}E{\isacharparenleft}cod\ y{\isacharparenright}\ \isactrlbold {\isasymrightarrow}\ E\ y{\isacharparenright}{\isachardoublequoteclose}\ \ \ \isanewline
\isadelimproof
\ \ \ \ %
\endisadelimproof
\isatagproof
\isacommand{using}\isamarkupfalse%
\ S{\isadigit{1}}\ S{\isadigit{2}}\ S{\isadigit{3}}\ \isacommand{by}\isamarkupfalse%
\ blast%
\endisatagproof
{\isafoldproof}%
\isadelimproof
\isanewline
\endisadelimproof
\ \ \isacommand{lemma}\isamarkupfalse%
\ E\isactrlsub i\isactrlsub vFromV{\isacharcolon}\ {\isachardoublequoteopen}E{\isacharparenleft}x{\isasymcdot}y{\isacharparenright}\ \isactrlbold {\isasymleftrightarrow}\ {\isacharparenleft}dom\ x\ {\isasymcong}\ cod\ y\ \isactrlbold {\isasymand}\ E{\isacharparenleft}cod\ y{\isacharparenright}{\isacharparenright}{\isachardoublequoteclose}\ \isanewline
\isadelimproof
\ \ \ \ %
\endisadelimproof
\isatagproof
\isacommand{using}\isamarkupfalse%
\ S{\isadigit{3}}\ \isacommand{by}\isamarkupfalse%
\ metis%
\endisatagproof
{\isafoldproof}%
\isadelimproof
\isanewline
\endisadelimproof
\ \ \isacommand{lemma}\isamarkupfalse%
\ A\isactrlsub i\isactrlsub vFromV{\isacharcolon}\ {\isachardoublequoteopen}x{\isasymcdot}{\isacharparenleft}y{\isasymcdot}z{\isacharparenright}\ {\isasymcong}\ {\isacharparenleft}x{\isasymcdot}y{\isacharparenright}{\isasymcdot}z{\isachardoublequoteclose}\ \isanewline
\isadelimproof
\ \ \ \ %
\endisadelimproof
\isatagproof
\isacommand{using}\isamarkupfalse%
\ S{\isadigit{4}}\ \isacommand{by}\isamarkupfalse%
\ blast%
\endisatagproof
{\isafoldproof}%
\isadelimproof
\isanewline
\endisadelimproof
\ \ \isacommand{lemma}\isamarkupfalse%
\ C\isactrlsub i\isactrlsub vFromV{\isacharcolon}\ {\isachardoublequoteopen}{\isacharparenleft}cod\ y{\isacharparenright}{\isasymcdot}y\ {\isasymcong}\ y{\isachardoublequoteclose}\ \isanewline
\isadelimproof
\ \ \ \ %
\endisadelimproof
\isatagproof
\isacommand{using}\isamarkupfalse%
\ S{\isadigit{6}}\ \isacommand{by}\isamarkupfalse%
\ blast%
\endisatagproof
{\isafoldproof}%
\isadelimproof
\isanewline
\endisadelimproof
\ \ \isacommand{lemma}\isamarkupfalse%
\ D\isactrlsub i\isactrlsub vFromV{\isacharcolon}\ {\isachardoublequoteopen}x{\isasymcdot}{\isacharparenleft}dom\ x{\isacharparenright}\ {\isasymcong}\ x{\isachardoublequoteclose}\ \isanewline
\isadelimproof
\ \ \ \ %
\endisadelimproof
\isatagproof
\isacommand{using}\isamarkupfalse%
\ S{\isadigit{5}}\ \isacommand{by}\isamarkupfalse%
\ blast%
\endisatagproof
{\isafoldproof}%
\isadelimproof
\endisadelimproof
\begin{isamarkuptext}%
Vice versa, Axiom Set IV implies Axiom Set V. Hence, both theories are
 equivalent.%
\end{isamarkuptext}\isamarkuptrue%
\ \ \isacommand{lemma}\isamarkupfalse%
\ S{\isadigit{1}}FromV{\isacharcolon}\ \ {\isachardoublequoteopen}E{\isacharparenleft}dom\ x{\isacharparenright}\ \isactrlbold {\isasymrightarrow}\ E\ x{\isachardoublequoteclose}\ \isanewline
\isadelimproof
\ \ \ \ %
\endisadelimproof
\isatagproof
\isacommand{using}\isamarkupfalse%
\ S\isactrlsub i\isactrlsub v\ \isacommand{by}\isamarkupfalse%
\ blast%
\endisatagproof
{\isafoldproof}%
\isadelimproof
\isanewline
\endisadelimproof
\ \ \isacommand{lemma}\isamarkupfalse%
\ S{\isadigit{2}}FromV{\isacharcolon}\ \ {\isachardoublequoteopen}E{\isacharparenleft}cod\ y{\isacharparenright}\ \isactrlbold {\isasymrightarrow}\ E\ y{\isachardoublequoteclose}\ \isanewline
\isadelimproof
\ \ \ \ %
\endisadelimproof
\isatagproof
\isacommand{using}\isamarkupfalse%
\ S\isactrlsub i\isactrlsub v\ \isacommand{by}\isamarkupfalse%
\ blast%
\endisatagproof
{\isafoldproof}%
\isadelimproof
\isanewline
\endisadelimproof
\ \ \isacommand{lemma}\isamarkupfalse%
\ S{\isadigit{3}}FromV{\isacharcolon}\ \ {\isachardoublequoteopen}E{\isacharparenleft}x{\isasymcdot}y{\isacharparenright}\ \isactrlbold {\isasymleftrightarrow}\ dom\ x\ {\isasymsimeq}\ cod\ y{\isachardoublequoteclose}\ \isanewline
\isadelimproof
\ \ \ \ %
\endisadelimproof
\isatagproof
\isacommand{using}\isamarkupfalse%
\ E\isactrlsub i\isactrlsub v\ \isacommand{by}\isamarkupfalse%
\ metis%
\endisatagproof
{\isafoldproof}%
\isadelimproof
\isanewline
\endisadelimproof
\ \ \isacommand{lemma}\isamarkupfalse%
\ S{\isadigit{4}}FromV{\isacharcolon}\ \ {\isachardoublequoteopen}x{\isasymcdot}{\isacharparenleft}y{\isasymcdot}z{\isacharparenright}\ {\isasymcong}\ {\isacharparenleft}x{\isasymcdot}y{\isacharparenright}{\isasymcdot}z{\isachardoublequoteclose}\ \isanewline
\isadelimproof
\ \ \ \ %
\endisadelimproof
\isatagproof
\isacommand{using}\isamarkupfalse%
\ A\isactrlsub i\isactrlsub v\ \isacommand{by}\isamarkupfalse%
\ blast%
\endisatagproof
{\isafoldproof}%
\isadelimproof
\isanewline
\endisadelimproof
\ \ \isacommand{lemma}\isamarkupfalse%
\ S{\isadigit{5}}FromV{\isacharcolon}\ \ {\isachardoublequoteopen}x{\isasymcdot}{\isacharparenleft}dom\ x{\isacharparenright}\ {\isasymcong}\ x{\isachardoublequoteclose}\ \isanewline
\isadelimproof
\ \ \ \ %
\endisadelimproof
\isatagproof
\isacommand{using}\isamarkupfalse%
\ D\isactrlsub i\isactrlsub v\ \isacommand{by}\isamarkupfalse%
\ blast%
\endisatagproof
{\isafoldproof}%
\isadelimproof
\isanewline
\endisadelimproof
\ \ \isacommand{lemma}\isamarkupfalse%
\ S{\isadigit{6}}FromV{\isacharcolon}\ \ {\isachardoublequoteopen}{\isacharparenleft}cod\ y{\isacharparenright}{\isasymcdot}y\ {\isasymcong}\ y{\isachardoublequoteclose}\ \isanewline
\isadelimproof
\ \ \ \ %
\endisadelimproof
\isatagproof
\isacommand{using}\isamarkupfalse%
\ C\isactrlsub i\isactrlsub v\ \isacommand{by}\isamarkupfalse%
\ blast\isanewline
\endisatagproof
{\isafoldproof}%
\isadelimproof
\endisadelimproof
\isamarkupsection{Axiom Sets VI and VII%
}
\isamarkuptrue%
\begin{isamarkuptext}%
The axiom set of Freyd and Scedrov from their textbook
 ``Categories, Allegories'' \cite{FreydScedrov90} becomes inconsistent in our free logic setting if we assume 
  non-existing objects of type \isa{i}, respectively, if we assume that the operations are 
  non-total.  Freyd and Scedrov employ a different notation for 
  \isa{dom\ x} and \isa{cod\ x}. They denote these operations by \isa{{\isasymbox}x} 
  and \isa{x{\isasymbox}}. Moreover, they employ diagrammatic composition \isa{{\isacharparenleft}f{\isasymcdot}g{\isacharparenright}\ x\ {\isasymcong}\ g{\isacharparenleft}f\ x{\isacharparenright}} 
  (functional composition from left to right) instead of the set-theoretic 
  definition \isa{{\isacharparenleft}f{\isasymcdot}g{\isacharparenright}\ x\ {\isasymcong}\ f{\isacharparenleft}g\ x{\isacharparenright}} (functional composition from right to left) used so far.
 
  We leave it to the reader to verify that their axiom system corresponds to the 
  axiom system given below modulo an appropriate conversion of notation.\footnote{A recipe for 
  this translation is as follows: (i) replace all \isa{x{\isasymcdot}y} by \isa{y{\isasymcdot}x}, 
(ii) rename the variables to get them again in alphabetical order,
(iii) replace \isa{{\isasymphi}{\isasymbox}} by \isa{cod\ {\isasymphi}} and \isa{{\isasymbox}{\isasymphi}}  by \isa{dom\ {\isasymphi}}, and finally
(iv) replace \isa{cod\ y\ {\isasymcong}\ dom\ x} (resp. \isa{cod\ y\ {\isasymsimeq}\ dom\ x}) 
   by \isa{dom\ x\ {\isasymcong}\ cod\ y} (resp. \isa{dom\ x\ {\isasymsimeq}\ cod\ y}).}
  In Subsection 9.2 we will also analyze their axiom system using their original notation.

  A main difference in the system by Freyd and Scedrov to our Axiom Set V from above concerns
  axiom \isa{S{\isadigit{3}}}. Namely, instead of the non-reflexive \isa{{\isasymsimeq}}, they use Kleene 
  equality \isa{{\isasymcong}}, cf. definition 1.11 on page 3 of \cite{FreydScedrov90}.\footnote{Def. 1.11 in Freyd 
  Scedrov: ``The ordinary equality sign \isa{{\isacharequal}} [i.e., our \isa{{\isasymcong}}] will be used in the
  symmetric sense, to wit: if either side is defined then so is the other and they are equal. \ldots''} 
  The difference seems minor, but in our free logic setting it has the effect to cause the mentioned
  constricted inconsistency issue. This could perhaps be an oversight, or it could indicate
  that Freyd and Scedrov actually mean the Axiom Set VIII below (where the variables in the axioms range 
  over defined objects only). However, in Axiom Set VIII we had to (re-)introduce explicit 
  strictness conditions to ensure equivalence to the Axiom Set V by Scott.%
\end{isamarkuptext}\isamarkuptrue%
\isamarkupsubsection{Axiom Set VI%
}
\isamarkuptrue%
\ \ A{\isadigit{1}}{\isacharcolon}\ {\isachardoublequoteopen}E{\isacharparenleft}x{\isasymcdot}y{\isacharparenright}\ \isactrlbold {\isasymleftrightarrow}\ dom\ x\ {\isasymsimeq}\ cod\ y{\isachardoublequoteclose}\ \isakeyword{and}\isanewline
\ A{\isadigit{2}}a{\isacharcolon}\ {\isachardoublequoteopen}cod{\isacharparenleft}dom\ x{\isacharparenright}\ {\isasymcong}\ dom\ x{\isachardoublequoteclose}\ \isakeyword{and}\ \ \isanewline
\ A{\isadigit{2}}b{\isacharcolon}\ {\isachardoublequoteopen}dom{\isacharparenleft}cod\ y{\isacharparenright}\ {\isasymcong}\ cod\ y{\isachardoublequoteclose}\ \isakeyword{and}\ \ \isanewline
\ A{\isadigit{3}}a{\isacharcolon}\ {\isachardoublequoteopen}x{\isasymcdot}{\isacharparenleft}dom\ x{\isacharparenright}\ {\isasymcong}\ x{\isachardoublequoteclose}\ \isakeyword{and}\ \isanewline
\ A{\isadigit{3}}b{\isacharcolon}\ {\isachardoublequoteopen}{\isacharparenleft}cod\ y{\isacharparenright}{\isasymcdot}y\ {\isasymcong}\ y{\isachardoublequoteclose}\ \isakeyword{and}\ \isanewline
\ A{\isadigit{4}}a{\isacharcolon}\ {\isachardoublequoteopen}dom{\isacharparenleft}x{\isasymcdot}y{\isacharparenright}\ {\isasymcong}\ dom{\isacharparenleft}{\isacharparenleft}dom\ x{\isacharparenright}{\isasymcdot}y{\isacharparenright}{\isachardoublequoteclose}\ \isakeyword{and}\ \isanewline
\ A{\isadigit{4}}b{\isacharcolon}\ {\isachardoublequoteopen}cod{\isacharparenleft}x{\isasymcdot}y{\isacharparenright}\ {\isasymcong}\ cod{\isacharparenleft}x{\isasymcdot}{\isacharparenleft}cod\ y{\isacharparenright}{\isacharparenright}{\isachardoublequoteclose}\ \isakeyword{and}\ \isanewline
\ \ A{\isadigit{5}}{\isacharcolon}\ {\isachardoublequoteopen}x{\isasymcdot}{\isacharparenleft}y{\isasymcdot}z{\isacharparenright}\ {\isasymcong}\ {\isacharparenleft}x{\isasymcdot}y{\isacharparenright}{\isasymcdot}z{\isachardoublequoteclose}%
\begin{isamarkuptext}%
The obligatory consistency checks are again positive. 
 But note that this only holds when we use \isa{{\isasymsimeq}} instead of  \isa{{\isasymcong}} in  \isa{A{\isadigit{1}}}.%
\end{isamarkuptext}\isamarkuptrue%
\ \ \isacommand{lemma}\isamarkupfalse%
\ True\ \ %
\isamarkupcmt{Nitpick finds a model%
}
\isanewline
\ \ \ \ \isacommand{nitpick}\isamarkupfalse%
\ {\isacharbrackleft}satisfy{\isacharcomma}\ user{\isacharunderscore}axioms{\isacharcomma}\ show{\isacharunderscore}all{\isacharcomma}\ format\ {\isacharequal}\ {\isadigit{2}}{\isacharcomma}\ expect\ {\isacharequal}\ genuine{\isacharbrackright}%
\isadelimproof
\ %
\endisadelimproof
\isatagproof
\isacommand{oops}\isamarkupfalse%
\endisatagproof
{\isafoldproof}%
\isadelimproof
\endisadelimproof
\isanewline
\ \ \isacommand{lemma}\isamarkupfalse%
\ \isakeyword{assumes}\ {\isachardoublequoteopen}{\isasymexists}x{\isachardot}\ \isactrlbold {\isasymnot}{\isacharparenleft}E\ x{\isacharparenright}{\isachardoublequoteclose}\ \isakeyword{shows}\ True\ \ \ %
\isamarkupcmt{Nitpick finds a model%
}
\ \ \isanewline
\ \ \ \ \isacommand{nitpick}\isamarkupfalse%
\ {\isacharbrackleft}satisfy{\isacharcomma}\ user{\isacharunderscore}axioms{\isacharcomma}\ show{\isacharunderscore}all{\isacharcomma}\ format\ {\isacharequal}\ {\isadigit{2}}{\isacharcomma}\ expect\ {\isacharequal}\ genuine{\isacharbrackright}%
\isadelimproof
\ %
\endisadelimproof
\isatagproof
\isacommand{oops}\isamarkupfalse%
\endisatagproof
{\isafoldproof}%
\isadelimproof
\endisadelimproof
\isanewline
\ \ \isacommand{lemma}\isamarkupfalse%
\ \isakeyword{assumes}\ {\isachardoublequoteopen}{\isacharparenleft}{\isasymexists}x{\isachardot}\ \isactrlbold {\isasymnot}{\isacharparenleft}E\ x{\isacharparenright}{\isacharparenright}\ {\isasymand}\ {\isacharparenleft}{\isasymexists}x{\isachardot}\ {\isacharparenleft}E\ x{\isacharparenright}{\isacharparenright}{\isachardoublequoteclose}\ \isakeyword{shows}\ True\ \ %
\isamarkupcmt{Nitpick finds a model%
}
\ \isanewline
\ \ \ \ \isacommand{nitpick}\isamarkupfalse%
\ {\isacharbrackleft}satisfy{\isacharcomma}\ user{\isacharunderscore}axioms{\isacharcomma}\ show{\isacharunderscore}all{\isacharcomma}\ format\ {\isacharequal}\ {\isadigit{2}}{\isacharcomma}\ expect\ {\isacharequal}\ genuine{\isacharbrackright}%
\isadelimproof
\ %
\endisadelimproof
\isatagproof
\isacommand{oops}\isamarkupfalse%
\endisatagproof
{\isafoldproof}%
\isadelimproof
\endisadelimproof
\begin{isamarkuptext}%
Axiom Set VI implies Axiom Set V.%
\end{isamarkuptext}\isamarkuptrue%
\ \ \isacommand{lemma}\isamarkupfalse%
\ S{\isadigit{1}}FromVI{\isacharcolon}\ {\isachardoublequoteopen}E{\isacharparenleft}dom\ x{\isacharparenright}\ \isactrlbold {\isasymrightarrow}\ E\ x{\isachardoublequoteclose}\ \isanewline
\isadelimproof
\ \ \ \ %
\endisadelimproof
\isatagproof
\isacommand{by}\isamarkupfalse%
\ {\isacharparenleft}metis\ A{\isadigit{1}}\ A{\isadigit{2}}a\ A{\isadigit{3}}a{\isacharparenright}%
\endisatagproof
{\isafoldproof}%
\isadelimproof
\isanewline
\endisadelimproof
\ \ \isacommand{lemma}\isamarkupfalse%
\ S{\isadigit{2}}FromVI{\isacharcolon}\ {\isachardoublequoteopen}E{\isacharparenleft}cod\ y{\isacharparenright}\ \isactrlbold {\isasymrightarrow}\ E\ y{\isachardoublequoteclose}\ \isanewline
\isadelimproof
\ \ \ \ %
\endisadelimproof
\isatagproof
\isacommand{using}\isamarkupfalse%
\ A{\isadigit{1}}\ A{\isadigit{2}}b\ A{\isadigit{3}}b\ \isacommand{by}\isamarkupfalse%
\ metis%
\endisatagproof
{\isafoldproof}%
\isadelimproof
\isanewline
\endisadelimproof
\ \ \isacommand{lemma}\isamarkupfalse%
\ S{\isadigit{3}}FromVI{\isacharcolon}\ {\isachardoublequoteopen}E{\isacharparenleft}x{\isasymcdot}y{\isacharparenright}\ \isactrlbold {\isasymleftrightarrow}\ dom\ x\ {\isasymsimeq}\ cod\ y{\isachardoublequoteclose}\ \isanewline
\isadelimproof
\ \ \ \ %
\endisadelimproof
\isatagproof
\isacommand{by}\isamarkupfalse%
\ {\isacharparenleft}metis\ A{\isadigit{1}}{\isacharparenright}%
\endisatagproof
{\isafoldproof}%
\isadelimproof
\isanewline
\endisadelimproof
\ \ \isacommand{lemma}\isamarkupfalse%
\ S{\isadigit{4}}FromVI{\isacharcolon}\ {\isachardoublequoteopen}x{\isasymcdot}{\isacharparenleft}y{\isasymcdot}z{\isacharparenright}\ {\isasymcong}\ {\isacharparenleft}x{\isasymcdot}y{\isacharparenright}{\isasymcdot}z{\isachardoublequoteclose}\ \isanewline
\isadelimproof
\ \ \ \ %
\endisadelimproof
\isatagproof
\isacommand{using}\isamarkupfalse%
\ A{\isadigit{5}}\ \isacommand{by}\isamarkupfalse%
\ blast%
\endisatagproof
{\isafoldproof}%
\isadelimproof
\isanewline
\endisadelimproof
\ \ \isacommand{lemma}\isamarkupfalse%
\ S{\isadigit{5}}FromVI{\isacharcolon}\ {\isachardoublequoteopen}x{\isasymcdot}{\isacharparenleft}dom\ x{\isacharparenright}\ {\isasymcong}\ x{\isachardoublequoteclose}\ \isanewline
\isadelimproof
\ \ \ \ %
\endisadelimproof
\isatagproof
\isacommand{using}\isamarkupfalse%
\ A{\isadigit{3}}a\ \isacommand{by}\isamarkupfalse%
\ blast%
\endisatagproof
{\isafoldproof}%
\isadelimproof
\isanewline
\endisadelimproof
\ \ \isacommand{lemma}\isamarkupfalse%
\ S{\isadigit{6}}FromVI{\isacharcolon}\ {\isachardoublequoteopen}{\isacharparenleft}cod\ y{\isacharparenright}{\isasymcdot}y\ {\isasymcong}\ y{\isachardoublequoteclose}\ \isanewline
\isadelimproof
\ \ \ \ %
\endisadelimproof
\isatagproof
\isacommand{using}\isamarkupfalse%
\ A{\isadigit{3}}b\ \isacommand{by}\isamarkupfalse%
\ blast%
\endisatagproof
{\isafoldproof}%
\isadelimproof
\endisadelimproof
\begin{isamarkuptext}%
Note, too, that Axiom Set VI is redundant. For example, axioms \isa{A{\isadigit{4}}a} and \isa{A{\isadigit{4}}b} are
  implied from the others. This kind of flaw in presenting axioms in our view is a more serious oversight.
  The automated theorem provers can quickly reveal such redundancies.%
\end{isamarkuptext}\isamarkuptrue%
\ \ \isacommand{lemma}\isamarkupfalse%
\ A{\isadigit{4}}aRedundant{\isacharcolon}\ {\isachardoublequoteopen}dom{\isacharparenleft}x{\isasymcdot}y{\isacharparenright}\ {\isasymcong}\ dom{\isacharparenleft}{\isacharparenleft}dom\ x{\isacharparenright}{\isasymcdot}y{\isacharparenright}{\isachardoublequoteclose}\ \isanewline
\isadelimproof
\ \ \ \ %
\endisadelimproof
\isatagproof
\isacommand{using}\isamarkupfalse%
\ A{\isadigit{1}}\ A{\isadigit{2}}a\ A{\isadigit{3}}a\ A{\isadigit{5}}\ \isacommand{by}\isamarkupfalse%
\ metis%
\endisatagproof
{\isafoldproof}%
\isadelimproof
\isanewline
\endisadelimproof
\ \ \isacommand{lemma}\isamarkupfalse%
\ A{\isadigit{4}}bRedundant{\isacharcolon}\ {\isachardoublequoteopen}cod{\isacharparenleft}x{\isasymcdot}y{\isacharparenright}\ {\isasymcong}\ cod{\isacharparenleft}x{\isasymcdot}{\isacharparenleft}cod\ y{\isacharparenright}{\isacharparenright}{\isachardoublequoteclose}\ \ \isanewline
\isadelimproof
\ \ \ \ %
\endisadelimproof
\isatagproof
\isacommand{using}\isamarkupfalse%
\ A{\isadigit{1}}\ A{\isadigit{2}}b\ A{\isadigit{3}}b\ A{\isadigit{5}}\ \isacommand{by}\isamarkupfalse%
\ metis%
\endisatagproof
{\isafoldproof}%
\isadelimproof
\endisadelimproof
\begin{isamarkuptext}%
Our attempts to further reduce the axioms set \isa{{\isacharparenleft}A{\isadigit{1}}\ A{\isadigit{2}}a\ A{\isadigit{2}}b\ A{\isadigit{3}}a\ A{\isadigit{3}}b\ A{\isadigit{5}}{\isacharparenright}}  were not successful.
Alternatively, we can e.g. keep \isa{A{\isadigit{4}}a} and \isa{A{\isadigit{4}}b} and show that axioms \isa{A{\isadigit{2}}a} 
and \isa{A{\isadigit{2}}b} are implied.%
\end{isamarkuptext}\isamarkuptrue%
\ \ \isacommand{lemma}\isamarkupfalse%
\ A{\isadigit{2}}aRedundant{\isacharcolon}\ {\isachardoublequoteopen}cod{\isacharparenleft}dom\ x{\isacharparenright}\ {\isasymcong}\ dom\ x{\isachardoublequoteclose}\ \isanewline
\isadelimproof
\ \ \ \ %
\endisadelimproof
\isatagproof
\isacommand{using}\isamarkupfalse%
\ A{\isadigit{1}}\ A{\isadigit{3}}a\ A{\isadigit{3}}b\ A{\isadigit{4}}a\ A{\isadigit{4}}b\ \isacommand{by}\isamarkupfalse%
\ smt%
\endisatagproof
{\isafoldproof}%
\isadelimproof
\isanewline
\endisadelimproof
\ \ \isacommand{lemma}\isamarkupfalse%
\ A{\isadigit{2}}bRedundant{\isacharcolon}\ {\isachardoublequoteopen}dom{\isacharparenleft}cod\ y{\isacharparenright}\ {\isasymcong}\ cod\ y{\isachardoublequoteclose}\ \isanewline
\isadelimproof
\ \ \ \ %
\endisadelimproof
\isatagproof
\isacommand{using}\isamarkupfalse%
\ \ A{\isadigit{1}}\ A{\isadigit{3}}a\ A{\isadigit{3}}b\ A{\isadigit{4}}a\ A{\isadigit{4}}b\ \isacommand{by}\isamarkupfalse%
\ smt%
\endisatagproof
{\isafoldproof}%
\isadelimproof
\endisadelimproof
\begin{isamarkuptext}%
Again, attempts to further reduce the set \isa{{\isacharparenleft}A{\isadigit{1}}\ A{\isadigit{3}}a\ A{\isadigit{3}}b\ A{\isadigit{4}}a\ A{\isadigit{4}}b\ A{\isadigit{5}}{\isacharparenright}} were not successful.
   Other reduced sets of axioms we identified in experiments are \isa{{\isacharparenleft}A{\isadigit{1}}\ A{\isadigit{2}}a\ A{\isadigit{3}}a\ A{\isadigit{3}}b\ A{\isadigit{4}}b\ A{\isadigit{5}}{\isacharparenright}} and
    \isa{{\isacharparenleft}A{\isadigit{1}}\ A{\isadigit{2}}b\ A{\isadigit{3}}a\ A{\isadigit{3}}b\ A{\isadigit{4}}a\ A{\isadigit{5}}{\isacharparenright}}. Attempts to remove axioms \isa{A{\isadigit{1}}}, \isa{A{\isadigit{3}}a}, 
    \isa{A{\isadigit{3}}b}, and \isa{A{\isadigit{5}}} from Axiom Set VI failed. Nitpick shows that they are independent. 

   However, when assuming strictness of \isa{dom} and \isa{cod}, the axioms \isa{A{\isadigit{2}}a}, 
   \isa{A{\isadigit{2}}b}, \isa{A{\isadigit{4}}a} and \isa{A{\isadigit{4}}b} are all implied. Hence, under this 
   assumptions, the reasoning tools quickly identify \isa{{\isacharparenleft}A{\isadigit{1}}\ A{\isadigit{3}}a\ A{\isadigit{3}}b\ A{\isadigit{5}}{\isacharparenright}} as a minimal axiom 
   set, which then exactly matches the Axiom Set V from above.\footnote{This minimal set of axioms 
   is also mentioned by Freyd in \cite{Freyd16} and attributed to Martin Knopman. However, the proof
   sketch presented there seems to fail when the adapted version of A1 (with \isa{{\isasymsimeq}}) is employed.}%
\end{isamarkuptext}\isamarkuptrue%
\begin{isamarkuptext}%
Axiom Set V implies Axiom Set VI. Hence, both theories are equivalent.%
\end{isamarkuptext}\isamarkuptrue%
\ \ \isacommand{lemma}\isamarkupfalse%
\ \ A{\isadigit{1}}FromV{\isacharcolon}\ {\isachardoublequoteopen}E{\isacharparenleft}x{\isasymcdot}y{\isacharparenright}\ \isactrlbold {\isasymleftrightarrow}\ dom\ x\ {\isasymsimeq}\ cod\ y{\isachardoublequoteclose}\ \isanewline
\isadelimproof
\ \ \ \ %
\endisadelimproof
\isatagproof
\isacommand{using}\isamarkupfalse%
\ S{\isadigit{3}}\ \isacommand{by}\isamarkupfalse%
\ blast%
\endisatagproof
{\isafoldproof}%
\isadelimproof
\isanewline
\endisadelimproof
\ \ \isacommand{lemma}\isamarkupfalse%
\ A{\isadigit{2}}aFromV{\isacharcolon}\ {\isachardoublequoteopen}cod{\isacharparenleft}dom\ x{\isacharparenright}\ {\isasymcong}\ dom\ x{\isachardoublequoteclose}\ \ \isanewline
\isadelimproof
\ \ \ \ %
\endisadelimproof
\isatagproof
\isacommand{by}\isamarkupfalse%
\ {\isacharparenleft}metis\ S{\isadigit{1}}\ S{\isadigit{2}}\ S{\isadigit{3}}\ S{\isadigit{5}}{\isacharparenright}%
\endisatagproof
{\isafoldproof}%
\isadelimproof
\isanewline
\endisadelimproof
\ \ \isacommand{lemma}\isamarkupfalse%
\ A{\isadigit{2}}bFromV{\isacharcolon}\ {\isachardoublequoteopen}dom{\isacharparenleft}cod\ y{\isacharparenright}\ {\isasymcong}\ cod\ y{\isachardoublequoteclose}\ \ \isanewline
\isadelimproof
\ \ \ \ %
\endisadelimproof
\isatagproof
\isacommand{using}\isamarkupfalse%
\ S{\isadigit{1}}\ S{\isadigit{2}}\ S{\isadigit{3}}\ S{\isadigit{6}}\ \isacommand{by}\isamarkupfalse%
\ metis%
\endisatagproof
{\isafoldproof}%
\isadelimproof
\isanewline
\endisadelimproof
\ \ \isacommand{lemma}\isamarkupfalse%
\ A{\isadigit{3}}aFromV{\isacharcolon}\ {\isachardoublequoteopen}x{\isasymcdot}{\isacharparenleft}dom\ x{\isacharparenright}\ {\isasymcong}\ x{\isachardoublequoteclose}\ \isanewline
\isadelimproof
\ \ \ \ %
\endisadelimproof
\isatagproof
\isacommand{using}\isamarkupfalse%
\ S{\isadigit{5}}\ \isacommand{by}\isamarkupfalse%
\ blast%
\endisatagproof
{\isafoldproof}%
\isadelimproof
\isanewline
\endisadelimproof
\ \ \isacommand{lemma}\isamarkupfalse%
\ A{\isadigit{3}}bFromV{\isacharcolon}\ {\isachardoublequoteopen}{\isacharparenleft}cod\ y{\isacharparenright}{\isasymcdot}y\ {\isasymcong}\ y{\isachardoublequoteclose}\ \isanewline
\isadelimproof
\ \ \ \ %
\endisadelimproof
\isatagproof
\isacommand{using}\isamarkupfalse%
\ S{\isadigit{6}}\ \isacommand{by}\isamarkupfalse%
\ blast%
\endisatagproof
{\isafoldproof}%
\isadelimproof
\isanewline
\endisadelimproof
\ \ \isacommand{lemma}\isamarkupfalse%
\ A{\isadigit{4}}aFromV{\isacharcolon}\ {\isachardoublequoteopen}dom{\isacharparenleft}x{\isasymcdot}y{\isacharparenright}\ {\isasymcong}\ dom{\isacharparenleft}{\isacharparenleft}dom\ x{\isacharparenright}{\isasymcdot}y{\isacharparenright}{\isachardoublequoteclose}\isanewline
\isadelimproof
\ \ \ \ %
\endisadelimproof
\isatagproof
\isacommand{by}\isamarkupfalse%
\ {\isacharparenleft}metis\ S{\isadigit{1}}\ S{\isadigit{3}}\ S{\isadigit{4}}\ S{\isadigit{5}}\ S{\isadigit{6}}{\isacharparenright}%
\endisatagproof
{\isafoldproof}%
\isadelimproof
\isanewline
\endisadelimproof
\ \ \isacommand{lemma}\isamarkupfalse%
\ A{\isadigit{4}}bFromV{\isacharcolon}\ {\isachardoublequoteopen}cod{\isacharparenleft}x{\isasymcdot}y{\isacharparenright}\ {\isasymcong}\ cod{\isacharparenleft}x{\isasymcdot}{\isacharparenleft}cod\ y{\isacharparenright}{\isacharparenright}{\isachardoublequoteclose}\ \isanewline
\isadelimproof
\ \ \ \ %
\endisadelimproof
\isatagproof
\isacommand{by}\isamarkupfalse%
\ {\isacharparenleft}metis\ S{\isadigit{2}}\ S{\isadigit{3}}\ S{\isadigit{4}}\ S{\isadigit{5}}\ S{\isadigit{6}}{\isacharparenright}%
\endisatagproof
{\isafoldproof}%
\isadelimproof
\isanewline
\endisadelimproof
\ \ \isacommand{lemma}\isamarkupfalse%
\ \ A{\isadigit{5}}FromV{\isacharcolon}\ {\isachardoublequoteopen}x{\isasymcdot}{\isacharparenleft}y{\isasymcdot}z{\isacharparenright}\ {\isasymcong}\ {\isacharparenleft}x{\isasymcdot}y{\isacharparenright}{\isasymcdot}z{\isachardoublequoteclose}\ \isanewline
\isadelimproof
\ \ \ \ %
\endisadelimproof
\isatagproof
\isacommand{using}\isamarkupfalse%
\ S{\isadigit{4}}\ \isacommand{by}\isamarkupfalse%
\ blast%
\endisatagproof
{\isafoldproof}%
\isadelimproof
\endisadelimproof
\isamarkupsubsection{Axiom Set VII%
}
\isamarkuptrue%
\begin{isamarkuptext}%
We now study the constricted inconsistency in Axiom Set VI when replacing  \isa{{\isasymsimeq}}  
 in  \isa{A{\isadigit{1}}} by  \isa{{\isasymcong}}. We call this Axiom Set VII. This set corresponds
 modulo representational transformation to the axioms as presented by Freyd and Scedrov. Remember, however,
 that the free variables are ranging here over all objects, defined or undefined. Below, when we study
 Axiom Set VIII, we will restrict the variables to range only over existing objects.%
\end{isamarkuptext}\isamarkuptrue%
\ \ A{\isadigit{1}}{\isacharcolon}\ {\isachardoublequoteopen}E{\isacharparenleft}x{\isasymcdot}y{\isacharparenright}\ \isactrlbold {\isasymleftrightarrow}\ dom\ x\ {\isasymcong}\ cod\ y{\isachardoublequoteclose}\ \isakeyword{and}\isanewline
\ A{\isadigit{2}}a{\isacharcolon}\ {\isachardoublequoteopen}cod{\isacharparenleft}dom\ x{\isacharparenright}\ {\isasymcong}\ dom\ x\ {\isachardoublequoteclose}\ \isakeyword{and}\ \ \isanewline
\ A{\isadigit{2}}b{\isacharcolon}\ {\isachardoublequoteopen}dom{\isacharparenleft}cod\ y{\isacharparenright}\ {\isasymcong}\ cod\ y{\isachardoublequoteclose}\ \isakeyword{and}\ \ \isanewline
\ A{\isadigit{3}}a{\isacharcolon}\ {\isachardoublequoteopen}x{\isasymcdot}{\isacharparenleft}dom\ x{\isacharparenright}\ {\isasymcong}\ x{\isachardoublequoteclose}\ \isakeyword{and}\ \isanewline
\ A{\isadigit{3}}b{\isacharcolon}\ {\isachardoublequoteopen}{\isacharparenleft}cod\ y{\isacharparenright}{\isasymcdot}y\ {\isasymcong}\ y{\isachardoublequoteclose}\ \isakeyword{and}\ \isanewline
\ A{\isadigit{4}}a{\isacharcolon}\ {\isachardoublequoteopen}dom{\isacharparenleft}x{\isasymcdot}y{\isacharparenright}\ {\isasymcong}\ dom{\isacharparenleft}{\isacharparenleft}dom\ x{\isacharparenright}{\isasymcdot}y{\isacharparenright}{\isachardoublequoteclose}\ \isakeyword{and}\ \isanewline
\ A{\isadigit{4}}b{\isacharcolon}\ {\isachardoublequoteopen}cod{\isacharparenleft}x{\isasymcdot}y{\isacharparenright}\ {\isasymcong}\ cod{\isacharparenleft}x{\isasymcdot}{\isacharparenleft}cod\ y{\isacharparenright}{\isacharparenright}{\isachardoublequoteclose}\ \isakeyword{and}\ \isanewline
\ \ A{\isadigit{5}}{\isacharcolon}\ {\isachardoublequoteopen}x{\isasymcdot}{\isacharparenleft}y{\isasymcdot}z{\isacharparenright}\ {\isasymcong}\ {\isacharparenleft}x{\isasymcdot}y{\isacharparenright}{\isasymcdot}z{\isachardoublequoteclose}%
\begin{isamarkuptext}%
A model can still be constructed if we do not make assumptions about non-existing
  objects. In fact, the model presented by Nitpick consists of a single, existing morphism.%
\end{isamarkuptext}\isamarkuptrue%
\ \ \isacommand{lemma}\isamarkupfalse%
\ True\ \isanewline
\ \ \ \ \isacommand{nitpick}\isamarkupfalse%
\ {\isacharbrackleft}satisfy{\isacharcomma}\ user{\isacharunderscore}axioms{\isacharcomma}\ show{\isacharunderscore}all{\isacharcomma}\ format\ {\isacharequal}\ {\isadigit{2}}{\isacharcomma}\ expect\ {\isacharequal}\ genuine{\isacharbrackright}%
\isadelimproof
\ %
\endisadelimproof
\isatagproof
\isacommand{oops}\isamarkupfalse%
\ %
\isamarkupcmt{Nitpick finds a model%
}
\endisatagproof
{\isafoldproof}%
\isadelimproof
\endisadelimproof
\begin{isamarkuptext}%
However, one can see directly that axiom  \isa{A{\isadigit{1}}} is problematic as written:
If  \isa{x} and  \isa{y} are undefined, then (presumably)  \isa{dom\ x} and 
\isa{cod\ y} are undefined as well, and by the definition of Kleene equality,
 \isa{dom\ x\ {\isasymcong}\ cod\ y}. \isa{A{\isadigit{1}}} stipulates that \isa{x{\isasymcdot}y}  
should be defined in this case, which appears unintended.

We shall see that the consequences of this version of the axiom are
even stronger. It implies that \emph{all} objects are defined,
that is, composition (as well as \isa{dom} and \isa{cod}) become total operations.
The theory described by these axioms ``collapses'' to the theory of
monoids. (If all objects are defined, then one can conclude from \isa{A{\isadigit{1}}} that 
\isa{dom\ x\ {\isasymcong}\ dom\ y} (resp.~\isa{dom\ x\ {\isasymcong}\ cod\ y} and \isa{cod\ x\ {\isasymcong}\ cod\ y}), 
and according to 1.14 of \cite{FreydScedrov90}, 
the category reduces to a monoid provided that it is not empty.)%
\end{isamarkuptext}\isamarkuptrue%
\ \ \isacommand{lemma}\isamarkupfalse%
\ \isakeyword{assumes}\ {\isachardoublequoteopen}{\isasymexists}x{\isachardot}\ \isactrlbold {\isasymnot}{\isacharparenleft}E\ x{\isacharparenright}{\isachardoublequoteclose}\ \isakeyword{shows}\ True\ \ \ %
\isamarkupcmt{Nitpick does *not* find a model%
}
\ \isanewline
\ \ \ \ \isacommand{nitpick}\isamarkupfalse%
\ {\isacharbrackleft}satisfy{\isacharcomma}\ user{\isacharunderscore}axioms{\isacharcomma}\ show{\isacharunderscore}all{\isacharcomma}\ format\ {\isacharequal}\ {\isadigit{2}}{\isacharcomma}\ expect\ {\isacharequal}\ none{\isacharbrackright}%
\isadelimproof
\ %
\endisadelimproof
\isatagproof
\isacommand{oops}\isamarkupfalse%
\endisatagproof
{\isafoldproof}%
\isadelimproof
\endisadelimproof
\begin{isamarkuptext}%
In fact, the automated theorem provers quickly prove falsity when assuming a 
 non-existing object of type \isa{i}. The provers identify the axioms \isa{A{\isadigit{1}}}, \isa{A{\isadigit{2}}a}
 and \isa{A{\isadigit{3}}a} to cause the problem under this assumption.%
\end{isamarkuptext}\isamarkuptrue%
\ \ \isacommand{lemma}\isamarkupfalse%
\ InconsistencyAutomaticVII{\isacharcolon}\ {\isachardoublequoteopen}{\isacharparenleft}{\isasymexists}x{\isachardot}\ \isactrlbold {\isasymnot}{\isacharparenleft}E\ x{\isacharparenright}{\isacharparenright}\ \isactrlbold {\isasymrightarrow}\ False{\isachardoublequoteclose}\ \isanewline
\isadelimproof
\ \ \ \ %
\endisadelimproof
\isatagproof
\isacommand{by}\isamarkupfalse%
\ {\isacharparenleft}metis\ A{\isadigit{1}}\ A{\isadigit{2}}a\ A{\isadigit{3}}a{\isacharparenright}%
\endisatagproof
{\isafoldproof}%
\isadelimproof
\endisadelimproof
\begin{isamarkuptext}%
Hence, all morphisms must be defined in theory of Axiom Set VII, or in other 
      words, all operations must be total.%
\end{isamarkuptext}\isamarkuptrue%
\ \ \isacommand{lemma}\isamarkupfalse%
\ {\isachardoublequoteopen}{\isasymforall}x{\isachardot}\ E\ x{\isachardoublequoteclose}%
\isadelimproof
\ %
\endisadelimproof
\isatagproof
\isacommand{using}\isamarkupfalse%
\ InconsistencyAutomaticVII\ \isacommand{by}\isamarkupfalse%
\ auto%
\endisatagproof
{\isafoldproof}%
\isadelimproof
\endisadelimproof
\begin{isamarkuptext}%
The constricted inconsistency proof can be turned into an interactive mathematical argument:%
\end{isamarkuptext}\isamarkuptrue%
\ \ \isacommand{lemma}\isamarkupfalse%
\ InconsistencyInteractiveVII{\isacharcolon}\ \isanewline
\ \ \ \ \isakeyword{assumes}\ NEx{\isacharcolon}\ {\isachardoublequoteopen}{\isasymexists}x{\isachardot}\ \isactrlbold {\isasymnot}{\isacharparenleft}E\ x{\isacharparenright}{\isachardoublequoteclose}\ \isakeyword{shows}\ False\ \isanewline
\isadelimproof
\ \ %
\endisadelimproof
\isatagproof
\isacommand{proof}\isamarkupfalse%
\ {\isacharminus}\isanewline
\ \ \ \ %
\isamarkupcmt{Let \isa{a} be an undefined object%
}
\isanewline
\ \ \ \isacommand{obtain}\isamarkupfalse%
\ a\ \isakeyword{where}\ {\isadigit{1}}{\isacharcolon}\ {\isachardoublequoteopen}\isactrlbold {\isasymnot}{\isacharparenleft}E\ a{\isacharparenright}{\isachardoublequoteclose}\ \isacommand{using}\isamarkupfalse%
\ NEx\ \isacommand{by}\isamarkupfalse%
\ auto\isanewline
\ \ \ \ %
\isamarkupcmt{We instantiate axiom \isa{A{\isadigit{3}}a} with \isa{a}.%
}
\isanewline
\ \ \ \isacommand{have}\isamarkupfalse%
\ {\isadigit{2}}{\isacharcolon}\ {\isachardoublequoteopen}a{\isasymcdot}{\isacharparenleft}dom\ a{\isacharparenright}\ {\isasymcong}\ a{\isachardoublequoteclose}\ \isacommand{using}\isamarkupfalse%
\ A{\isadigit{3}}a\ \isacommand{by}\isamarkupfalse%
\ blast\ \ \isanewline
\ \ \ \ %
\isamarkupcmt{By unfolding the definition of \isa{{\isasymcong}} we get from 1 that
        \isa{a{\isasymcdot}{\isacharparenleft}dom\ a{\isacharparenright}} is not defined. This is 
        easy to see, since if \isa{a{\isasymcdot}{\isacharparenleft}dom\ a{\isacharparenright}} were defined, we also had that \isa{a} is 
        defined, which is not the case by assumption.%
}
\isanewline
\ \ \ \isacommand{have}\isamarkupfalse%
\ {\isadigit{3}}{\isacharcolon}\ {\isachardoublequoteopen}\isactrlbold {\isasymnot}{\isacharparenleft}E{\isacharparenleft}a{\isasymcdot}{\isacharparenleft}dom\ a{\isacharparenright}{\isacharparenright}{\isacharparenright}{\isachardoublequoteclose}\ \isacommand{using}\isamarkupfalse%
\ {\isadigit{1}}\ {\isadigit{2}}\ \isacommand{by}\isamarkupfalse%
\ metis\isanewline
\ \ \ \ %
\isamarkupcmt{We instantiate axiom \isa{A{\isadigit{1}}} with \isa{a} and \isa{dom\ a}.%
}
\isanewline
\ \ \ \isacommand{have}\isamarkupfalse%
\ {\isadigit{4}}{\isacharcolon}\ {\isachardoublequoteopen}E{\isacharparenleft}a{\isasymcdot}{\isacharparenleft}dom\ a{\isacharparenright}{\isacharparenright}\ \isactrlbold {\isasymleftrightarrow}\ dom\ a\ {\isasymcong}\ cod{\isacharparenleft}dom\ a{\isacharparenright}{\isachardoublequoteclose}\ \isacommand{using}\isamarkupfalse%
\ A{\isadigit{1}}\ \isacommand{by}\isamarkupfalse%
\ blast\isanewline
\ \ \ \ %
\isamarkupcmt{We instantiate axiom \isa{A{\isadigit{2}}a} with \isa{a}.%
}
\isanewline
\ \ \ \isacommand{have}\isamarkupfalse%
\ {\isadigit{5}}{\isacharcolon}\ {\isachardoublequoteopen}cod{\isacharparenleft}dom\ a{\isacharparenright}\ {\isasymcong}\ dom\ a{\isachardoublequoteclose}\ \isacommand{using}\isamarkupfalse%
\ A{\isadigit{2}}a\ \isacommand{by}\isamarkupfalse%
\ blast\ \isanewline
\ \ \ \ %
\isamarkupcmt{We use 5 (and symmetry and transitivity of \isa{{\isasymcong}}) to rewrite the 
         right-hand of the equivalence 4 into \isa{dom\ a\ {\isasymcong}\ dom\ a}.%
}
\ \isanewline
\ \ \ \isacommand{have}\isamarkupfalse%
\ {\isadigit{6}}{\isacharcolon}\ {\isachardoublequoteopen}E{\isacharparenleft}a{\isasymcdot}{\isacharparenleft}dom\ a{\isacharparenright}{\isacharparenright}\ \isactrlbold {\isasymleftrightarrow}\ dom\ a\ {\isasymcong}\ dom\ a{\isachardoublequoteclose}\ \isacommand{using}\isamarkupfalse%
\ {\isadigit{4}}\ {\isadigit{5}}\ \isacommand{by}\isamarkupfalse%
\ auto\isanewline
\ \ \ \ %
\isamarkupcmt{By reflexivity of \isa{{\isasymcong}} we get that \isa{a{\isasymcdot}{\isacharparenleft}dom\ a{\isacharparenright}} must be defined.%
}
\isanewline
\ \ \ \isacommand{have}\isamarkupfalse%
\ {\isadigit{7}}{\isacharcolon}\ {\isachardoublequoteopen}E{\isacharparenleft}a{\isasymcdot}{\isacharparenleft}dom\ a{\isacharparenright}{\isacharparenright}{\isachardoublequoteclose}\ \isacommand{using}\isamarkupfalse%
\ {\isadigit{6}}\ \isacommand{by}\isamarkupfalse%
\ blast\isanewline
\ \ \ \ %
\isamarkupcmt{We have shown in 7 that \isa{a{\isasymcdot}{\isacharparenleft}dom\ a{\isacharparenright}} is defined, and in 3 
         that it is undefined. Contradiction.%
}
\isanewline
\ \ \ \isacommand{then}\isamarkupfalse%
\ \isacommand{show}\isamarkupfalse%
\ {\isacharquery}thesis\ \isacommand{using}\isamarkupfalse%
\ {\isadigit{7}}\ {\isadigit{3}}\ \isacommand{by}\isamarkupfalse%
\ blast\isanewline
\ \ \isacommand{qed}\isamarkupfalse%
\endisatagproof
{\isafoldproof}%
\isadelimproof
\endisadelimproof
\begin{isamarkuptext}%
We present the constricted inconsistency argument once again, but this time in the original
  notation of Freyd and Scedrov.%
\end{isamarkuptext}\isamarkuptrue%
\isacommand{consts}\isamarkupfalse%
\ \ \isanewline
\ \ \ source{\isacharcolon}{\isacharcolon}\ {\isachardoublequoteopen}i{\isasymRightarrow}i{\isachardoublequoteclose}\ {\isacharparenleft}{\isachardoublequoteopen}{\isasymbox}{\isacharunderscore}{\isachardoublequoteclose}\ {\isacharbrackleft}{\isadigit{1}}{\isadigit{0}}{\isadigit{8}}{\isacharbrackright}\ {\isadigit{1}}{\isadigit{0}}{\isadigit{9}}{\isacharparenright}\ \isanewline
\ \ \ target{\isacharcolon}{\isacharcolon}\ {\isachardoublequoteopen}i{\isasymRightarrow}i{\isachardoublequoteclose}\ {\isacharparenleft}{\isachardoublequoteopen}{\isacharunderscore}{\isasymbox}{\isachardoublequoteclose}\ {\isacharbrackleft}{\isadigit{1}}{\isadigit{1}}{\isadigit{0}}{\isacharbrackright}\ {\isadigit{1}}{\isadigit{1}}{\isadigit{1}}{\isacharparenright}\ \isanewline
\ \ \ compositionF{\isacharcolon}{\isacharcolon}\ {\isachardoublequoteopen}i{\isasymRightarrow}i{\isasymRightarrow}i{\isachardoublequoteclose}\ {\isacharparenleft}\isakeyword{infix}\ {\isachardoublequoteopen}\isactrlbold {\isasymcdot}{\isachardoublequoteclose}\ {\isadigit{1}}{\isadigit{1}}{\isadigit{0}}{\isacharparenright}\isanewline
\isanewline
\ \ A{\isadigit{1}}{\isacharcolon}\ {\isachardoublequoteopen}E{\isacharparenleft}x\isactrlbold {\isasymcdot}y{\isacharparenright}\ \isactrlbold {\isasymleftrightarrow}\ {\isacharparenleft}x{\isasymbox}\ {\isasymcong}\ {\isasymbox}y{\isacharparenright}{\isachardoublequoteclose}\ \isakeyword{and}\ \isanewline
\ A{\isadigit{2}}a{\isacharcolon}\ {\isachardoublequoteopen}{\isacharparenleft}{\isacharparenleft}{\isasymbox}x{\isacharparenright}{\isasymbox}{\isacharparenright}\ {\isasymcong}\ {\isasymbox}x{\isachardoublequoteclose}\ \isakeyword{and}\ \isanewline
\ A{\isadigit{2}}b{\isacharcolon}\ {\isachardoublequoteopen}{\isasymbox}{\isacharparenleft}x{\isasymbox}{\isacharparenright}\ {\isasymcong}\ {\isasymbox}x{\isachardoublequoteclose}\ \isakeyword{and}\ \isanewline
\ A{\isadigit{3}}a{\isacharcolon}\ {\isachardoublequoteopen}{\isacharparenleft}{\isasymbox}x{\isacharparenright}\isactrlbold {\isasymcdot}x\ {\isasymcong}\ x{\isachardoublequoteclose}\ \isakeyword{and}\ \isanewline
\ A{\isadigit{3}}b{\isacharcolon}\ {\isachardoublequoteopen}x\isactrlbold {\isasymcdot}{\isacharparenleft}x{\isasymbox}{\isacharparenright}\ {\isasymcong}\ x{\isachardoublequoteclose}\ \isakeyword{and}\ \isanewline
\ A{\isadigit{4}}a{\isacharcolon}\ {\isachardoublequoteopen}{\isasymbox}{\isacharparenleft}x\isactrlbold {\isasymcdot}y{\isacharparenright}\ {\isasymcong}\ {\isasymbox}{\isacharparenleft}x\isactrlbold {\isasymcdot}{\isacharparenleft}{\isasymbox}y{\isacharparenright}{\isacharparenright}{\isachardoublequoteclose}\ \isakeyword{and}\ \isanewline
\ A{\isadigit{4}}b{\isacharcolon}\ {\isachardoublequoteopen}{\isacharparenleft}x\isactrlbold {\isasymcdot}y{\isacharparenright}{\isasymbox}\ {\isasymcong}\ {\isacharparenleft}{\isacharparenleft}x{\isasymbox}{\isacharparenright}\isactrlbold {\isasymcdot}y{\isacharparenright}{\isasymbox}{\isachardoublequoteclose}\ \isakeyword{and}\ \isanewline
\ \ A{\isadigit{5}}{\isacharcolon}\ {\isachardoublequoteopen}x\isactrlbold {\isasymcdot}{\isacharparenleft}y\isactrlbold {\isasymcdot}z{\isacharparenright}\ {\isasymcong}\ {\isacharparenleft}x\isactrlbold {\isasymcdot}y{\isacharparenright}\isactrlbold {\isasymcdot}z{\isachardoublequoteclose}%
\begin{isamarkuptext}%
\label{subsec:FreydNotation} Again, the automated theorem provers via Sledgehammer 
       find the constricted inconsistency very quickly and they identify the  exact dependencies.%
\end{isamarkuptext}\isamarkuptrue%
\isacommand{lemma}\isamarkupfalse%
\ InconsistencyAutomatic{\isacharcolon}\ {\isachardoublequoteopen}{\isacharparenleft}{\isasymexists}x{\isachardot}\ \isactrlbold {\isasymnot}{\isacharparenleft}E\ x{\isacharparenright}{\isacharparenright}\ \isactrlbold {\isasymrightarrow}\ False{\isachardoublequoteclose}\ \isanewline
\isadelimproof
\ \ %
\endisadelimproof
\isatagproof
\isacommand{by}\isamarkupfalse%
\ {\isacharparenleft}metis\ A{\isadigit{1}}\ A{\isadigit{2}}a\ A{\isadigit{3}}a{\isacharparenright}%
\endisatagproof
{\isafoldproof}%
\isadelimproof
\endisadelimproof
\begin{isamarkuptext}%
The following alternative interactive proof is slightly shorter than the one 
        presented above.%
\end{isamarkuptext}\isamarkuptrue%
\ \ \isacommand{lemma}\isamarkupfalse%
\ InconsistencyInteractive{\isacharcolon}\ \isakeyword{assumes}\ NEx{\isacharcolon}\ {\isachardoublequoteopen}{\isasymexists}x{\isachardot}\ \isactrlbold {\isasymnot}{\isacharparenleft}E\ x{\isacharparenright}{\isachardoublequoteclose}\ \isakeyword{shows}\ False\ \isanewline
\isadelimproof
\ \ %
\endisadelimproof
\isatagproof
\isacommand{proof}\isamarkupfalse%
\ {\isacharminus}\isanewline
\ \ \ \ %
\isamarkupcmt{Let \isa{a} be an undefined object%
}
\isanewline
\ \ \ \isacommand{obtain}\isamarkupfalse%
\ a\ \isakeyword{where}\ {\isadigit{1}}{\isacharcolon}\ {\isachardoublequoteopen}\isactrlbold {\isasymnot}{\isacharparenleft}E\ a{\isacharparenright}{\isachardoublequoteclose}\ \isacommand{using}\isamarkupfalse%
\ assms\ \isacommand{by}\isamarkupfalse%
\ auto\isanewline
\ \ \ \ %
\isamarkupcmt{We instantiate axiom \isa{A{\isadigit{3}}a} with \isa{a}.%
}
\isanewline
\ \ \ \isacommand{have}\isamarkupfalse%
\ {\isadigit{2}}{\isacharcolon}\ {\isachardoublequoteopen}{\isacharparenleft}{\isasymbox}a{\isacharparenright}\isactrlbold {\isasymcdot}a\ {\isasymcong}\ a{\isachardoublequoteclose}\ \isacommand{using}\isamarkupfalse%
\ A{\isadigit{3}}a\ \isacommand{by}\isamarkupfalse%
\ blast\isanewline
\ \ \ \ %
\isamarkupcmt{By unfolding the definition of \isa{{\isasymcong}} we get from 1 that \isa{{\isacharparenleft}{\isasymbox}a{\isacharparenright}\isactrlbold {\isasymcdot}a} is 
         not defined. This is 
         easy to see, since if \isa{{\isacharparenleft}{\isasymbox}a{\isacharparenright}\isactrlbold {\isasymcdot}a} were defined, we also had that  \isa{a} is 
         defined, which is not the case by assumption.%
}
\isanewline
\ \ \ \isacommand{have}\isamarkupfalse%
\ {\isadigit{3}}{\isacharcolon}\ {\isachardoublequoteopen}\isactrlbold {\isasymnot}{\isacharparenleft}E{\isacharparenleft}{\isacharparenleft}{\isasymbox}a{\isacharparenright}\isactrlbold {\isasymcdot}a{\isacharparenright}{\isacharparenright}{\isachardoublequoteclose}\ \isacommand{using}\isamarkupfalse%
\ {\isadigit{1}}\ {\isadigit{2}}\ \isacommand{by}\isamarkupfalse%
\ metis\isanewline
\ \ \ \ %
\isamarkupcmt{We instantiate axiom \isa{A{\isadigit{1}}} with \isa{{\isasymbox}a} and \isa{a}.%
}
\isanewline
\ \ \ \isacommand{have}\isamarkupfalse%
\ {\isadigit{4}}{\isacharcolon}\ {\isachardoublequoteopen}E{\isacharparenleft}{\isacharparenleft}{\isasymbox}a{\isacharparenright}\isactrlbold {\isasymcdot}a{\isacharparenright}\ \isactrlbold {\isasymleftrightarrow}\ {\isacharparenleft}{\isasymbox}a{\isacharparenright}{\isasymbox}\ {\isasymcong}\ {\isasymbox}a{\isachardoublequoteclose}\ \isacommand{using}\isamarkupfalse%
\ A{\isadigit{1}}\ \isacommand{by}\isamarkupfalse%
\ blast\isanewline
\ \ \ \ %
\isamarkupcmt{We instantiate axiom \isa{A{\isadigit{2}}a} with \isa{a}.%
}
\isanewline
\ \ \ \isacommand{have}\isamarkupfalse%
\ {\isadigit{5}}{\isacharcolon}\ {\isachardoublequoteopen}{\isacharparenleft}{\isasymbox}a{\isacharparenright}{\isasymbox}\ {\isasymcong}\ {\isasymbox}a{\isachardoublequoteclose}\ \isacommand{using}\isamarkupfalse%
\ A{\isadigit{2}}a\ \isacommand{by}\isamarkupfalse%
\ blast\ \isanewline
\ \ \ \ %
\isamarkupcmt{From 4 and 5 we obtain \isa{\isactrlbold {\isacharparenleft}E{\isacharparenleft}{\isacharparenleft}{\isasymbox}a{\isacharparenright}\isactrlbold {\isasymcdot}a{\isacharparenright}{\isacharparenright}} by propositional logic.%
}
\ \isanewline
\ \ \ \isacommand{have}\isamarkupfalse%
\ {\isadigit{6}}{\isacharcolon}\ {\isachardoublequoteopen}E{\isacharparenleft}{\isacharparenleft}{\isasymbox}a{\isacharparenright}\isactrlbold {\isasymcdot}a{\isacharparenright}{\isachardoublequoteclose}\ \isacommand{using}\isamarkupfalse%
\ {\isadigit{4}}\ {\isadigit{5}}\ \isacommand{by}\isamarkupfalse%
\ blast\ \isanewline
\ \ \ \ %
\isamarkupcmt{We have \isa{\isactrlbold {\isasymnot}{\isacharparenleft}E{\isacharparenleft}{\isacharparenleft}{\isasymbox}a{\isacharparenright}\isactrlbold {\isasymcdot}a{\isacharparenright}{\isacharparenright}} and \isa{E{\isacharparenleft}{\isacharparenleft}{\isasymbox}a{\isacharparenright}\isactrlbold {\isasymcdot}a{\isacharparenright}}, hence Falsity.%
}
\isanewline
\ \ \ \isacommand{then}\isamarkupfalse%
\ \isacommand{show}\isamarkupfalse%
\ {\isacharquery}thesis\ \isacommand{using}\isamarkupfalse%
\ {\isadigit{6}}\ {\isadigit{3}}\ \isacommand{by}\isamarkupfalse%
\ blast\isanewline
\ \ \isacommand{qed}\isamarkupfalse%
\endisatagproof
{\isafoldproof}%
\isadelimproof
\endisadelimproof
\begin{isamarkuptext}%
Obviously Axiom Set VII is also redundant, and we have previously reported 
on respective redundancies \cite{C57}. However, this was before the discovery of the above 
constricted inconsistency issue, which tells us that the system (in our setting) can even be reduced 
to \isa{A{\isadigit{1}}}, \isa{A{\isadigit{2}}a} and \isa{A{\isadigit{3}}a} (when we additionally assume \isa{NEx}).%
\end{isamarkuptext}\isamarkuptrue%
\isamarkupsection{Axiom Set VIII%
}
\isamarkuptrue%
\begin{isamarkuptext}%
We study the axiom system by Freyd and Scedrov once again. However, this time we restrict 
the free variables in their system to range over existing objects only. By employing the free logic 
universal quantifier \isa{\isactrlbold {\isasymforall}} we thus modify Axiom Set VII as follows:%
\end{isamarkuptext}\isamarkuptrue%
\ \ B{\isadigit{1}}{\isacharcolon}\ {\isachardoublequoteopen}\isactrlbold {\isasymforall}x{\isachardot}\isactrlbold {\isasymforall}y{\isachardot}\ E{\isacharparenleft}x{\isasymcdot}y{\isacharparenright}\ \isactrlbold {\isasymleftrightarrow}\ dom\ x\ {\isasymcong}\ cod\ y{\isachardoublequoteclose}\ \isakeyword{and}\isanewline
\ B{\isadigit{2}}a{\isacharcolon}\ {\isachardoublequoteopen}\isactrlbold {\isasymforall}x{\isachardot}\ cod{\isacharparenleft}dom\ x{\isacharparenright}\ {\isasymcong}\ dom\ x\ {\isachardoublequoteclose}\ \isakeyword{and}\ \ \isanewline
\ B{\isadigit{2}}b{\isacharcolon}\ {\isachardoublequoteopen}\isactrlbold {\isasymforall}y{\isachardot}\ dom{\isacharparenleft}cod\ y{\isacharparenright}\ {\isasymcong}\ cod\ y{\isachardoublequoteclose}\ \isakeyword{and}\ \ \isanewline
\ B{\isadigit{3}}a{\isacharcolon}\ {\isachardoublequoteopen}\isactrlbold {\isasymforall}x{\isachardot}\ x{\isasymcdot}{\isacharparenleft}dom\ x{\isacharparenright}\ {\isasymcong}\ x{\isachardoublequoteclose}\ \isakeyword{and}\ \isanewline
\ B{\isadigit{3}}b{\isacharcolon}\ {\isachardoublequoteopen}\isactrlbold {\isasymforall}y{\isachardot}\ {\isacharparenleft}cod\ y{\isacharparenright}{\isasymcdot}y\ {\isasymcong}\ y{\isachardoublequoteclose}\ \isakeyword{and}\ \isanewline
\ B{\isadigit{4}}a{\isacharcolon}\ {\isachardoublequoteopen}\isactrlbold {\isasymforall}x{\isachardot}\isactrlbold {\isasymforall}y{\isachardot}\ dom{\isacharparenleft}x{\isasymcdot}y{\isacharparenright}\ {\isasymcong}\ dom{\isacharparenleft}{\isacharparenleft}dom\ x{\isacharparenright}{\isasymcdot}y{\isacharparenright}{\isachardoublequoteclose}\ \isakeyword{and}\ \isanewline
\ B{\isadigit{4}}b{\isacharcolon}\ {\isachardoublequoteopen}\isactrlbold {\isasymforall}x{\isachardot}\isactrlbold {\isasymforall}y{\isachardot}\ cod{\isacharparenleft}x{\isasymcdot}y{\isacharparenright}\ {\isasymcong}\ cod{\isacharparenleft}x{\isasymcdot}{\isacharparenleft}cod\ y{\isacharparenright}{\isacharparenright}{\isachardoublequoteclose}\ \isakeyword{and}\ \isanewline
\ \ B{\isadigit{5}}{\isacharcolon}\ {\isachardoublequoteopen}\isactrlbold {\isasymforall}x{\isachardot}\isactrlbold {\isasymforall}y{\isachardot}\isactrlbold {\isasymforall}z{\isachardot}\ x{\isasymcdot}{\isacharparenleft}y{\isasymcdot}z{\isacharparenright}\ {\isasymcong}\ {\isacharparenleft}x{\isasymcdot}y{\isacharparenright}{\isasymcdot}z{\isachardoublequoteclose}%
\begin{isamarkuptext}%
Now, the two consistency checks succeed.%
\end{isamarkuptext}\isamarkuptrue%
\ \ \isacommand{lemma}\isamarkupfalse%
\ True\ \ %
\isamarkupcmt{Nitpick finds a model%
}
\isanewline
\ \ \ \ \isacommand{nitpick}\isamarkupfalse%
\ {\isacharbrackleft}satisfy{\isacharcomma}\ user{\isacharunderscore}axioms{\isacharcomma}\ show{\isacharunderscore}all{\isacharcomma}\ format\ {\isacharequal}\ {\isadigit{2}}{\isacharcomma}\ expect\ {\isacharequal}\ genuine{\isacharbrackright}%
\isadelimproof
\ %
\endisadelimproof
\isatagproof
\isacommand{oops}\isamarkupfalse%
\endisatagproof
{\isafoldproof}%
\isadelimproof
\endisadelimproof
\isanewline
\ \ \isacommand{lemma}\isamarkupfalse%
\ \isakeyword{assumes}\ {\isachardoublequoteopen}{\isasymexists}x{\isachardot}\ \isactrlbold {\isasymnot}{\isacharparenleft}E\ x{\isacharparenright}{\isachardoublequoteclose}\ \isakeyword{shows}\ True\ \ \ %
\isamarkupcmt{Nitpick finds a model%
}
\ \ \isanewline
\ \ \ \ \isacommand{nitpick}\isamarkupfalse%
\ {\isacharbrackleft}satisfy{\isacharcomma}\ user{\isacharunderscore}axioms{\isacharcomma}\ show{\isacharunderscore}all{\isacharcomma}\ format\ {\isacharequal}\ {\isadigit{2}}{\isacharcomma}\ expect\ {\isacharequal}\ genuine{\isacharbrackright}%
\isadelimproof
\ %
\endisadelimproof
\isatagproof
\isacommand{oops}\isamarkupfalse%
\endisatagproof
{\isafoldproof}%
\isadelimproof
\endisadelimproof
\isanewline
\ \ \isacommand{lemma}\isamarkupfalse%
\ \isakeyword{assumes}\ {\isachardoublequoteopen}{\isacharparenleft}{\isasymexists}x{\isachardot}\ \isactrlbold {\isasymnot}{\isacharparenleft}E\ x{\isacharparenright}{\isacharparenright}\ {\isasymand}\ {\isacharparenleft}{\isasymexists}x{\isachardot}\ {\isacharparenleft}E\ x{\isacharparenright}{\isacharparenright}{\isachardoublequoteclose}\ \isakeyword{shows}\ True\ \ %
\isamarkupcmt{Nitpick finds a model%
}
\ \isanewline
\ \ \ \ \isacommand{nitpick}\isamarkupfalse%
\ {\isacharbrackleft}satisfy{\isacharcomma}\ user{\isacharunderscore}axioms{\isacharcomma}\ show{\isacharunderscore}all{\isacharcomma}\ format\ {\isacharequal}\ {\isadigit{2}}{\isacharcomma}\ expect\ {\isacharequal}\ genuine{\isacharbrackright}%
\isadelimproof
\ %
\endisadelimproof
\isatagproof
\isacommand{oops}\isamarkupfalse%
\endisatagproof
{\isafoldproof}%
\isadelimproof
\endisadelimproof
\begin{isamarkuptext}%
However, this axiom set is obviously weaker than our Axiom Set V. In fact, none of 
the \isa{V}-axioms are implied:%
\end{isamarkuptext}\isamarkuptrue%
\ \ \isacommand{lemma}\isamarkupfalse%
\ S{\isadigit{1}}{\isacharcolon}\ {\isachardoublequoteopen}E{\isacharparenleft}dom\ x{\isacharparenright}\ \isactrlbold {\isasymrightarrow}\ E\ x{\isachardoublequoteclose}\ \ %
\isamarkupcmt{Nitpick finds a countermodel%
}
\ \ \isanewline
\ \ \ \ \isacommand{nitpick}\isamarkupfalse%
\ {\isacharbrackleft}user{\isacharunderscore}axioms{\isacharcomma}\ show{\isacharunderscore}all{\isacharcomma}\ format\ {\isacharequal}\ {\isadigit{2}}{\isacharbrackright}%
\isadelimproof
\ %
\endisadelimproof
\isatagproof
\isacommand{oops}\isamarkupfalse%
\endisatagproof
{\isafoldproof}%
\isadelimproof
\endisadelimproof
\ \isanewline
\ \ \isacommand{lemma}\isamarkupfalse%
\ S{\isadigit{2}}{\isacharcolon}\ {\isachardoublequoteopen}E{\isacharparenleft}cod\ y{\isacharparenright}\ \isactrlbold {\isasymrightarrow}\ E\ y{\isachardoublequoteclose}\ \ %
\isamarkupcmt{Nitpick finds a countermodel%
}
\ \ \isanewline
\ \ \ \ \isacommand{nitpick}\isamarkupfalse%
\ {\isacharbrackleft}user{\isacharunderscore}axioms{\isacharcomma}\ show{\isacharunderscore}all{\isacharcomma}\ format\ {\isacharequal}\ {\isadigit{2}}{\isacharbrackright}%
\isadelimproof
\ %
\endisadelimproof
\isatagproof
\isacommand{oops}\isamarkupfalse%
\endisatagproof
{\isafoldproof}%
\isadelimproof
\endisadelimproof
\ \isanewline
\ \ \isacommand{lemma}\isamarkupfalse%
\ S{\isadigit{3}}{\isacharcolon}\ {\isachardoublequoteopen}E{\isacharparenleft}x{\isasymcdot}y{\isacharparenright}\ \isactrlbold {\isasymleftrightarrow}\ dom\ x\ {\isasymsimeq}\ cod\ y{\isachardoublequoteclose}\ \ \ %
\isamarkupcmt{Nitpick finds a countermodel%
}
\ \isanewline
\ \ \ \ \isacommand{nitpick}\isamarkupfalse%
\ {\isacharbrackleft}user{\isacharunderscore}axioms{\isacharcomma}\ show{\isacharunderscore}all{\isacharcomma}\ format\ {\isacharequal}\ {\isadigit{2}}{\isacharbrackright}%
\isadelimproof
\ %
\endisadelimproof
\isatagproof
\isacommand{oops}\isamarkupfalse%
\endisatagproof
{\isafoldproof}%
\isadelimproof
\endisadelimproof
\ \isanewline
\ \ \isacommand{lemma}\isamarkupfalse%
\ S{\isadigit{4}}{\isacharcolon}\ {\isachardoublequoteopen}x{\isasymcdot}{\isacharparenleft}y{\isasymcdot}z{\isacharparenright}\ {\isasymcong}\ {\isacharparenleft}x{\isasymcdot}y{\isacharparenright}{\isasymcdot}z{\isachardoublequoteclose}\ \ \ %
\isamarkupcmt{Nitpick finds a countermodel%
}
\ \isanewline
\ \ \ \ \isacommand{nitpick}\isamarkupfalse%
\ {\isacharbrackleft}user{\isacharunderscore}axioms{\isacharcomma}\ show{\isacharunderscore}all{\isacharcomma}\ format\ {\isacharequal}\ {\isadigit{2}}{\isacharbrackright}%
\isadelimproof
\ %
\endisadelimproof
\isatagproof
\isacommand{oops}\isamarkupfalse%
\endisatagproof
{\isafoldproof}%
\isadelimproof
\endisadelimproof
\ \isanewline
\ \ \isacommand{lemma}\isamarkupfalse%
\ S{\isadigit{5}}{\isacharcolon}\ {\isachardoublequoteopen}x{\isasymcdot}{\isacharparenleft}dom\ x{\isacharparenright}\ {\isasymcong}\ x{\isachardoublequoteclose}\ \ \ %
\isamarkupcmt{Nitpick finds a countermodel%
}
\ \isanewline
\ \ \ \ \isacommand{nitpick}\isamarkupfalse%
\ {\isacharbrackleft}user{\isacharunderscore}axioms{\isacharcomma}\ show{\isacharunderscore}all{\isacharcomma}\ format\ {\isacharequal}\ {\isadigit{2}}{\isacharbrackright}%
\isadelimproof
\ %
\endisadelimproof
\isatagproof
\isacommand{oops}\isamarkupfalse%
\endisatagproof
{\isafoldproof}%
\isadelimproof
\endisadelimproof
\ \isanewline
\ \ \isacommand{lemma}\isamarkupfalse%
\ S{\isadigit{6}}{\isacharcolon}\ {\isachardoublequoteopen}{\isacharparenleft}cod\ y{\isacharparenright}{\isasymcdot}y\ {\isasymcong}\ y{\isachardoublequoteclose}\ \ \ %
\isamarkupcmt{Nitpick finds a countermodel%
}
\ \isanewline
\ \ \ \ \isacommand{nitpick}\isamarkupfalse%
\ {\isacharbrackleft}user{\isacharunderscore}axioms{\isacharcomma}\ show{\isacharunderscore}all{\isacharcomma}\ format\ {\isacharequal}\ {\isadigit{2}}{\isacharbrackright}%
\isadelimproof
\ %
\endisadelimproof
\isatagproof
\isacommand{oops}\isamarkupfalse%
\endisatagproof
{\isafoldproof}%
\isadelimproof
\endisadelimproof
\begin{isamarkuptext}%
The situation changes when we explicitly postulate strictness of \isa{dom},
\isa{cod} and \isa{{\isasymcdot}}. We thus obtain our Axiom Set VIII:%
\end{isamarkuptext}\isamarkuptrue%
\ B{\isadigit{0}}a{\isacharcolon}\ {\isachardoublequoteopen}E{\isacharparenleft}x{\isasymcdot}y{\isacharparenright}\ \isactrlbold {\isasymrightarrow}\ {\isacharparenleft}E\ x\ \isactrlbold {\isasymand}\ E\ y{\isacharparenright}{\isachardoublequoteclose}\ \isakeyword{and}\isanewline
\ B{\isadigit{0}}b{\isacharcolon}\ {\isachardoublequoteopen}E{\isacharparenleft}dom\ x{\isacharparenright}\ \isactrlbold {\isasymrightarrow}\ E\ x{\isachardoublequoteclose}\ \isakeyword{and}\isanewline
\ B{\isadigit{0}}c{\isacharcolon}\ {\isachardoublequoteopen}E{\isacharparenleft}cod\ x{\isacharparenright}\ \isactrlbold {\isasymrightarrow}\ E\ x{\isachardoublequoteclose}\ \isakeyword{and}\isanewline
\ \ B{\isadigit{1}}{\isacharcolon}\ {\isachardoublequoteopen}\isactrlbold {\isasymforall}x{\isachardot}\isactrlbold {\isasymforall}y{\isachardot}\ E{\isacharparenleft}x{\isasymcdot}y{\isacharparenright}\ \isactrlbold {\isasymleftrightarrow}\ dom\ x\ {\isasymcong}\ cod\ y{\isachardoublequoteclose}\ \isakeyword{and}\isanewline
\ B{\isadigit{2}}a{\isacharcolon}\ {\isachardoublequoteopen}\isactrlbold {\isasymforall}x{\isachardot}\ cod{\isacharparenleft}dom\ x{\isacharparenright}\ {\isasymcong}\ dom\ x\ {\isachardoublequoteclose}\ \isakeyword{and}\ \ \isanewline
\ B{\isadigit{2}}b{\isacharcolon}\ {\isachardoublequoteopen}\isactrlbold {\isasymforall}y{\isachardot}\ dom{\isacharparenleft}cod\ y{\isacharparenright}\ {\isasymcong}\ cod\ y{\isachardoublequoteclose}\ \isakeyword{and}\ \ \isanewline
\ B{\isadigit{3}}a{\isacharcolon}\ {\isachardoublequoteopen}\isactrlbold {\isasymforall}x{\isachardot}\ x{\isasymcdot}{\isacharparenleft}dom\ x{\isacharparenright}\ {\isasymcong}\ x{\isachardoublequoteclose}\ \isakeyword{and}\ \isanewline
\ B{\isadigit{3}}b{\isacharcolon}\ {\isachardoublequoteopen}\isactrlbold {\isasymforall}y{\isachardot}\ {\isacharparenleft}cod\ y{\isacharparenright}{\isasymcdot}y\ {\isasymcong}\ y{\isachardoublequoteclose}\ \isakeyword{and}\ \isanewline
\ B{\isadigit{4}}a{\isacharcolon}\ {\isachardoublequoteopen}\isactrlbold {\isasymforall}x{\isachardot}\isactrlbold {\isasymforall}y{\isachardot}\ dom{\isacharparenleft}x{\isasymcdot}y{\isacharparenright}\ {\isasymcong}\ dom{\isacharparenleft}{\isacharparenleft}dom\ x{\isacharparenright}{\isasymcdot}y{\isacharparenright}{\isachardoublequoteclose}\ \isakeyword{and}\ \isanewline
\ B{\isadigit{4}}b{\isacharcolon}\ {\isachardoublequoteopen}\isactrlbold {\isasymforall}x{\isachardot}\isactrlbold {\isasymforall}y{\isachardot}\ cod{\isacharparenleft}x{\isasymcdot}y{\isacharparenright}\ {\isasymcong}\ cod{\isacharparenleft}x{\isasymcdot}{\isacharparenleft}cod\ y{\isacharparenright}{\isacharparenright}{\isachardoublequoteclose}\ \isakeyword{and}\ \isanewline
\ \ B{\isadigit{5}}{\isacharcolon}\ {\isachardoublequoteopen}\isactrlbold {\isasymforall}x{\isachardot}\isactrlbold {\isasymforall}y{\isachardot}\isactrlbold {\isasymforall}z{\isachardot}\ x{\isasymcdot}{\isacharparenleft}y{\isasymcdot}z{\isacharparenright}\ {\isasymcong}\ {\isacharparenleft}x{\isasymcdot}y{\isacharparenright}{\isasymcdot}z{\isachardoublequoteclose}%
\begin{isamarkuptext}%
Again, the two consistency checks succeed%
\end{isamarkuptext}\isamarkuptrue%
\ \ \isacommand{lemma}\isamarkupfalse%
\ True\ \ %
\isamarkupcmt{Nitpick finds a model%
}
\isanewline
\ \ \ \ \isacommand{nitpick}\isamarkupfalse%
\ {\isacharbrackleft}satisfy{\isacharcomma}\ user{\isacharunderscore}axioms{\isacharcomma}\ show{\isacharunderscore}all{\isacharcomma}\ format\ {\isacharequal}\ {\isadigit{2}}{\isacharcomma}\ expect\ {\isacharequal}\ genuine{\isacharbrackright}%
\isadelimproof
\ %
\endisadelimproof
\isatagproof
\isacommand{oops}\isamarkupfalse%
\endisatagproof
{\isafoldproof}%
\isadelimproof
\endisadelimproof
\isanewline
\ \ \isacommand{lemma}\isamarkupfalse%
\ \isakeyword{assumes}\ {\isachardoublequoteopen}{\isasymexists}x{\isachardot}\ \isactrlbold {\isasymnot}{\isacharparenleft}E\ x{\isacharparenright}{\isachardoublequoteclose}\ \isakeyword{shows}\ True\ \ \ %
\isamarkupcmt{Nitpick finds a model%
}
\ \ \isanewline
\ \ \ \ \isacommand{nitpick}\isamarkupfalse%
\ {\isacharbrackleft}satisfy{\isacharcomma}\ user{\isacharunderscore}axioms{\isacharcomma}\ show{\isacharunderscore}all{\isacharcomma}\ format\ {\isacharequal}\ {\isadigit{2}}{\isacharcomma}\ expect\ {\isacharequal}\ genuine{\isacharbrackright}%
\isadelimproof
\ %
\endisadelimproof
\isatagproof
\isacommand{oops}\isamarkupfalse%
\endisatagproof
{\isafoldproof}%
\isadelimproof
\endisadelimproof
\isanewline
\ \ \isacommand{lemma}\isamarkupfalse%
\ \isakeyword{assumes}\ {\isachardoublequoteopen}{\isacharparenleft}{\isasymexists}x{\isachardot}\ \isactrlbold {\isasymnot}{\isacharparenleft}E\ x{\isacharparenright}{\isacharparenright}\ {\isasymand}\ {\isacharparenleft}{\isasymexists}x{\isachardot}\ {\isacharparenleft}E\ x{\isacharparenright}{\isacharparenright}{\isachardoublequoteclose}\ \isakeyword{shows}\ True\ \ %
\isamarkupcmt{Nitpick finds a model%
}
\ \isanewline
\ \ \ \ \isacommand{nitpick}\isamarkupfalse%
\ {\isacharbrackleft}satisfy{\isacharcomma}\ user{\isacharunderscore}axioms{\isacharcomma}\ show{\isacharunderscore}all{\isacharcomma}\ format\ {\isacharequal}\ {\isadigit{2}}{\isacharcomma}\ expect\ {\isacharequal}\ genuine{\isacharbrackright}%
\isadelimproof
\ %
\endisadelimproof
\isatagproof
\isacommand{oops}\isamarkupfalse%
\endisatagproof
{\isafoldproof}%
\isadelimproof
\endisadelimproof
\begin{isamarkuptext}%
Now Axiom Set V is implied.%
\end{isamarkuptext}\isamarkuptrue%
\ \ \isacommand{lemma}\isamarkupfalse%
\ S{\isadigit{1}}FromVIII{\isacharcolon}\ {\isachardoublequoteopen}E{\isacharparenleft}dom\ x{\isacharparenright}\ \isactrlbold {\isasymrightarrow}\ E\ x{\isachardoublequoteclose}%
\isadelimproof
\ \ %
\endisadelimproof
\isatagproof
\isacommand{using}\isamarkupfalse%
\ B{\isadigit{0}}b\ \isacommand{by}\isamarkupfalse%
\ blast%
\endisatagproof
{\isafoldproof}%
\isadelimproof
\endisadelimproof
\isanewline
\ \ \isacommand{lemma}\isamarkupfalse%
\ S{\isadigit{2}}FromVIII{\isacharcolon}\ {\isachardoublequoteopen}E{\isacharparenleft}cod\ y{\isacharparenright}\ \isactrlbold {\isasymrightarrow}\ E\ y{\isachardoublequoteclose}%
\isadelimproof
\ \ %
\endisadelimproof
\isatagproof
\isacommand{using}\isamarkupfalse%
\ B{\isadigit{0}}c\ \isacommand{by}\isamarkupfalse%
\ blast%
\endisatagproof
{\isafoldproof}%
\isadelimproof
\endisadelimproof
\ \isanewline
\ \ \isacommand{lemma}\isamarkupfalse%
\ S{\isadigit{3}}FromVIII{\isacharcolon}\ {\isachardoublequoteopen}E{\isacharparenleft}x{\isasymcdot}y{\isacharparenright}\ \isactrlbold {\isasymleftrightarrow}\ dom\ x\ {\isasymsimeq}\ cod\ y{\isachardoublequoteclose}%
\isadelimproof
\ %
\endisadelimproof
\isatagproof
\isacommand{by}\isamarkupfalse%
\ {\isacharparenleft}metis\ B{\isadigit{0}}a\ B{\isadigit{0}}b\ B{\isadigit{0}}c\ B{\isadigit{1}}\ B{\isadigit{3}}a{\isacharparenright}%
\endisatagproof
{\isafoldproof}%
\isadelimproof
\endisadelimproof
\isanewline
\ \ \isacommand{lemma}\isamarkupfalse%
\ S{\isadigit{4}}FromVIII{\isacharcolon}\ {\isachardoublequoteopen}x{\isasymcdot}{\isacharparenleft}y{\isasymcdot}z{\isacharparenright}\ {\isasymcong}\ {\isacharparenleft}x{\isasymcdot}y{\isacharparenright}{\isasymcdot}z{\isachardoublequoteclose}%
\isadelimproof
\ %
\endisadelimproof
\isatagproof
\isacommand{by}\isamarkupfalse%
\ {\isacharparenleft}meson\ B{\isadigit{0}}a\ B{\isadigit{5}}{\isacharparenright}%
\endisatagproof
{\isafoldproof}%
\isadelimproof
\endisadelimproof
\ \isanewline
\ \ \isacommand{lemma}\isamarkupfalse%
\ S{\isadigit{5}}FromVIII{\isacharcolon}\ {\isachardoublequoteopen}x{\isasymcdot}{\isacharparenleft}dom\ x{\isacharparenright}\ {\isasymcong}\ x{\isachardoublequoteclose}%
\isadelimproof
\ %
\endisadelimproof
\isatagproof
\isacommand{using}\isamarkupfalse%
\ B{\isadigit{0}}a\ B{\isadigit{3}}a\ \isacommand{by}\isamarkupfalse%
\ blast%
\endisatagproof
{\isafoldproof}%
\isadelimproof
\endisadelimproof
\ \ \isanewline
\ \ \isacommand{lemma}\isamarkupfalse%
\ S{\isadigit{6}}FromVIII{\isacharcolon}\ {\isachardoublequoteopen}{\isacharparenleft}cod\ y{\isacharparenright}{\isasymcdot}y\ {\isasymcong}\ y{\isachardoublequoteclose}%
\isadelimproof
\ %
\endisadelimproof
\isatagproof
\isacommand{using}\isamarkupfalse%
\ B{\isadigit{0}}a\ B{\isadigit{3}}b\ \isacommand{by}\isamarkupfalse%
\ blast%
\endisatagproof
{\isafoldproof}%
\isadelimproof
\endisadelimproof
\begin{isamarkuptext}%
Vive versa, Axiom Set V implies Axiom Set VIII. Hence, both theories are equivalent.%
\end{isamarkuptext}\isamarkuptrue%
\ S{\isadigit{1}}{\isacharcolon}\ %
\isamarkupcmt{\makebox[2cm][l]{Strictness:}%
}
\ {\isachardoublequoteopen}E{\isacharparenleft}dom\ x{\isacharparenright}\ \isactrlbold {\isasymrightarrow}\ E\ x{\isachardoublequoteclose}\ \isakeyword{and}\isanewline
\ S{\isadigit{2}}{\isacharcolon}\ %
\isamarkupcmt{\makebox[2cm][l]{Strictness:}%
}
\ {\isachardoublequoteopen}E{\isacharparenleft}cod\ y{\isacharparenright}\ \isactrlbold {\isasymrightarrow}\ E\ y{\isachardoublequoteclose}\ \isakeyword{and}\isanewline
\ S{\isadigit{3}}{\isacharcolon}\ %
\isamarkupcmt{\makebox[2cm][l]{Existence:}%
}
\ {\isachardoublequoteopen}E{\isacharparenleft}x{\isasymcdot}y{\isacharparenright}\ \isactrlbold {\isasymleftrightarrow}\ dom\ x\ {\isasymsimeq}\ cod\ y{\isachardoublequoteclose}\ \isakeyword{and}\ \isanewline
\ S{\isadigit{4}}{\isacharcolon}\ %
\isamarkupcmt{\makebox[2cm][l]{Associativity:}%
}
\ {\isachardoublequoteopen}x{\isasymcdot}{\isacharparenleft}y{\isasymcdot}z{\isacharparenright}\ {\isasymcong}\ {\isacharparenleft}x{\isasymcdot}y{\isacharparenright}{\isasymcdot}z{\isachardoublequoteclose}\ \isakeyword{and}\isanewline
\ S{\isadigit{5}}{\isacharcolon}\ %
\isamarkupcmt{\makebox[2cm][l]{Domain:}%
}
\ {\isachardoublequoteopen}x{\isasymcdot}{\isacharparenleft}dom\ x{\isacharparenright}\ {\isasymcong}\ x{\isachardoublequoteclose}\ \isakeyword{and}\isanewline
\ S{\isadigit{6}}{\isacharcolon}\ %
\isamarkupcmt{\makebox[2cm][l]{Codomain:}%
}
\ {\isachardoublequoteopen}{\isacharparenleft}cod\ y{\isacharparenright}{\isasymcdot}y\ {\isasymcong}\ y{\isachardoublequoteclose}\ \isanewline
\isanewline
\ \ \isacommand{lemma}\isamarkupfalse%
\ B{\isadigit{0}}a{\isacharcolon}\ {\isachardoublequoteopen}E{\isacharparenleft}x{\isasymcdot}y{\isacharparenright}\ \isactrlbold {\isasymrightarrow}\ {\isacharparenleft}E\ x\ \isactrlbold {\isasymand}\ E\ y{\isacharparenright}{\isachardoublequoteclose}%
\isadelimproof
\ %
\endisadelimproof
\isatagproof
\isacommand{using}\isamarkupfalse%
\ S{\isadigit{1}}\ S{\isadigit{2}}\ S{\isadigit{3}}\ \isacommand{by}\isamarkupfalse%
\ blast%
\endisatagproof
{\isafoldproof}%
\isadelimproof
\endisadelimproof
\isanewline
\ \ \isacommand{lemma}\isamarkupfalse%
\ B{\isadigit{0}}b{\isacharcolon}\ {\isachardoublequoteopen}E{\isacharparenleft}dom\ x{\isacharparenright}\ \isactrlbold {\isasymrightarrow}\ E\ x{\isachardoublequoteclose}%
\isadelimproof
\ %
\endisadelimproof
\isatagproof
\isacommand{using}\isamarkupfalse%
\ S{\isadigit{1}}\ \isacommand{by}\isamarkupfalse%
\ blast%
\endisatagproof
{\isafoldproof}%
\isadelimproof
\endisadelimproof
\isanewline
\ \ \isacommand{lemma}\isamarkupfalse%
\ B{\isadigit{0}}c{\isacharcolon}\ {\isachardoublequoteopen}E{\isacharparenleft}cod\ x{\isacharparenright}\ \isactrlbold {\isasymrightarrow}\ E\ x{\isachardoublequoteclose}%
\isadelimproof
\ %
\endisadelimproof
\isatagproof
\isacommand{using}\isamarkupfalse%
\ S{\isadigit{2}}\ \isacommand{by}\isamarkupfalse%
\ blast%
\endisatagproof
{\isafoldproof}%
\isadelimproof
\endisadelimproof
\isanewline
\ \ \isacommand{lemma}\isamarkupfalse%
\ \ B{\isadigit{1}}{\isacharcolon}\ {\isachardoublequoteopen}\isactrlbold {\isasymforall}x{\isachardot}\isactrlbold {\isasymforall}y{\isachardot}\ E{\isacharparenleft}x{\isasymcdot}y{\isacharparenright}\ \isactrlbold {\isasymleftrightarrow}\ dom\ x\ {\isasymcong}\ cod\ y{\isachardoublequoteclose}%
\isadelimproof
\ %
\endisadelimproof
\isatagproof
\isacommand{by}\isamarkupfalse%
\ {\isacharparenleft}metis\ S{\isadigit{3}}\ S{\isadigit{5}}{\isacharparenright}%
\endisatagproof
{\isafoldproof}%
\isadelimproof
\endisadelimproof
\isanewline
\ \ \isacommand{lemma}\isamarkupfalse%
\ B{\isadigit{2}}a{\isacharcolon}\ {\isachardoublequoteopen}\isactrlbold {\isasymforall}x{\isachardot}\ cod{\isacharparenleft}dom\ x{\isacharparenright}\ {\isasymcong}\ dom\ x\ {\isachardoublequoteclose}%
\isadelimproof
\ %
\endisadelimproof
\isatagproof
\isacommand{by}\isamarkupfalse%
\ {\isacharparenleft}metis\ S{\isadigit{3}}\ S{\isadigit{5}}{\isacharparenright}%
\endisatagproof
{\isafoldproof}%
\isadelimproof
\endisadelimproof
\isanewline
\ \ \isacommand{lemma}\isamarkupfalse%
\ B{\isadigit{2}}b{\isacharcolon}\ {\isachardoublequoteopen}\isactrlbold {\isasymforall}y{\isachardot}\ dom{\isacharparenleft}cod\ y{\isacharparenright}\ {\isasymcong}\ cod\ y{\isachardoublequoteclose}%
\isadelimproof
\ %
\endisadelimproof
\isatagproof
\isacommand{by}\isamarkupfalse%
\ {\isacharparenleft}metis\ S{\isadigit{3}}\ S{\isadigit{6}}{\isacharparenright}%
\endisatagproof
{\isafoldproof}%
\isadelimproof
\endisadelimproof
\ \ \isanewline
\ \ \isacommand{lemma}\isamarkupfalse%
\ B{\isadigit{3}}a{\isacharcolon}\ {\isachardoublequoteopen}\isactrlbold {\isasymforall}x{\isachardot}\ x{\isasymcdot}{\isacharparenleft}dom\ x{\isacharparenright}\ {\isasymcong}\ x{\isachardoublequoteclose}%
\isadelimproof
\ %
\endisadelimproof
\isatagproof
\isacommand{using}\isamarkupfalse%
\ S{\isadigit{5}}\ \isacommand{by}\isamarkupfalse%
\ auto%
\endisatagproof
{\isafoldproof}%
\isadelimproof
\endisadelimproof
\isanewline
\ \ \isacommand{lemma}\isamarkupfalse%
\ B{\isadigit{3}}b{\isacharcolon}\ {\isachardoublequoteopen}\isactrlbold {\isasymforall}y{\isachardot}\ {\isacharparenleft}cod\ y{\isacharparenright}{\isasymcdot}y\ {\isasymcong}\ y{\isachardoublequoteclose}%
\isadelimproof
\ %
\endisadelimproof
\isatagproof
\isacommand{using}\isamarkupfalse%
\ S{\isadigit{6}}\ \isacommand{by}\isamarkupfalse%
\ blast%
\endisatagproof
{\isafoldproof}%
\isadelimproof
\endisadelimproof
\isanewline
\ \ \isacommand{lemma}\isamarkupfalse%
\ B{\isadigit{4}}a{\isacharcolon}\ {\isachardoublequoteopen}\isactrlbold {\isasymforall}x{\isachardot}\isactrlbold {\isasymforall}y{\isachardot}\ dom{\isacharparenleft}x{\isasymcdot}y{\isacharparenright}\ {\isasymcong}\ dom{\isacharparenleft}{\isacharparenleft}dom\ x{\isacharparenright}{\isasymcdot}y{\isacharparenright}{\isachardoublequoteclose}%
\isadelimproof
\ %
\endisadelimproof
\isatagproof
\isacommand{by}\isamarkupfalse%
\ {\isacharparenleft}metis\ S{\isadigit{1}}\ S{\isadigit{3}}\ S{\isadigit{4}}\ S{\isadigit{5}}{\isacharparenright}%
\endisatagproof
{\isafoldproof}%
\isadelimproof
\endisadelimproof
\isanewline
\ \ \isacommand{lemma}\isamarkupfalse%
\ B{\isadigit{4}}b{\isacharcolon}\ {\isachardoublequoteopen}\isactrlbold {\isasymforall}x{\isachardot}\isactrlbold {\isasymforall}y{\isachardot}\ cod{\isacharparenleft}x{\isasymcdot}y{\isacharparenright}\ {\isasymcong}\ cod{\isacharparenleft}x{\isasymcdot}{\isacharparenleft}cod\ y{\isacharparenright}{\isacharparenright}{\isachardoublequoteclose}%
\isadelimproof
\ %
\endisadelimproof
\isatagproof
\isacommand{by}\isamarkupfalse%
\ {\isacharparenleft}metis\ S{\isadigit{2}}\ S{\isadigit{3}}\ S{\isadigit{4}}\ S{\isadigit{6}}{\isacharparenright}%
\endisatagproof
{\isafoldproof}%
\isadelimproof
\endisadelimproof
\isanewline
\ \ \isacommand{lemma}\isamarkupfalse%
\ \ B{\isadigit{5}}{\isacharcolon}\ {\isachardoublequoteopen}\isactrlbold {\isasymforall}x{\isachardot}\isactrlbold {\isasymforall}y{\isachardot}\isactrlbold {\isasymforall}z{\isachardot}\ x{\isasymcdot}{\isacharparenleft}y{\isasymcdot}z{\isacharparenright}\ {\isasymcong}\ {\isacharparenleft}x{\isasymcdot}y{\isacharparenright}{\isasymcdot}z{\isachardoublequoteclose}%
\isadelimproof
\ %
\endisadelimproof
\isatagproof
\isacommand{using}\isamarkupfalse%
\ S{\isadigit{4}}\ \isacommand{by}\isamarkupfalse%
\ blast%
\endisatagproof
{\isafoldproof}%
\isadelimproof
\endisadelimproof
\begin{isamarkuptext}%
Axiom Set VIII is redundant (as expected from previous observations).
The theorem provers quickly confirm that axioms \isa{B{\isadigit{2}}a{\isacharcomma}\ B{\isadigit{2}}b{\isacharcomma}\ B{\isadigit{4}}a{\isacharcomma}\ B{\isadigit{4}}b} are implied.%
\end{isamarkuptext}\isamarkuptrue%
\ B{\isadigit{0}}a{\isacharcolon}\ {\isachardoublequoteopen}E{\isacharparenleft}x{\isasymcdot}y{\isacharparenright}\ \isactrlbold {\isasymrightarrow}\ {\isacharparenleft}E\ x\ \isactrlbold {\isasymand}\ E\ y{\isacharparenright}{\isachardoublequoteclose}\ \isakeyword{and}\isanewline
\ B{\isadigit{0}}b{\isacharcolon}\ {\isachardoublequoteopen}E{\isacharparenleft}dom\ x{\isacharparenright}\ \isactrlbold {\isasymrightarrow}\ E\ x{\isachardoublequoteclose}\ \isakeyword{and}\isanewline
\ B{\isadigit{0}}c{\isacharcolon}\ {\isachardoublequoteopen}E{\isacharparenleft}cod\ x{\isacharparenright}\ \isactrlbold {\isasymrightarrow}\ E\ x{\isachardoublequoteclose}\ \isakeyword{and}\isanewline
\ \ B{\isadigit{1}}{\isacharcolon}\ {\isachardoublequoteopen}\isactrlbold {\isasymforall}x{\isachardot}\isactrlbold {\isasymforall}y{\isachardot}\ E{\isacharparenleft}x{\isasymcdot}y{\isacharparenright}\ \isactrlbold {\isasymleftrightarrow}\ dom\ x\ {\isasymcong}\ cod\ y{\isachardoublequoteclose}\ \isakeyword{and}\isanewline
\ B{\isadigit{3}}a{\isacharcolon}\ {\isachardoublequoteopen}\isactrlbold {\isasymforall}x{\isachardot}\ x{\isasymcdot}{\isacharparenleft}dom\ x{\isacharparenright}\ {\isasymcong}\ x{\isachardoublequoteclose}\ \isakeyword{and}\ \isanewline
\ B{\isadigit{3}}b{\isacharcolon}\ {\isachardoublequoteopen}\isactrlbold {\isasymforall}y{\isachardot}\ {\isacharparenleft}cod\ y{\isacharparenright}{\isasymcdot}y\ {\isasymcong}\ y{\isachardoublequoteclose}\ \isakeyword{and}\ \isanewline
\ \ B{\isadigit{5}}{\isacharcolon}\ {\isachardoublequoteopen}\isactrlbold {\isasymforall}x{\isachardot}\isactrlbold {\isasymforall}y{\isachardot}\isactrlbold {\isasymforall}z{\isachardot}\ x{\isasymcdot}{\isacharparenleft}y{\isasymcdot}z{\isacharparenright}\ {\isasymcong}\ {\isacharparenleft}x{\isasymcdot}y{\isacharparenright}{\isasymcdot}z{\isachardoublequoteclose}\ \ \isanewline
\isanewline
\ \ \isacommand{lemma}\isamarkupfalse%
\ B{\isadigit{2}}aRedundant{\isacharcolon}\ {\isachardoublequoteopen}\isactrlbold {\isasymforall}x{\isachardot}\ cod{\isacharparenleft}dom\ x{\isacharparenright}\ {\isasymcong}\ dom\ x\ {\isachardoublequoteclose}%
\isadelimproof
\ %
\endisadelimproof
\isatagproof
\isacommand{by}\isamarkupfalse%
\ {\isacharparenleft}metis\ B{\isadigit{0}}a\ B{\isadigit{1}}\ B{\isadigit{3}}a{\isacharparenright}%
\endisatagproof
{\isafoldproof}%
\isadelimproof
\endisadelimproof
\ \isanewline
\ \ \isacommand{lemma}\isamarkupfalse%
\ B{\isadigit{2}}bRedundant{\isacharcolon}\ {\isachardoublequoteopen}\isactrlbold {\isasymforall}y{\isachardot}\ dom{\isacharparenleft}cod\ y{\isacharparenright}\ {\isasymcong}\ cod\ y{\isachardoublequoteclose}%
\isadelimproof
\ %
\endisadelimproof
\isatagproof
\isacommand{by}\isamarkupfalse%
\ {\isacharparenleft}metis\ B{\isadigit{0}}a\ B{\isadigit{1}}\ B{\isadigit{3}}b{\isacharparenright}%
\endisatagproof
{\isafoldproof}%
\isadelimproof
\endisadelimproof
\ \isanewline
\ \ \isacommand{lemma}\isamarkupfalse%
\ B{\isadigit{4}}aRedundant{\isacharcolon}\ {\isachardoublequoteopen}\isactrlbold {\isasymforall}x{\isachardot}\isactrlbold {\isasymforall}y{\isachardot}\ dom{\isacharparenleft}x{\isasymcdot}y{\isacharparenright}\ {\isasymcong}\ dom{\isacharparenleft}{\isacharparenleft}dom\ x{\isacharparenright}{\isasymcdot}y{\isacharparenright}{\isachardoublequoteclose}%
\isadelimproof
\ %
\endisadelimproof
\isatagproof
\isacommand{by}\isamarkupfalse%
\ {\isacharparenleft}metis\ B{\isadigit{0}}a\ B{\isadigit{0}}b\ B{\isadigit{1}}\ B{\isadigit{3}}a\ B{\isadigit{5}}{\isacharparenright}%
\endisatagproof
{\isafoldproof}%
\isadelimproof
\endisadelimproof
\ \isanewline
\ \ \isacommand{lemma}\isamarkupfalse%
\ B{\isadigit{4}}bRedundant{\isacharcolon}\ {\isachardoublequoteopen}\isactrlbold {\isasymforall}x{\isachardot}\isactrlbold {\isasymforall}y{\isachardot}\ cod{\isacharparenleft}x{\isasymcdot}y{\isacharparenright}\ {\isasymcong}\ cod{\isacharparenleft}x{\isasymcdot}{\isacharparenleft}cod\ y{\isacharparenright}{\isacharparenright}{\isachardoublequoteclose}%
\isadelimproof
\ %
\endisadelimproof
\isatagproof
\isacommand{by}\isamarkupfalse%
\ {\isacharparenleft}metis\ B{\isadigit{0}}a\ B{\isadigit{0}}c\ B{\isadigit{1}}\ B{\isadigit{3}}b\ B{\isadigit{5}}{\isacharparenright}%
\endisatagproof
{\isafoldproof}%
\isadelimproof
\endisadelimproof
\begin{isamarkuptext}%
Again, note the relation and similarity of the reduced Axiom Set VIII to Axiom Set V by Scott, 
which we prefer, since it avoids a mixed use of free and bound variables in the encoding and 
since it is smaller.%
\end{isamarkuptext}\isamarkuptrue%
\isamarkupparagraph{Acknowledgements%
}
\isamarkuptrue%
\begin{isamarkuptext}%
We thank G\"unter Rote and Lutz Schr\"oder for their valuable comments to earlier drafts of this paper.%
\end{isamarkuptext}\isamarkuptrue%
\isadelimtheory
\endisadelimtheory
\isatagtheory
\endisatagtheory
{\isafoldtheory}%
\isadelimtheory
\endisadelimtheory
\end{isabellebody}%
%%% Local Variables:
%%% mode: latex
%%% TeX-master: "root"
%%% End: